\documentclass[manuscript]{acmart}
\usepackage{hyperref}
\usepackage{hyperxmp}
\usepackage{booktabs}
\usepackage{subcaption}
\usepackage{graphicx}
\usepackage{dblfloatfix}
\usepackage{algorithm}
\usepackage{algpseudocode}
\usepackage{array}
\usepackage{multirow}
\usepackage{dialogue}
\usepackage{listings}
\usepackage[T1]{fontenc}
\usepackage[utf8]{inputenc}

\AtBeginDocument{%
  \providecommand\BibTeX{{%
    \normalfont B\kern-0.5em{\scshape i\kern-0.25em b}\kern-0.8em\TeX}}}

\setcopyright{acmcopyright}
\copyrightyear{2018}
\acmYear{2018}
\acmDOI{XXXXXXX.XXXXXXX}

\acmConference[Conference acronym 'XX]{Make sure to enter the correct
  conference title from your rights confirmation emai}{June 03--05,
  2018}{Woodstock, NY}
\acmBooktitle{Woodstock '18: ACM Symposium on Neural Gaze Detection,
 June 03--05, 2018, Woodstock, NY} 
\acmPrice{15.00}
\acmISBN{978-1-4503-XXXX-X/18/06}

\begin{document}

\title[CA+: Cognition Augmented Counselor Agent Framework]{CA+: Cognition Augmented Counselor Agent Framework for Long-term Dynamic Client Engagement}

\author{Yuanrong Tang}
\email{tyr24@mails.tsinghua.edu.cn}
\orcid{0009-0008-9714-4321}
\affiliation{%
  \institution{Tsinghua University, Institute for AI Industry Research}
  \city{Beijing}
  \country{China}
}

\author{Yu Kang}
\email{christokyu@gmail.com}
\orcid{0009-0006-9688-4622}
\affiliation{%
  \institution{Tsinghua University, Institute for AI Industry Research}
  \city{Beijing}
  \country{China}
}

\author{Yifan Wang}
\orcid{0009-0009-2544-0479}
\affiliation{%
  \institution{Tsinghua University, Institute for AI Industry Research}
  \city{Beijing}
  \country{China}
}
\email{wyf2002002@gmail.com}

\author{Tianhong Wang}
\orcid{0000-0002-9008-1582}
\affiliation{%
  \institution{Tsinghua University, Institute for AI Industry Research}
  \city{Beijing}
  \country{China}
}
\email{zxhydlx@163.com}

\author{Chen Zhong}
\orcid{0009-0001-7503-3064}
\affiliation{%
  \institution{Tsinghua University, Institute for AI Industry Research}
  \city{Beijing}
  \country{China}
}
\email{myheimu@gmail.com}

\author{Jiangtao Gong}
\authornote{Corresponding author.}

\orcid{0000-0002-4310-1894}
\affiliation{%
  \institution{Tsinghua University, Institute for AI Industry Research}
  \city{Beijing}
  \country{China}
}
\email{gongjiangtao2@gmail.com}

\renewcommand{\shortauthors}{Yuanrong Tang, et al.}

% \begin{abstract}
%   The global mental health crisis demands innovative solutions. AI-based psychological counseling has emerged as a potential solution, but low user engagement and retention hinder its effectiveness. We propose CA+, a Cognition Augmented Counselor Agent framework that incorporates embodied theory, cognitive model, and co-creation with expert counselors to achieve long-term dynamic engagement. The framework introduces a hierarchical planning and embodied reasoning workflow, enabling structured, deep understanding, and contextually relevant responses. We employ a co-creation approach to align expert knowledge and ensure professionally compliant counseling ideas. Additionally, we design a multi-turn conversation user attention retention mechanism utilizing dynamic psychological monitoring, implicit client profiles, and personalized recommendations. Two empirical studies demonstrate the framework's effectiveness in user retention, perceived helpfulness, and adherence to professional counseling standards, as evaluated by clients and licensed counselors. This research contributes to the development of more engaging and effective AI-based mental health counseling systems.
% \end{abstract}

\begin{abstract}
% Current psychological counseling systems struggle with long-term client engagement due to limited multi-turn contextual understanding and evolving support. We propose CA+, a \underline{\textbf{C}}ognition \underline{\textbf{A}}ugmented \underline{\textbf{C}}ounselor \underline{\textbf{A}}gent framework to enhance client engagement the integration of human embodied cognition theory. This novel framework comprises three key components: 
% (1) Therapy Strategies Module: Implements hierarchical Goals-Session-Action planning with bidirectional adaptation based on client feedback;
% (2) Communication Form Module: Orchestrates parallel guidance and empathy pathways for balanced therapeutic progress and emotional resonance;
% (3) Information Management: Utilizes client profile and therapeutic knowledge databases for dynamic, context-aware interventions.
%(1) Cognition Augmented Embodied Reasoning, which enhances contextual understanding through a hierarchical cognitive model; (2) Dynamically Aligned Expertise, which ensures professional counseling standards by integrating domain knowledge and standard operating procedures; and (3) Multi-Turn Dialogic Engagement, which maintains client engagement through psychological monitoring and personalization.
%CA+ enables complex dialogue comprehension and evolving responses over multiple sessions. 

Current AI counseling systems struggle with maintaining effective long-term client engagement. Through formative research with counselors and a systematic literature review, we identified five key design considerations for AI counseling interactions. Based on these insights, we propose CA+, a \underline{\textbf{C}}ognition \underline{\textbf{A}}ugmented counselor framework enhancing contextual understanding through three components: 
%a hierarchical Therapy Strategies Module with bidirectional planning, a Communication Form Module integrating guidance and empathy pathways, and a comprehensive Information Management system. 
(1) Therapy Strategies Module: Implements hierarchical Goals-Session-Action planning with bidirectional adaptation based on client feedback;
(2) Communication Form Module: Orchestrates parallel guidance and empathy pathways for balanced therapeutic progress and emotional resonance;
(3) Information Management: Utilizes client profile and therapeutic knowledge databases for dynamic, context-aware interventions.
A three-day longitudinal study with 24 clients demonstrates CA+'s significant improvements in client engagement, perceived empathy, and overall satisfaction compared to a baseline system. Besides, two licensed counselors confirm its high professionalism. 
Our research demonstrates the potential for enhancing LLM engagement in psychological counseling dialogues through cognitive theory, which may inspire further innovations in computational interaction in the future.

\end{abstract}




\begin{teaserfigure}
  \includegraphics[width=\textwidth]{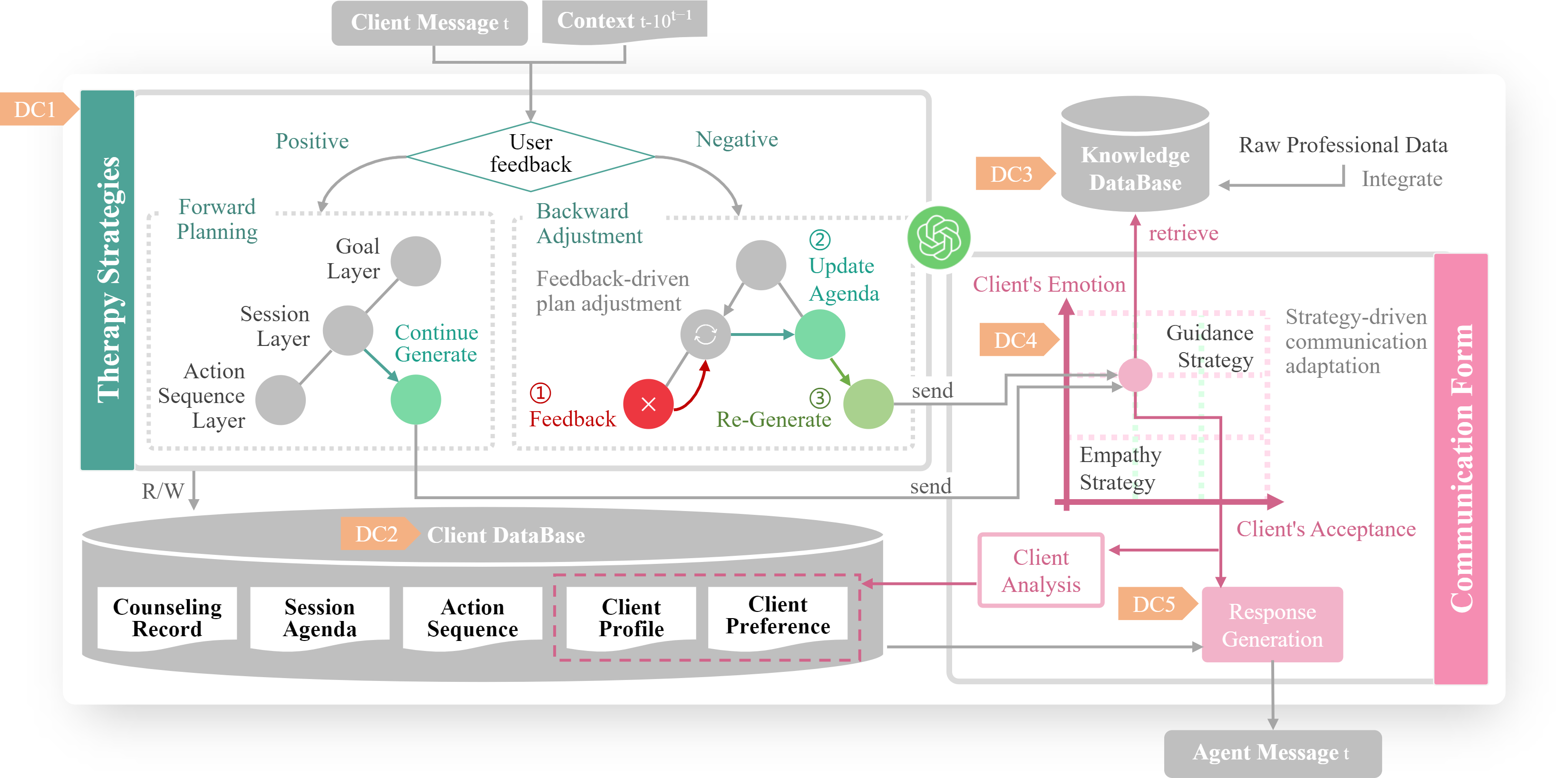}
  \caption{\textbf{Cognition Augmented Engaging AI Counselor Agent Framework (CA+).} The framework introduces recursive self-evolving decision-making for LLM-based counseling through synergistic Therapy Strategies and Communication Form modules, continuously adapting therapeutic approaches via real-time feedback.}
  \Description{The diagram outlines a comprehensive AI framework designed to augment counseling through cognitive and emotional intelligence. It consists of three main components: Multi-Turn Dialogic Engagement, Cognition Augmented Reasoning, and Dynamically Aligned Expertise. Multi-Turn Dialogic Engagement: This component focuses on continuous interaction, accommodating both long-term preferences and short-term changes through adaptive empathy and personalized strategies, considering the client's emotions and attitudes. Cognition Augmented Reasoning: Integrates user inputs and conceptualizes them to plan and execute appropriate actions. It employs a bi-level reflective mechanism for optimizing responses based on session context, client needs, and therapy goals. Dynamically Aligned Expertise: Specializes in conceptualizing the client's case, forming explanatory hypotheses, and developing treatment plans. It ensures that therapeutic techniques and problem-solving tips are dynamically integrated with professional data to guide intervention strategies.}
  \label{fig:teaser}
\end{teaserfigure}

\received{20 February 2007}
\received[revised]{12 March 2009}
\received[accepted]{5 June 2009}

\maketitle
\section{Introduction}

Artificial intelligence (AI) powered mental health counseling systems have, in recent years, emerged as a promising solution to address the global shortage of mental health professionals, offering more accessible psychological support options \cite{shum2018eliza, prochaska2021therapeutic}. Multiple studies have shown that these systems can extend service reach to previously underserved populations and reduce barriers to initiating help-seeking behaviors. Building on these advancements, improvements in natural language processing, especially with the advent of large language models (LLMs) have enhanced dialogue capabilities, enabling more nuanced emotional responses and better context handling in therapeutic interactions~\cite{chen2024structured, algherairy2025prompting}. For instance, these systems now demonstrate improved abilities in recognizing emotional patterns and maintaining conversation flow in ways that earlier rule-based approaches could not, thereby facilitating more natural and effective counseling sessions.

Despite these technological gains, a fundamental challenge remains: sustaining long-term client\footnote{In psychological counseling, a client is an individual seeking professional mental health support or guidance to address personal, emotional, or behavioral concerns.} engagement. Multiple studies report retention rates falling between 20\% and 52\%, which often result in treatment discontinuation and diminished therapeutic outcomes \cite{gual2022using, burns2011harnessing}.

Recent innovations in both research and commercial domains are actively addressing the engagement gap—the challenge of maintaining continuous, personalized, and professional interactions over extended periods. On the research front, models like PsycoLLM \cite{hu2024psycollm} enhance cognitive and emotional analysis through clinical dialogue data, directly promoting sustained client engagement. Likewise, MemoryBank employs advanced long-term memory mechanisms for ongoing personalization across sessions; Todak leverages multi-modal feedback (verbal and non-verbal) to preserve therapeutic coherence; and Trust through Words quantifies how system interactions affect user understanding and the therapeutic alliance.

In the commercial arena, platforms are scaling key elements of professional counseling to boost engagement. For example, Woebot Health\footnote{\url{https://woebothealth.com/}} positions itself as an “ally that's with you through it all,” using evidence-based CBT techniques and emphasizing relationship continuity to build lasting therapeutic bonds. Manifest\footnote{\url{https://www.manifestapp.xyz/}} reimagines formal self-care as daily challenges that encourage steady participation, while Hume AI\footnote{\url{https://dev.hume.ai/intro}} \cite{brooks2023deep} employs its empathic voice interface to capture subtle emotional cues, enriching therapeutic assessments and deepening overall engagement.

Parallel to these advances, the human-computer interaction (HCI) community has made significant strides in addressing various aspects of this engagement challenge in AI-based mental health interventions. Research has explored user acceptance and interface design to improve initial uptake and usability \cite{chandrashekar2018mental, 10.1145/3555146}. Studies have demonstrated that these systems can provide basic psychological support and have the potential to reach underserved populations, potentially increasing engagement through improved accessibility \cite{vaidyam2019chatbots}. Investigations have also shown that AI systems are capable of handling basic multi-turn interactions, offering a foundation for more engaging and complex counseling scenarios \cite{fitzpatrick2017delivering}.

% Effective counseling requires sustained contextual awareness and adaptive support across multiple sessions—aspects that current AI systems struggle to maintain consistently \cite{chandrashekar2018mental}. Existing systems often fail to provide the depth and continuity required for effective long-term psychological support \cite{miner2016smartphone}.
Even with these considerable advances, maintaining long-term client engagement remains a critical and unresolved issue in the HCI field. A significant limitation is that current AI systems struggle to maintain the sustained contextual awareness and adaptive support required for effective counseling across multiple sessions, which are vital for addressing the evolving needs of clients \cite{vaidyam2019chatbots, chandrashekar2018mental, miner2016smartphone}. Although LLMs have demonstrated remarkable performance in single-turn, task-oriented generation, they still fall short of replicating the adaptive, multi-turn dialogue management observed in human counseling sessions \cite{wang2023large}.

Therefore, to address the identified limitations in current AI counseling systems, we conducted a comprehensive preparatory investigation combining practical expertise with theoretical foundations (Section \ref{sec:Formative_Research}). Our research followed two complementary paths: a formative study involving interviews with human counselors to capture implicit professional knowledge, and a thematic review of AI counseling systems examining technical approaches and research gaps. This dual approach yielded five critical design considerations that guided our system development: (1) Cognition Augmented Hierarchical Thinking, (2) Conceptualization-Driven Implicit Client Profiling, (3) Book-style Data Generation and Retrieval, (4) Adaptive Empathy and Ecological Self, and (5) Adapting to AI-Specific Client Expectations.

Based on these design considerations, we propose CA+, a \underline{\textbf{C}}ognition \underline{\textbf{A}}ugmented \underline{\textbf{C}}ounselor \underline{\textbf{A}}gent framework (Section \ref{sec:framework}). CA+ introduces a novel cognition-augmented reasoning approach, significantly enhancing the LLM-based agent counselor's ability to maintain multi-turn contextual awareness and provide engaging, personalized, and professionally aligned support over extended periods. 
We evaluated our proposed framework through two comprehensive studies (Section \ref{sec:study1} \& \ref{sec:study2}): a three-day, multi-turn interaction experiment with 24 clients comparing CA+ to a baseline system, and an assessment by 2 licensed counselors evaluating the system's professionalism and adherence to counseling standards.
Our evaluations show that CA+ significantly outperforms baseline systems across counseling quality metrics, client engagement indicators, and user feedback.
CA+ framework comprises three innovative components: (1) \textbf{Therapy Strategies Module}, which implements a hierarchical three-layer structure with bidirectional planning mechanisms for dynamic therapeutic adaptation; (2) \textbf{Communication Form Module}, which orchestrates strategy-driven communication by integrating guidance and empathy pathways to ensure both therapeutic effectiveness and emotional resonance; and (3) \textbf{Information Management and Storage}, which maintains comprehensive client records and structured therapeutic knowledge for efficient real-time retrieval during counseling sessions. 
% We evaluated our proposed framework through two comprehensive studies: a three-day, multi-turn interaction experiment with 24 clients comparing CA+ to a baseline system, and an assessment by 2 licensed counselors evaluating the system's professionalism and adherence to counseling standards.

In summary, the contribution of this paper is as follows:
\begin{enumerate}
    \item CA+, a novel LLM-based agent counselor framework employing cognition augmented reasoning to enhance multi-turn contextual awareness and provide personalized, professionally aligned mental health support.
    \item A controlled experimental study demonstrating significant improvements in client engagement and satisfaction, with quantitative metrics showing a 95.4\% increase in average session length (from 14.38 to 28.10 minutes, $p < 0.001$) and a 46.7\% improvement in overall counseling relationship ratings compared to a baseline system.
    \item A qualitative assessment by licensed counselors validating the CA+ system's adherence to professional counseling standards.
    \item Design implications and guidelines for the integration of LLMs into mental health support systems, based on comprehensive system development and empirical findings.
\end{enumerate}

\section{Related Work}
\label{sec:rw}

The following sections review existing literature on AI-assisted mental health interventions (Section \ref{sec:rw_evolution}), examining their evolution, user engagement challenges (Section \ref{sec:engage}), and approaches for improving therapeutic effectiveness (Section \ref{sec:rw_engage}).

% \subsection{Current Challenges in AI-Assisted Mental Health Interventions}
\subsection{Evolution and Promise of AI-Assisted Mental Health Interventions}

\label{sec:rw_evolution}

AI counselors can be broadly defined as computational systems that engage in dialogue with users to offer psychological support. Early efforts in this domain relied on rule-based systems that used predefined scripts and decision trees to simulate counseling sessions \cite{ABDALRAZAQ2019103978}. For example, systems such as ELIZA, Woebot, TeenChat, and LOUISE leveraged structured conversational flows to deliver therapeutic interventions and emotional support \cite{weizenbaum1966eliza, fitzpatrick2017delivering, 10.1007/978-3-319-19156-0_14, Wargnier2018LOUISE}. These systems were limited in their flexibility and adaptability, which sometimes resulted in responses that lacked the nuance and empathy required for effective therapy. \cite{vaidyam2019chatbots}. 

The development of AI counselors underwent a significant shift with the introduction of machine learning, particularly deep learning, enabling models to better understand and generate human-like language \cite{devlin2019bert}. Transformer-based models, such as GPT (Generative Pre-trained Transformer), further pushed these capabilities by being trained on vast amounts of text data, allowing them to generate contextually relevant and coherent responses suited for dynamic counseling conversations \cite{brown2020language}. LLMs, like GPT-3, have demonstrated an ability to engage in extended, empathetic dialogues, offering personalized mental health support. 
% Systems such as Tess, Youper, and Replika exemplify the application of LLMs in AI counseling \cite{fulmer2018using, Mehta2021Youper}.

Despite their promise, current research also highlights significant limitations when compared with human counselors \cite{luxton2014artificial,miner2017talking}. While these AI systems can generate contextually relevant responses, they lack the robustness, reliability, and emotional intelligence of human professionals \cite{miner2017talking,brown2020language}. These models struggle with maintaining coherent long-term context across extended interactions and often produce inconsistent or inappropriate responses \cite{brown2020language,vaidyam2019chatbots}. Although their outputs may seem plausible on the surface, they frequently lack the depth and sensitivity needed to address complex emotional issues \cite{devlin2019bert}. Despite their promise, these technologies continue to fall short of the comprehensive capabilities exhibited by human counselors. While AI counseling technology has made significant advances, it remains largely in the research and development phase and is not currently ready for widespread practical implementation in mental health settings \cite{luxton2014artificial,vaidyam2019chatbots}. These limitations underscore the need for continued research and development to bridge the gap between AI capabilities and the nuanced requirements of effective mental health counseling. For example, improving long-term context retention and emotional responsiveness

\subsection{User Engagement in AI Psychological Therapy}
\label{sec:engage}

Engagement is crucial for successful psychological therapy, influencing the therapeutic relationship, treatment adherence, and outcomes \cite{holdsworth2014client, tetley2011systematic, thompson2007treatment}. However, AI-based psychological counseling systems often struggle with low user engagement, as evident in the low retention rates of popular AI chatbots like Woebot and Wysa \cite{fitzpatrick2017delivering, inkster2018empathy}.

Despite the rapid advancement of large language models (LLMs), contemporary AI counseling systems face persistent engagement challenges. A significant limitation is their inability to develop a deep, nuanced understanding of users' experiences over time. LLM-based applications often produce responses that appear relevant on the surface yet remain shallow, failing to address the root causes of users' problems \cite{horvath1991relation}. Furthermore, their stateless design prevents them from effectively tracking the dynamic evolution of user states across interactions, a critical shortcoming in maintaining continuity in therapy \cite{greenberg1986psychotherapeutic}. This compounds the difficulty in building effective therapeutic relationships.

The professional competency of AI counselors also remains a concern. Current systems struggle to meaningfully apply psychological theories and therapeutic procedures when addressing mental health difficulties, often resulting in generic and superficial solutions—such as standardized responses that do not account for individual emotional nuances \cite{miner2017talking, zhang2022natural}.
Moreover, the challenge of effectively integrating specialized domain knowledge with AI systems \cite{wang2023large} contributes to users' lack of trust and reduced adherence to therapeutic suggestions.

Emotional connection, a fundamental driver of behavioral change in human psychology \cite{ryan2000self, elliott2018therapist}, represents another area where AI counselors face substantial limitations. Users often perceive AI assistance as inauthentic due to the system's inability to connect through shared experiences or genuine empathy \cite{liu2018should}. Without honest self-disclosure, a core element of human counseling relationships, AI systems often fail to establish deep trust, resulting in users remaining guarded during therapeutic interactions \cite{collins1994self, sannon2018personification}.

Furthermore, the motivational impact of AI counselors is often weakened by a one-dimensional approach that relies on standardized, non-personalized feedback. Research indicates that individualized motivational feedback significantly enhances user loyalty in AI-based interventions \cite{kocielnik2018reflection}, yet current systems typically offer standardized encouragement that fails to address unique personal needs. This generic approach disregards individual differences in motivation, reinforcing users' perception that the AI counselor lacks genuine understanding of their specific situation.

These interconnected challenges highlight the complex nature of creating engaging AI counseling experiences. As the field continues to evolve, addressing these limitations will likely require innovative approaches that extend beyond incremental improvements in language capabilities, towards developing AI systems capable of fostering more meaningful therapeutic connections.

\subsection{Approaches in AI-Assisted Counseling for Enhanced Engagement and Effectiveness}
\label{sec:rw_engage}

Recent advancements in AI-assisted psychological counseling have been driven by the goal of creating more human-like (e.g., exhibiting natural conversational flow and empathetic responses), effective, and engaging systems. 
% This pursuit has led researchers to draw inspiration from various fields, including human cognition, reasoning agents, and computational neuroscience. 
The overarching aim has been to develop AI counselors that can offer more natural interactions, demonstrate professional competence, and provide personalized support. Wiltshire et al. \cite{wiltshire2017enabling} highlighted the potential of integrating cognitive architectures and computational neuroscience models to create more human-like AI systems. Building on this foundation, researchers like Hudlicka \cite{hudlicka2013virtual} and Bickmore et al. \cite{bickmore2009taking} have explored the application of embodied behavior in virtual health coaching and patient education, demonstrating the value of anthropomorphic interfaces in fostering user trust and engagement in healthcare settings.

In the realm of cognition-augmented approaches—that is, methods aimed at replicating human-like reasoning and decision-making—significant strides have been made. Laird et al. \cite{laird2017standard} proposed cognitive architectures that aim to replicate human-like reasoning and decision-making processes in AI systems. These architectures form a foundation for more sophisticated AI counselors capable of complex cognitive tasks. Additionally, Sun \cite{sun2007importance} introduced the CLARION cognitive architecture, which integrates implicit and explicit cognitive processes, potentially enhancing AI systems' ability to address the subtle and complex dynamics inherent in psychological counseling.

Despite these advancements, current AI counseling systems, including those with cognitive augmentation, face significant challenges in replicating the nuanced understanding and adaptive responses of human counselors. While researchers have made progress in integrating expert knowledge—as demonstrated by Bickmore et al. \cite{bickmore2013automated} and Fitzpatrick et al. \cite{fitzpatrick2017delivering}—and recent work with LLMs \cite{li2023chatdoctor, hu2024psycollm} has enhanced AI's capacity to process medical information, these systems still fall short of replicating the deep, contextually rich understanding typically exhibited by human professionals. Efforts to maintain long-term user engagement, such as those by Elgedawy et al. \cite{hu2024psycollm}, have shown promise, but AI counselors struggle to adapt to evolving user needs and maintain meaningful therapeutic relationships over extended periods.

Personalization has seen advancements through improved user profiling techniques, as demonstrated by Kadariya et al. \cite{kadariya2019kbot}, Lisetti et al. \cite{lisetti2013can}, and De Rosis et al. \cite{de2006user}, with Yan et al. \cite{xu2024can} extending this work to LLM-based companions. However, these systems still fall short in deeply understanding individual user contexts and dynamically adapting their approach. Similarly, while Morris et al. \cite{morris2018towards} and Ghandeharioun et al. \cite{ghandeharioun2019emma} have made strides in creating empathetic AI agents, current systems struggle to accurately perceive and respond to subtle emotional cues, a critical aspect of effective counseling. The challenge of creating a coherent and trustworthy AI persona, as investigated by Go and Sundar \cite{go2019humanizing}, remains significant, particularly in maintaining a strong therapeutic alliance.

A more holistic AI counselling paradigm is needed due to these constraints. Although cognition-augmented techniques have shown potential in improving AI's reasoning, they have yet to incorporate the emotional intelligence and adaptive behaviors that are essential for effective counseling. Even with advanced cognitive architectures, current AI systems still struggle to concurrently implement robust logical decision-making, exhibit genuinely sympathetic responses, and sustain long-term relational capacities. This gap underscores the need for a novel strategy that not only enhances cognitive processing but also integrates sophisticated emotional modeling and adaptive interaction techniques, thereby more closely emulating the multifaceted expertise of human therapists.

\section{Formative Research: Understanding Human and AI Counseling Engagement}
\label{sec:Formative_Research}

This chapter presents our formative investigation into human counselor practices and AI counseling approaches. We employ a mixed-methods design, first examining therapeutic interventions through practitioner interviews that focus on both therapeutic procedures and client engagement (Section \ref{sec:formative_Counselor}). We then systematically review existing AI counseling systems to identify critical implementation gaps (Section \ref{sec:formative_Review}).
Our methodology yields five specific design considerations (Section \ref{sec:formative_Considerations}). 
% The investigation of human counselor practices directly informs DC1 (Embodied Cognitive Formulation) and DC4 (Therapeutic Persona Design), drawing principles from established therapeutic practices. DC3 (Actionable Knowledge Base) emerges primarily from prior AI counseling research, focusing on essential knowledge structures for effective interventions. DC2 (Proactive Client Psychological Profiling) and DC5 (Meeting Stereotypical Expectations) synthesize insights from both research components, addressing the intersection of human counseling expertise and AI implementation challenges.
This integrated approach bridges human therapeutic expertise with technological capabilities, establishing a foundation for our counseling agent architecture that will be elaborated in subsequent chapters.

\subsection{Formative Investigation of Human Counselor Practices}
\label{sec:formative_Counselor}

As the first component of our dual research approach, this formative study examined the expertise, cognitive processes, and decision-making strategies of human counselors. Our investigation focused on three critical dimensions of counseling practice that directly inform AI design requirements: (1) therapeutic procedures (Section \ref{sec:understanding}), (2) strategic self-disclosure coupled with client engagement (Section \ref{sec:interview_empathy}), addressing how counselors build rapport and trust, and (3) professional documentation practices (Section \ref{sec:interview_doc}). Understanding these elements informed the foundational requirements for our AI counselor design.

% As the first component of our dual research approach, this formative study aimed to extract the professional expertise, cognitive processes, and decision-making strategies employed by human counselors during therapeutic interactions. By understanding how experienced practitioners navigate complex counseling scenarios, we sought to identify key elements that should be incorporated into our AI counselor design. 
% This approach addresses a critical gap in current AI counseling systems, which often lack the nuanced understanding of counseling dynamics that human professionals develop through training and experience. 

\subsubsection{\textbf{Methods and Participants}}

\begin{table*}
\caption{Demographic Information of the Participants}
\label{tab:formative_demographics}
\begin{tabular}{@{}llllll@{}}
\toprule
         ID&  Occupation&  Years of Experience& Current Role\\
         \midrule
         P1&  Supervising Therapist/Professor&  20& Active Practice \& Supervision\\
         P2&  Licensed Counselor&  10& Active Practice\\
         P3&  Licensed Counselor&  8& Active Practice\\
         P4&  Licensed Counselor&  5& Active Practice\\
         P5&  Licensed Counselor&  3& Active Practice\\
\bottomrule
\end{tabular}
\end{table*}

% \subsubsection{Findings}
% DC1,DC4
% We conducted a formative study to understand the professional practices and standard operating procedures (SOPs) of human counselors in psychological counseling. This understanding would inform the design of a more grounded framework for therapeutic AI agents. 
We recruited five licensed psychological counselors through professional networks, including one supervising counselor/professor and four practicing counselors, as shown in Table \ref{tab:formative_demographics}.
All participants were actively practicing psychological counseling and were well-versed in therapeutic processes and practical implementations. The supervising counselor (P1) brought additional expertise in training and evaluating other counselors, offering valuable insights into both practice and professional development.

In the formative study, we first explained our research objective: to deconstruct the working procedures of human counselors and understand how they engage clients in the counseling process. We emphasized that this understanding would be crucial for developing a more evidence-based framework for therapeutic AI agents. The participants then engaged in semi-structured interviews where they detailed their therapeutic approaches, decision-making processes, and professional practices. Each interview lasted approximately 60 minutes. The moderator's guide for the semi-structured interviews can be found in the appendix \ref{sec:APP_interview}.

After completing all interviews, we transcribed the audio recordings verbatim. The analysis process followed a systematic coding approach: First, one researcher performed initial coding on the transcripts, identifying recurring patterns in therapeutic procedures and decision-making processes. A second researcher then conducted independent coding and reviewed the first researcher's findings. Finally, a third researcher joined to discuss and resolve any coding disagreements with the previous two researchers. 

Our analysis primarily focuses on CBT and psychodynamic approaches predominant in the Chinese therapeutic context, though the identified cognitive patterns and operational procedures potentially extend to other therapeutic modalities, as they reflect fundamental aspects of therapeutic interaction and decision-making.
Through this rigorous analysis process, we identified several key components in counselors' standard operating procedures, which we detail in the following section.

\subsubsection{\textbf{Understanding Therapeutic Procedures}}
\label{sec:understanding}
Our analysis of counselors' standard operating procedures revealed a hierarchical structure of cognitive processes and operational procedures, organized into three interconnected layers that seamlessly integrate high-level planning with moment-to-moment adjustments. These layers reflect how counselors organize their therapeutic work from overarching treatment phases down to moment-by-moment interactions, creating a recursive planning system that remains both structured and adaptable.

% Our analysis primarily focuses on CBT and psychodynamic approaches predominant in the Chinese therapeutic context, though the identified cognitive patterns and operational procedures potentially extend to other therapeutic modalities, as they reflect fundamental aspects of therapeutic interaction and decision-making.

\textbf{(1) Goal Layer: Overarching Therapeutic Phases}
\begin{itemize}
    \item \textbf{Initial Assessment Phase (typically lasting 1-2 Sessions):} The therapeutic journey begins with a careful assessment phase, where counselors establish the foundation for treatment. As P1 explained: \textit{``The initial phase is crucial for building our case conceptualization. While digital tools help us collect basic information efficiently, we use the face-to-face sessions to deepen our understanding of the client's narrative and develop a comprehensive clinical formulation."} 
    During these initial sessions, counselors focus on three key objectives: validating and expanding upon pre-collected information, establishing a collaborative therapeutic relationship, and working with clients to define specific therapeutic goals.

    \item \textbf{Middle Phase (typically 8-40+ Sessions):} The middle phase constitutes the bulk of the therapeutic work. This phase focuses on implementing core therapeutic interventions, developing clients' coping skills, processing challenging emotions, and tracking progress toward established goals. Counselors regularly revisit treatment plans and adjust strategies based on client responses and evolving needs.

    \item \textbf{Termination Phase (2-3 Sessions):} The termination phase begins when therapeutic goals have been substantially achieved. As P2 explained: \textit{``Termination is about consolidating gains and ensuring clients' readiness for independence."} This phase emphasizes both celebrating progress and strengthening clients' confidence in maintaining their therapeutic gains independently.

    % \item \textbf{Goal Layer Adaptation:} Counselors continuously evaluate the client's progress and may recursively adjust which therapeutic phase is appropriate. As P3 noted: \textit{``Sometimes what appears to be a straightforward case reveals deeper issues during the middle phase, requiring us to temporarily step back to additional assessment before moving forward again."}
\end{itemize}

    Counselors continuously evaluate the client's progress and may adjust or even revisit earlier therapeutic phases if new issues emerge. As P3 noted: \textit{``Sometimes what appears to be a straightforward case reveals deeper issues during the middle phase, requiring us to temporarily step back to additional assessment before moving forward again."}

\textbf{(2) Session Layer: Session-Specific Agenda Planning}

Within each therapeutic phase, counselors develop detailed session agendas that serve specific purposes aligned with the current phase. As P2 noted: \textit{``We follow a planned agenda while remaining responsive to emerging themes that might offer deeper therapeutic value."} 
Each session typically follows a structured format designed to optimize therapeutic progress:
    \begin{itemize}
        \item Brief check-in and agenda-setting (5-10 minutes)
        \item Main therapeutic work (30-35 minutes) focusing on in-depth exploration and intervention
        \item Progress review and homework planning (5-10 minutes)
        \item Brief closure (5 minutes) to synthesize key insights
    \end{itemize}

% \textbf{Session Agenda Adaptation:}
Counselors regularly evaluate the effectiveness of their session agendas and make recursive adjustments when needed. P3 described: \textit{``While we come to each session with a clear agenda, we must remain flexible when client needs or emerging insights suggest a different direction would be more beneficial. These adjustments still need to serve the overall phase objectives."}

\textbf{(3) Action Sequence Layer: Moment-to-Moment Therapeutic Responses}

% \begin{itemize}
    % \item \textbf{Tactical Implementation:} 
    Within each session agenda item, counselors execute specific therapeutic actions through a sophisticated recursive process. As P2 described: \textit{``I'm constantly integrating multiple streams of information, what the client is saying now, how it connects to our previous discussions, their non-verbal cues, and our overall therapeutic goals."} 

    This action sequencing process involves:
    \begin{itemize}
        \item \textbf{Perceptive Client Message:} Gathering comprehensive information from verbal content, non-verbal cues, and historical context.
        
        \item \textbf{Strategic Response Selection:} Choosing specific therapeutic techniques that implement the current agenda item. P3 noted: \textit{``It's about selecting the most appropriate intervention for this specific moment - whether that's deepening exploration, offering validation, or introducing a new perspective."}
        
        \item \textbf{Therapeutic Delivery:} Executing the chosen response with appropriate timing and phrasing.
        
        \item \textbf{Reflective Integration:} Evaluating this information against the current agenda item and therapeutic direction.
        
    \end{itemize}

    % \item \textbf{Immediate Action Adjustment:} 
    When client responses indicate the current action sequence is ineffective, counselors immediately adapt their approach. As P3 emphasized: \textit{``The effectiveness often lies in how we deliver our response, the timing, the phrasing, even knowing when silence might be most therapeutic."}
% \end{itemize}

% \textbf{Bidirectional Planning Integration:} 
These three layers operate in a continuous bidirectional relationship. The therapeutic phase (Goal Layer) determines appropriate session agendas (Session Layer), which guide specific therapeutic actions (Action Sequence Layer), which in turn ensures the therapy remains both coherent and responsive to clients' evolving needs. Simultaneously, observations and client responses at the Action Sequence Layer may trigger adjustments to the Session Layer, which can sometimes propagate up to necessitate changes at the Goal Layer. As P1 summarized: \textit{``Effective therapy requires both structure and adaptability. We need clear direction while remaining responsive to what emerges in the therapeutic process."} 
This hierarchical, recursive planning system allows counselors to maintain therapeutic continuity while adapting to evolving client needs, creating a balance between structured progression and responsive flexibility that is essential for effective counseling.

\subsubsection{\textbf{Therapeutic Self-disclosure and Personalized Empathic Engagement}}
\label{sec:interview_empathy}
This subsection examines how counselors establish authentic therapeutic relationships through strategic self-disclosure and personalized empathic engagement.
   
\textbf{Strategic Self-disclosure in Therapeutic Context: }
Our interviews revealed that counselors strategically employ self-disclosure to establish an authentic and empathetic connection with their clients. As P3 described: \textit{``There are moments when being just a professional isn't enough, clients need to know you're a human who understands. I might share a brief personal experience, not to make it about me, but to convey `I know this terrain not just from books.'"} This judicious self-disclosure serves to normalize client experiences while maintaining appropriate boundaries.

\textbf{Timing and Personalization of Empathic Responses: }
Counselors emphasized the importance of timing and personalization in empathic engagement. P4 explained: \textit{``Some clients expect immediate solutions, but often what they truly need first is to feel completely seen. I create space for this emotional connection before moving to action steps."} Additionally, counselors develop individualized empathic approaches based on client preferences and backgrounds. P8 noted: \textit{``What reads as genuine empathy varies between clients. Discovering each person's unique `empathy language' is essential to therapeutic effectiveness."}

% 管理工具：文档（record、agenda、case conceptualization；）
\subsubsection{\textbf{Client Documentation Practices in Professional Counseling}}
\label{sec:interview_doc}

Our investigation revealed that human counselors typically maintain three critical types of documentation that form the foundation of effective case management and therapeutic progression:

\begin{itemize}
    \item \textbf{Session Records:} These are detailed accounts of each counseling session, documenting client statements, emotional responses, interventions applied, and significant developments \cite{american2007record}. Counselors maintain these records chronologically, often including a summary section highlighting key insights and progress markers. These records serve both as legal documentation and as reference material for tracking therapeutic progress over time.
    
    \item \textbf{Session Agendas:} Counselors develop structured plans for upcoming sessions based on previous interactions and client needs. These agendas outline therapeutic goals, specific techniques to implement, topics to explore, and potential challenges to address \cite{beck2020cognitive}. Agendas are dynamic documents that evolve based on session records and case conceptualizations, ensuring therapeutic continuity while accommodating emerging client needs.
    
    \item \textbf{Case Conceptualizations:} Perhaps the most critical cognitive tool in a counselor's practice, case conceptualizations are comprehensive theoretical frameworks that organize client information into coherent explanatory models. These living documents integrate client history, presenting issues, underlying psychological mechanisms, strengths, challenges, and treatment directions \cite{sperry2020case}. A well-developed case conceptualization guides intervention selection, helps predict client responses, and provides a theoretical foundation for therapeutic decisions.
\end{itemize}

The interrelationship between these documentation practices creates a cognitive scaffolding that supports therapeutic reasoning. Session records inform case conceptualizations, which in turn shape session agendas, creating a continuous feedback loop that enables responsive and personalized therapeutic progression.

\subsection{Formative Review of AI Counseling Approaches}
\label{sec:formative_Review}

% Having examined human counselors' practices through interviews, we gained crucial insights into the fundamental elements of professional counseling services 
% % , including therapeutic goal setting, standardized operational procedures, and continuous improvement through supervision. These findings provide valuable reference points for conceptualizing AI-assisted mental health services
% . 
As LLMs rapidly evolve, AI counselors are emerging in diverse forms—ranging from text-based conversational agents to virtual therapy assistants—across the mental health domain. To design more effective AI counseling frameworks, it is essential to systematically analyze the interaction patterns and technical implementations of current AI counseling systems. By dissecting the strengths and limitations of existing AI counselors, while incorporating key insights from human counseling practices, we can establish a theoretical foundation for designing LLM-based mental health counseling systems.

\subsubsection{\textbf{Review Methodology}}
\begin{figure}
  \includegraphics[width=0.8\textwidth]{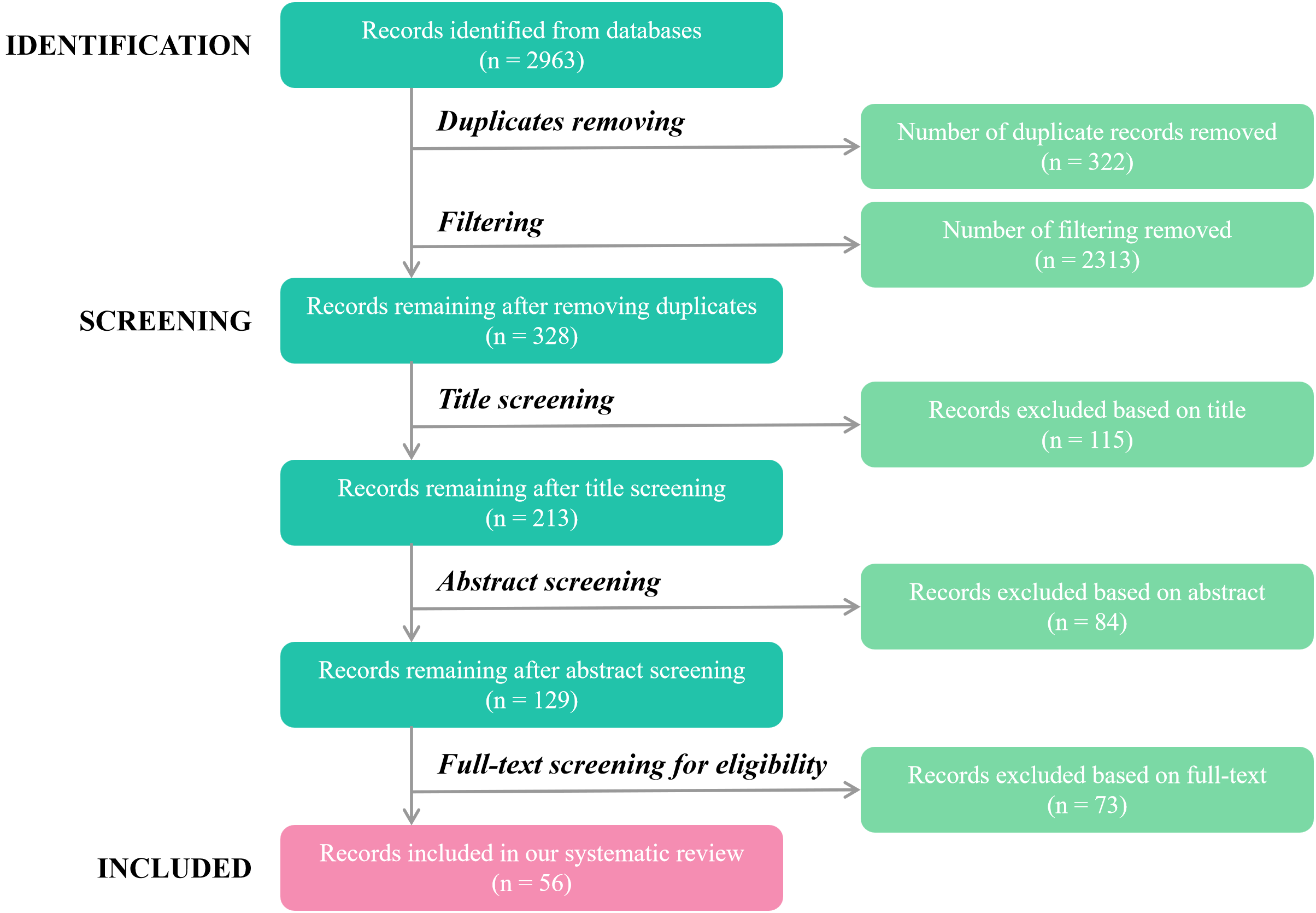}
  \caption{PRISMA flow diagram illustrating the literature selection process.}
  \Description{PRISMA flow diagram illustrating the literature selection process: from 2963 initially identified records to 56 studies included in the systematic review after duplicate removal, filtering, and sequential screening of titles, abstracts, and full texts.}
  \label{fig:review}
\end{figure}

% \subsection{Method}
We used the Preferred Reporting Items for Systematic Reviews and Meta-Analyses (PRISMA) \cite{page2021prisma} for surveying research papers. There were two successive searches and analyses based on hierarchical research questions. Figure \ref{fig:review} provides an overview of our searching process and related search results.

\textbf{Search and Automated Filtering:}

\begin{itemize}
    \item \textbf{Database:} Our database search was conducted from April to May 2024. Although four databases were initially considered—ACM Digital Library, ScienceDirect (Elsevier), PsycInfo, and Scopus—the final search results predominantly originated from Scopus, the ACM Digital Library, and APA PsycInfo. 
    
    %and arXiv (because this field is developing rapidly and there is a lot of new excellent research that has not yet been published).

    \item \textbf{Search Scope and Strings:} We constructed our search query using the following terms: \textit{(``wellbeing" OR ``health" OR ``medical" OR ``mental" OR ``psychological" OR ``counseling" OR ``therapy" OR ``consulting" ) AND ( ``AI Agent" OR ``conversational agent" OR ``chatbot")}.

    We restricted our search results to Articles and Conference Papers found in reputable journals and proceedings, and excluded all non-English publications. A total of 2963 entries were obtained from these queries: 2600 from Scopus and 188 from the ACM Digital Library, 175 from the APA PsycInfo.

    We exported the results to a table, standardized the columns containing publication details, eliminated duplicates, and removed articles that did not meet our formal criteria (e.g., workshop papers). Our initial corpus ultimately comprised 2641 papers.

    \item \textbf{Filtering:} We used a strategy based on the association strength \cite{jenkins2004evaluation} of normalized citations to filter papers. Association strength, denoted by $S_{\mathrm{A}}(c_{ij},s_{i},s_{j})$, calculates the strength of association between variables $s_{i}$ and $s_{j}$ given a certain context $c_{i j}$ Mathematically, it is expressed as:
    \begin{equation}
    S_{\mathrm{A}}\left(c_{i j}, s_{i}, s_{j}\right)=\frac{c_{i j}}{s_{i} s_{j}},
    \end{equation}
    Here, $c_{i j}$ represents the context in which the association is being measured, while $s_{i}$ and $s_{j}$ denote the scores of variables $i$ and $j$ respectively. This measure enhances the objectivity and quantitative rigor of the selection process by prioritizing studies with stronger empirical associations. Papers whose normalized citation score fell below $0.3$ were subsequently excluded.
    We were left with a total of 328 highly-cited core papers.

\end{itemize}

\textbf{Human Screening:}

We screened all search outcomes using a three-step process:

\begin{itemize}

\item \textbf{Title Screening:} During the initial screening phase, one researcher evaluated the titles of every paper and distinguished between relevant and irrelevant papers based on their alignment with our theme. Titles were considered relevant if they contained terms related to conversational agent and well-being. 
In all, 213 papers with related titles were to be further screened 

\item \textbf{Abstract Screening:} One researcher carefully examined and analyzed the abstracts of papers to determine their relevance to the subject area. A different researcher evaluated the papers that underwent screening. 
The abstract review removed 129 papers, leaving 84 appropriate articles for full-text screening.

\item \textbf{Full-text Screening:} Two researchers conducted a thorough examination of the remaining articles to establish the definitive inclusion of papers based on the inclusion and exclusion criteria specified by the researchers. We left 56 papers.

Studies that satisfied at least one of the following exclusion criteria were eliminated: 
\begin{enumerate}
    \item The article did not undergo peer review, meaning it was not evaluated by experts in the field.
    \item The article was either an abstract or an expanded abstract, providing a concise summary of the research.
    \item The article was based on secondary research, such as review papers.
\end{enumerate}

\end{itemize}

\textbf{Quality Assessment: }
\begin{list}{}{
  \setlength{\leftmargin}{\leftmargini}
  \setlength{\rightmargin}{0pt}
  \setlength{\itemindent}{0pt}
  \setlength{\labelwidth}{0pt}
  \setlength{\labelsep}{0pt}
}
    \item Then, we evaluated the quality of each article in the remaining papers of our corpus using critical appraisal tools. These tools offer analytical assessments of the quality of each study through freely available online checklists and spreadsheets \cite{crombie2022pocket, katrak2004systematic}. 
    Two researchers individually assessed each publication. The article's quality was thoroughly debated until a consensus was reached, resolving any disagreements.
    Among the 56 full-text screened papers, 1 paper was classified as ``poor” using our critical evaluation tools and was thus excluded. Consequently, a total of 55 papers were selected for inclusion in our review. Following the completion of the inclusion and exclusion criteria procedures, a total of 55 papers were selected for inclusion in our research.
\end{list}

\textbf{Data Extraction and Analysis: }
\begin{list}{}{
  \setlength{\leftmargin}{\leftmargini}
  \setlength{\rightmargin}{0pt}
  \setlength{\itemindent}{0pt}
  \setlength{\labelwidth}{0pt}
  \setlength{\labelsep}{0pt}
}
    \item Using thematic analysis~\cite{clarke2017thematic}, we developed a set of criteria that clearly defined the specific data that we require to collect from the papers. Afterward, we conducted a preliminary examination on 43 papers to determine if these criteria are rational and if the data can truly be extracted. Throughout this process and after multiple conversations among three researchers, the criteria gradually developed. In Section 2.2, we provide the ultimate version of their shape, together with the associated outcome. 

\end{list}

\subsubsection{\textbf{Findings}}

% 总结段
% finding了啥，介绍整个怎么读和理解；

Our review of the literature reveals that current research in AI counseling systems has identified limitations across several key therapeutic domains. Researchers have highlighted deficiencies in case conceptualization \cite{oh2017chatbot}, agent character designing \cite{lee2024influence}, and treatment planning \cite{rathnayaka2022mental, denecke2020mental} capabilities. Additional work has focused on limitations in therapeutic technique implementation, psychoeducational delivery, and empathetic understanding. The literature also addresses challenges in creating fulfilling experiences, personalization, preserving user autonomy, building therapeutic alliances, and maintaining transparency. The following sections examine these identified limitations in greater detail, revealing fundamental gaps that must be addressed to advance AI counseling capabilities.

\textbf{Limitations in Cognitive Processing and Planning: }
\label{Sec:review_DC1}
Prior automated mental health systems have demonstrated significant cognitive processing limitations that hinder their therapeutic effectiveness. Systems built on decision-tree algorithms \cite{quinlan1986induction, yasavur2014let, shah2022development} or scripted dialogues \cite{bickmore2009taking, bickmore2010maintaining, fitzpatrick2017delivering, fadhil2019assistive, chung2019chatbot} inherently operate within rigid, predetermined interaction pathways \cite{gardner2002multidimensional}. These constraints fundamentally limit their ability to engage in the adaptive cognitive processing essential for effective counseling.

Even with recent advancements in language model capabilities, current LLM-based counseling applications still demonstrate critical planning deficiencies. While these systems can successfully mimic counselors' communication styles and retrieve relevant professional knowledge through various techniques including fine-tuning \cite{jo2024understanding, xu2024can}, retrieval-augmented generation (RAG) \cite{hu2024psycollm, li2023chatdoctor}, and prompt engineering \cite{singh2024revolutionizing}, they primarily operate as sophisticated response generators rather than cognitive agents capable of therapeutic planning.

Our analysis reveals that these systems typically process conversational turns in isolation, lacking the ability to maintain coherent therapeutic progression across interactions \cite{abd2019overview}. This fragmented approach stands in stark contrast to human counselors' integrated cognitive processes, which seamlessly coordinate between session-level therapeutic goals and moment-to-moment interventions based on evolving client needs. This fundamental limitation in cognitive processing and planning capability represents a critical gap that must be addressed in developing more effective AI counseling systems.

\textbf{Limitations in Client Assessment and Profiling: }
\label{sec:review_dc2}
Traditional automated mental health systems have typically relied on static questionnaires or predetermined assessment pathways \cite{quinlan1986induction, yasavur2014let, shah2022development, bickmore2009taking, bickmore2010maintaining, bickmore2013automated, fitzpatrick2017delivering, fadhil2019assistive, chung2019chatbot, hauser2020smartphone, battineni2020ai}. Even with recent advances in LLM technology, most systems still lack the capability for ongoing, iterative client understanding that characterizes effective human counseling practice \cite{hu2024psycollm, 10.1145/3613904.3642420}. These approaches typically collect client information through fixed protocols at specific points, failing to capture the dynamic nature of clients' psychological states and evolving needs throughout the therapeutic process.

\textbf{Limitations in Therapeutic Knowledge Utilization: }
While knowledge base retrieval is commonly used in counseling applications \cite{hu2024psycollm, li2023chatdoctor, jo2024understanding}, current approaches lack systematic organization and operational specificity. Existing systems typically employ raw information retrieval without the necessary transformation into actionable therapeutic interventions. This creates a significant implementation gap between information access and therapeutic execution \cite{luxton2014artificial}. Most current systems retrieve and present relevant information but fail to structure this knowledge in ways that enable contextually appropriate therapeutic interventions. This limitation compromises the ability of AI counseling systems to effectively apply therapeutic techniques in response to dynamic client needs.

\textbf{Therapeutic Relationship Building in AI Counseling Systems: }
Research in AI-based counseling has made notable progress in developing systems that can express empathy and build therapeutic alliances. Systems incorporating expressions of empathy \cite{casas2021enhancing} and positive regard \cite{rogers1957necessary} have shown improved client engagement. Fitzpatrick et al. \cite{fitzpatrick2017delivering} demonstrated promising results with approaches that balance therapeutic presence with technological transparency. However, these implementations typically address specific components of the therapeutic relationship rather than offering a comprehensive framework for creating an authentic AI therapeutic persona. Building on these advances, a comprehensive approach to therapeutic relationship building is needed to complement the cognitive and knowledge-processing aspects of AI counseling systems.

\subsection{Design Considerations from Formative Research}
\label{sec:formative_Considerations}

While acknowledging that LLM-based counseling systems also require robust safety protocols, strict data protection, and comprehensive ethical guidelines, which are beyond the scope of this paper, this work focuses specifically on designing AI counseling agents that effectively emulate therapeutic interactions. Safety concerns, regulatory compliance, and comprehensive risk mitigation strategies warrant further investigation in subsequent research. Our formative research yields five interconnected design considerations:

 DC1 defines the cognitive architecture that underpins and guides therapeutic processes. DC2 addresses the need for continuous psychological assessment—ensuring that dynamic client states inform appropriate goal-setting, while DC3 provides a robust knowledge infrastructure that organizes and delivers the information necessary for effective therapeutic interventions. DC4 and DC5 tackle complementary aspects of AI counseling: while DC4 focuses on designing a therapeutic persona capable of genuine empathy and effective relationship-building, DC5 leverages AI's computational advantages to provide robust knowledge support and facilitate psychoeducation. Together, these considerations form a framework for creating AI counseling agents that balance authentic therapeutic relationships with technological capabilities.

\subsubsection{\textbf{DC1: Cognitive Formulation of Counseling Processes}}

Building on our analysis of the cognitive planning limitations present in current AI systems (Section \ref{Sec:review_DC1}), we draw upon our investigation of human counselors’ approaches to therapeutic processes to derive this design consideration. Our analysis of counselors' standard operating procedures (Section \ref{sec:understanding}) reveals that effective therapeutic practice requires a hierarchical goal-oriented planning framework with recursive updating mechanisms.

The foundation of this planning framework begins with systematic objective setting. Building on comprehensive client profiles, counselors develop structured intervention strategies that translate understanding into actionable plans. Our observations indicate this planning unfolds across three interconnected layers:

\begin{itemize}
    \item At the \textbf{Goal Layer}, counselors establish overarching therapeutic objectives that span multiple sessions. These high-level goals define the complete therapeutic journey from initial assessment through termination. As detailed in Section \ref{sec:understanding}, these goals remain relatively stable but can be recursively updated when persistent changes in client behavior or significant therapeutic shifts are detected.
    
    \item At the \textbf{Session Layer}, counselors develop session-specific agendas derived from the higher-level therapeutic goals. Each session is deliberately planned to serve specific purposes within the broader therapeutic progression. This intermediate planning layer gets recursively adjusted when accumulated feedback indicates significant deviation from expected progress.
    
    \item At the \textbf{Action Sequence Layer}, counselors plan their moment-to-moment therapeutic responses within individual sessions. As P2 described, \textit{``I'm constantly integrating multiple streams of information - what the client is saying now, how it connects to our previous discussions, their non-verbal cues, and our overall therapeutic goals."} This layer involves continuous assessment and strategic response generation, with immediate recursive adjustments when client messages indicate a need for strategy shifts.
\end{itemize}

These planning layers operate through a bidirectional mechanism: top-down content generation where higher layers inform lower layers, and bottom-up feedback propagation where client responses trigger recursive adjustments at appropriate levels. This bidirectional approach enables structured yet adaptable therapeutic progression, allowing AI counselors to maintain overarching therapeutic goals while flexibly responding to emerging client needs.

Critical to this process is the integration of systematic reflection. Our interviews with both counselors and supervisors emphasize that effective therapeutic practice requires ongoing metacognitive evaluation of both current and historical interactions. As P5 noted, \textit{``Regular reflection on my therapeutic choices helps me recognize patterns I might otherwise miss and informs how I approach similar situations in the future."} This reflective dimension enables counselors to critically examine their own reasoning, identify potential biases or ineffective approaches, and continuously refine their therapeutic strategies based on accumulated experience. For AI counseling systems, incorporating this reflective capacity is essential for moving beyond simple response generation toward more sophisticated therapeutic reasoning that evolves through experience.

\subsubsection{\textbf{DC2: Proactive and Continuous Client Psychological Profiling}}
% 咨询的计划目标，首先来自于对client的精准评估；
% 在主动、长线收集和维护；涉及到现在存在的问题目标、偏好；
% 怎么做的profiling，不是被动接受，而是主动收集和维护信息；提供最恰当的干预；

% As revealed in our analysis of counselors' practices (Sec \ref{sec:understanding}), client assessment is a continuous and proactive process throughout the therapeutic journey. While initial profiling begins with pre-session questionnaires and early interactions, counselors actively deepen their understanding during each therapeutic encounter. 

% Prior automated mental health assessments, whether using decision-tree algorithms \cite{quinlan1986induction, yasavur2014let, shah2022development} or scripted dialogues \cite{bickmore2009taking, bickmore2010maintaining, bickmore2013automated, fitzpatrick2017delivering, fadhil2019assistive, chung2019chatbot, hauser2020smartphone, battineni2020ai}, were limited by their reactive nature and predetermined pathways \cite{gardner2002multidimensional}. 

% Traditional automated mental health systems have typically relied on static questionnaires or predetermined assessment pathways \cite{quinlan1986induction, yasavur2014let, shah2022development,bickmore2009taking, bickmore2010maintaining, bickmore2013automated, fitzpatrick2017delivering, fadhil2019assistive, chung2019chatbot, hauser2020smartphone, battineni2020ai}. Even with recent advances in LLM technology, most systems still lack the capability for ongoing, iterative client understanding that characterizes effective human counseling practice \cite{hu2024psycollm,10.1145/3613904.3642420}.

As revealed in our analysis of counselors' practices (Sec \ref{sec:understanding}), client assessment is a continuous and proactive process throughout the therapeutic journey. While initial profiling begins with pre-session questionnaires and early interactions, counselors actively deepen their understanding during each therapeutic encounter.

To address these limitations (Shown in Section \ref{sec:review_dc2}) and advance the continuous deepening of client understanding throughout the counseling intervention process, our design consideration emphasizes two key aspects of systematic psychological profiling:

\begin{itemize}
    \item \textbf{Proactive Understanding:} Rather than passively receiving information, systems must actively explore and validate various aspects of clients' situations throughout each session. As evidenced in counselors' practice, ongoing assessment involves deliberate exploration of emerging themes and active validation of developing insights. This dynamic understanding process ensures that therapeutic interventions remain responsive to clients' evolving needs.

    \item \textbf{Progressive Refinement:} Client profiles should be continuously updated as new insights emerge. This involves maintaining a dynamic psychological representation that synthesizes both immediate concerns and evolving therapeutic needs. Regular refinement of client understanding enables more precise intervention selection and ensures therapeutic alignment with clients' changing circumstances.
\end{itemize}

\subsubsection{\textbf{DC3: Actionable Knowledge Base for Agent Intervention}} 

% 陈述性知识、操作性知识？

% agent行为、治疗策略选择、行为规划？- 操作性知识的调控；
% 给到用户 info；
% formulate的方式，不是单纯搜索推荐；如何回复一句话？

% 1.操作性知识，根据心理学有关的干预手段，是个陈述性知识，使得agent能够行为-action的知识，程序性知识的数据库。给出思路是，提取agent可用的陈述性知识；

% While knowledge base retrieval is commonly used in counseling applications \cite{hu2024psycollm,li2023chatdoctor,jo2024understanding}, current approaches lack systematic organization and operational specificity. To bridge this gap between information access and therapeutic execution \cite{luxton2014artificial}, we identify three critical aspects in knowledge design for therapeutic systems:

To bridge the identified gap, where therapeutic knowledge is accessible but not effectively translated into practical interventions, we propose a design consideration focused on transforming therapeutic knowledge into actionable interventions:

\begin{itemize}
    \item \textbf{Knowledge Preprocessing:} The foundation of effective therapeutic intervention lies in the systematic processing and structuring of therapeutic knowledge—including extraction, classification, and standardization of therapeutic content. Raw therapeutic materials must be systematically transformed into standardized, retrievable operational formats that support efficient real-time intervention. This involves extracting specific intervention procedures and techniques, structuring them into fine-grained components with contextual tags, and organizing information to support rapid retrieval during live interactions. This preprocessing ensures that therapeutic knowledge is not just stored, but prepared for operational use in counseling scenarios.

    \item \textbf{Context-Aware Retrieval:} Beyond simple information storage, the system must intelligently determine when and what knowledge to retrieve based on therapeutic context. This requires sophisticated mapping between various conversation states—including emotional tone and semantic cues—and corresponding knowledge needs, thereby enabling the system to identify appropriate moments for intervention. The retrieval mechanism must align selected knowledge components with current therapeutic goals and conversation dynamics. As P4 noted: \textit{``Reading the right moment for intervention is just as important as the intervention itself."}

    \item \textbf{Adaptive Knowledge Application:} Retrieved knowledge must be thoughtfully adapted to serve individual client needs. This final stage involves integrating client profiles and session history to tailor interventions appropriately. The system must assess client readiness, based on indicators such as engagement level and emotional stability, and adjust intervention intensity accordingly, while always preserving the therapeutic alliance As P7 described: \textit{``Each therapeutic technique needs careful adaptation, what works for one client might need significant modification for another."} This adaptive application ensures that therapeutic knowledge translates into personalized, contextually appropriate interventions.

\end{itemize}

% 利用知识库查询知识是对LLM输出内容做优化的常用途径\cite{hu2024psycollm,li2023chatdoctor,jo2024understanding}；我们尤其注重如何从领域内已有的其他材料提取体系化的知识、以什么形式进行处理存储（尤其是什么样的处理和内容形式是适合被agent在咨询过程中进行检索的）；每次能被调用的体系的知识应当是末端的（可操作的知识）用以支持细粒度的输出；
% While knowledge base retrieval is commonly used to enhance LLM outputs in counseling applications \cite{hu2024psycollm,li2023chatdoctor,jo2024understanding}, current approaches often lack systematic organization and operational specificity. To bridge the gap between AI capabilities and human expertise in counseling \cite{luxton2014artificial}, we propose developing a knowledge base.

% \textbf{AI Guidance Knowledge:} The knowledge base must be structured to support fine-grained retrieval during counseling interactions. This requires systematic extraction and organization of domain knowledge from existing therapeutic materials, with information processed and stored in formats optimized for agent retrieval during counseling. The focus should be on operational, endpoint knowledge that directly supports specific counseling interventions, while maintaining clear links between theoretical frameworks and practical applications. Such structured organization ensures that the AI system can access and apply relevant therapeutic knowledge at appropriate moments in the counseling process.

% \subsubsection{\textbf{DC4: Designing for Therapeutic Relationship Development}}

\subsubsection{\textbf{DC4: Therapeutic Persona Design for Authentic Empathy}}

% chatbot要扮演的两个角色？
% 4 像人，提供的empathy才有意义。
% persona - 》 共情鼓励机制-）

% 5 AI（重复的知识和信息，能够在用户已掌握的信息不足的情况下）；满足用户的角色期待；用户（提供的知识，信息供给；
% AI counseling agents face unique challenges in meeting clients' dual expectations: they are expected to function both as empathetic counselors capable of human-like therapeutic relationships, and as AI systems that can provide comprehensive information support. This dual-role expectation creates fundamental design challenges in establishing authentic therapeutic connections.

Unlike human counselors who build therapeutic relationships through their nuanced subjectivity and rich lived experiences (Described in Section \ref{sec:interview_empathy}), AI agents lack the fundamental human qualities that facilitate authentic connection. To bridge this gap, we focus on designing human-like therapeutic interactions that can foster genuine therapeutic relationships. Our design is informed by key findings from our literature review, which highlighted the critical role of empathy and self-disclosure in effective therapeutic interactions.

\begin{itemize}

    \item {\textbf{Credible Presence:} The system must establish a consistent and credible therapeutic persona that projects professionalism and empathy, while simultaneously being transparent about its AI nature. This requires being upfront about its technological identity while cultivating a well-defined therapeutic presence that consistently demonstrates key professional qualities—such as empathy, reliability, and ethical conduct \cite{horvath1993role}. This includes building trust through consistent behaviors and clear communication about its capabilities and limitations.}

    \item {\textbf{Empathetic Response:} Beyond basic therapeutic presence, the system must develop authentic ways of expressing empathy and understanding \cite{casas2021enhancing,adikari2022empathic,lee2024influence}. Rather than mimicking human behaviors, this involves crafting response patterns that genuinely acknowledge client emotions while respecting the boundaries of AI capabilities. The focus should be on developing AI-appropriate methods for conveying understanding and support.}
% 分类不均衡

    \item {\textbf{Alliance Building:} The system must maintain consistent therapeutic alliance building across interactions through emotional congruence and unconditional positive regard \cite{rogers1957necessary}. This requires tailoring engagement styles to individual client needs—such as adjusting communication tone and timing based on client feedback \cite{fitzpatrick2017delivering}—while demonstrating reliable emotional awareness. The ultimate goal is to cultivate a sustainable therapeutic relationship in which the interface transparently integrates human-like empathy with the distinct capabilities and limitations of AI, thereby fostering long-term client trust and engagement.}

\end{itemize}

% The core challenge lies in creating an AI persona that exhibits human-like therapeutic qualities while maintaining authenticity. This requires carefully crafted responses that mirror human counselors' empathetic understanding, active listening, and emotional attunement \cite{horvath1993role}. The system should demonstrate human-like emotional intelligence in recognizing client emotions and responding with appropriate warmth and support.

% Beyond individual responses, the system must maintain consistent human-like interaction patterns that build trust over time. This involves developing a stable personality that clients can relate to and depend on, similar to how human counselors establish therapeutic rapport \cite{rogers1957necessary}.

\subsubsection{\textbf{DC5: Meeting Stereotypical Expectations Toward AI Agents}}

% stereotypical，对于LLM的一种预设；如果不能做到，就会违背预期。
% 人们已经有大模型使用经验，对此使用也是有预期；

%=========================

% 满足的是一种刻板印象？

% 标题可能需要修改
% 与对标真人咨询师，提供更类人（共情、人设等）之外，AI咨询师还面临着client对AI咨询师“你是一个AI（agent）”的预设。
% 因此AI咨询师的设计还需要满足client对agent AI的预设，行为要符合；与真人是不同的视角。client不会去预设agent像人一样共情，会预设agent是：

% （可以再进行聚类和阐述）信息的、理性的、提供大量精确数据量信息的（+解释性）；下位的（作为非人提供服务的）；不会有“麻烦他人”的；容易进行社交距离的突破；隐私的无负担，对非人内容的信息分享，不会有泄露隐私的动机；

While AI counseling agents strive to emulate human-like qualities such as empathy and persona development, they must also actively address clients' inherent preconceptions—such as expecting highly analytical information and a subordinate service role—about interacting with artificial intelligence. Clients approach AI counselors with the fundamental awareness that they are engaging with non-human entities, creating expectation patterns distinct from those in human therapeutic relationships. These differing expectations manifest in several key dimensions:

\begin{itemize}
    \item First, clients expect AI counselors to be \textbf{informational and analytical}. Evaluations of therapeutic chatbots reveal that users anticipate and value responses offering concrete information and evidence-based suggestions \cite{fitzpatrick2017delivering}. Similar research documents expectations for AI health agents to provide comprehensive, precise information with greater detail than human counselors \cite{abd2019overview}.

    \item Second, clients position AI counselors in a \textbf{subordinate role}. Studies indicate that users perceive AI systems primarily as service providers rather than equal participants. This hierarchy affects interaction patterns, with users demonstrating more directive communication styles with AI agents \cite{pradhan2019phantom}.

    \item Third, users experience \textbf{reduced social burden} when interacting with AI counselors. Research shows that users report significantly less concern about ``bothering" AI systems compared to human counselors \cite{miner2016smartphone}.

    \item Fourth, clients demonstrate \textbf{accelerated intimacy development} with AI systems. Observations reveal participants bypassing normal social distance barriers more rapidly with AI systems, often sharing sensitive information within the first few interactions \cite{ho2018psychological}.

    \item Finally, clients perceive \textbf{reduced privacy concerns} with AI counselors. Studies confirm users report greater comfort in sharing personal information with AI systems due to reduced fears of human judgment \cite{ly2017fully}.
\end{itemize}

These findings indicate that effective AI counseling systems must be designed not only to mimic human counselors but also to accommodate and leverage these AI-specific expectations. As Bickmore et al. \cite{bickmore2010response} argue, acknowledging and adapting to non-human status may enhance rather than detract from therapeutic effectiveness in AI counseling contexts.

\section{CA+ Agent Framework}
\label{sec:framework}

% 技术壁垒？AIUI层面上，进一步需要有壁垒/价值的层面，要积累的东西是什么？

% 流程图？if else??
% 操作序号、选择判断、实体关系；

% 但我觉得心理咨询师其实是在解决问题的，我大胆暴论一下这个问题，首先是疾病层面：病人有（a,b,c）不舒服，可能的原因是（d,e,f）,潜在解决方案（g,i,h）。我们需要找到三者的关系。其次是沟通层面，诊断阶段，主要是通过提问，找到一个病因x使p(x|a,b,c)最大。然后是治疗阶段，主要是通过沟通，找到一个解决方案y使p(y|x)和用户接受度多目标优化。
% 我再大胆暴论一下AIUI，用户有（a,b,c）的模糊不开心，可能的需求意图是（d,e,f）,潜在解决方案（g,i,h）。我们需要找到三者的关系。其次是沟通层面，确定需求意图，主要是通过提问，找到一个/多个意图x使p(x|a,b,c)最大。然后是解决阶段，主要是通过沟通，找到一个/多个解决方案y使p(y|x1，x2)和用户接受度多目标优化。

% 1. 依赖人类咨询师的行业流程和处理个案的思考模式，我们设计并开发了一个Agent咨询师的框架。在内容、对来访进行的操作上，我们基本依赖和咨询师讨论得出的流程，包括如何开始接待个案、如何进行个案概念化、如何进行治疗计划、如何通过每个步骤进行干预等。但是，由于AI心理咨询师相关的研究已经进展了很多年，因此我们提出framework时也会囊括研究以往AI\HCI领域提出的咨询师框架或者某些特点功能的实现方式（比如共情、个案理解概念化、个性化治疗和治疗技巧信息使用等等）。

% 2. 问题定义；按照section

This section introduces the CA+ framework, an innovative framework for LLM-based psychological counseling. In the first part (Section \ref{sec:module}), we will explore the main components of the CA+ framework. The second part (Section \ref{sec:DC1} to \ref{sec:DC5}) of this section focuses on the practical implementation of the framework's design goals. Here, we will elaborate on five specific approaches. Through iterative refinement with participant feedback and professional oversight (Section \ref{sec:implementation}), we developed a system that addresses both technical challenges and interaction requirements unique to AI-facilitated counseling.

% \subsection{Framework Architecture and Module Design}
% \label{sec:module}
% The CA+ framework introduces an innovative approach to LLM-based agent design for AI-assisted psychological counseling. This framework enhances LLMs with specialized components to create a more engaging AI counselor. 

% As illustrated in Figure X, the framework consists of two main modules: "What to do" and "How to do." The "What to do" module implements a three-layer hierarchical planning structure (Goals, Session, and Action Sequence) with both forward planning and backtracking capabilities. Based on the current state $(St)$, context, and client input, the system either proceeds with forward planning if the current strategy is valid or initiates backtracking when adjustment is needed. The "How to do" module focuses on response generation by considering both the client's emotional state and attitude. It employs dual strategies - guidance and empathy - while retrieving relevant information from the counseling knowledge base. This bi-modular design enables the AI counselor to maintain both strategic coherence in therapeutic progression and appropriate emotional engagement in immediate interactions, creating a more comprehensive and adaptive counseling experience.
\subsection{Framework Architecture and Module Design}
%需要推敲，有的地方还是没说的足够清楚；比如S这块；名字就要看得出是什么；
%这里讲模块作用和输入输出的overview；

\label{sec:module}
The proposed framework introduces a novel approach to LLM-based psychological counseling by dynamically constructing a recursive, self-evolving decision graph that navigates complex problem-solving scenarios. As illustrated in Figure~\ref{fig:teaser}, our system leverages therapeutic knowledge and professional counseling patterns to implement this dynamic decision-making process through two main modules—the Therapy Strategies Module and the Communication Form Module—while supporting real-time operations with an integrated Information Management and Storage component. The core innovation lies in the system’s recursive decision-making process, which continuously refines the counseling solution by adapting its therapeutic strategies based on real-time client feedback and evolving contextual cues.

\begin{itemize}
    \item \textbf{Therapy Strategies Module:}
The Therapy Strategies module operates through a hierarchical three-layer structure comprising Goals, Session, and Action Sequence layers, implementing a bidirectional planning mechanism based on user feedback. When receiving client messages and historical context, the system first evaluates feedback to determine the planning direction. Positive feedback triggers forward planning, maintaining the current therapeutic trajectory while generating new actions. Conversely, negative feedback initiates a backward adjustment process, where the system collects feedback, updates the session agenda, and regenerates action sequences. This adaptive mechanism ensures therapeutic coherence while allowing for dynamic strategy adjustment based on real-time client responses.

    \item \textbf{Communication Form Module: }
The Communication Form module orchestrates strategy-driven communication adaptation by integrating guidance and empathy strategies. Through continuous client analysis based on profiles and preferences, the system maintains two parallel strategic pathways: a guidance strategy directing therapeutic progression and an empathy strategy managing emotional engagement. These strategies converge in the response generation component, which synthesizes appropriate responses while considering both client emotions and acceptance levels. This dual-strategy approach ensures both therapeutic effectiveness and emotional resonance in client interactions.

    \item \textbf{Information Management and Storage: }
The framework's information architecture centers on two primary databases working in concert. The client database maintains comprehensive records, including counseling histories, session agendas, action sequences, and detailed client profiles with preferences, that support real-time read/write operations and facilitate dynamic decision-making. The knowledge database incorporates professional therapeutic knowledge transformed through our designed mechanism, converting certified counseling literature and materials into structured data pairs. These knowledge pairs are systematically stored for efficient retrieval during counseling sessions. 

\end{itemize}

\subsection{DC1: Cognition Augmented Hierarchical Thinking}
\label{sec:DC1}

Consider a psychological counseling scenario where a client initially presents with ``work stress", but through interaction reveals deeper issues with work-life balance, career development anxiety, and family expectations. Our framework's multi-layer recursive architecture enables the system to dynamically navigate such complex, interrelated issues. When the client expresses immediate work pressure, the system can respond with specific coping strategies while simultaneously updating its understanding of underlying issues and adjusting longer-term therapeutic goals - much like how an experienced counselor would both address immediate concerns and develop deeper therapeutic insights.

\subsubsection{\textbf{Recursive Layer Architecture}}

The system implements a three-layer architecture where each layer defines rules for both node generation and conditional updates. Each layer maintains its planned content (Shown in Fig. \ref{fig:dc1_CASE}) until specific conditions trigger recursive updates:

\begin{figure}
  \includegraphics[width=\textwidth]{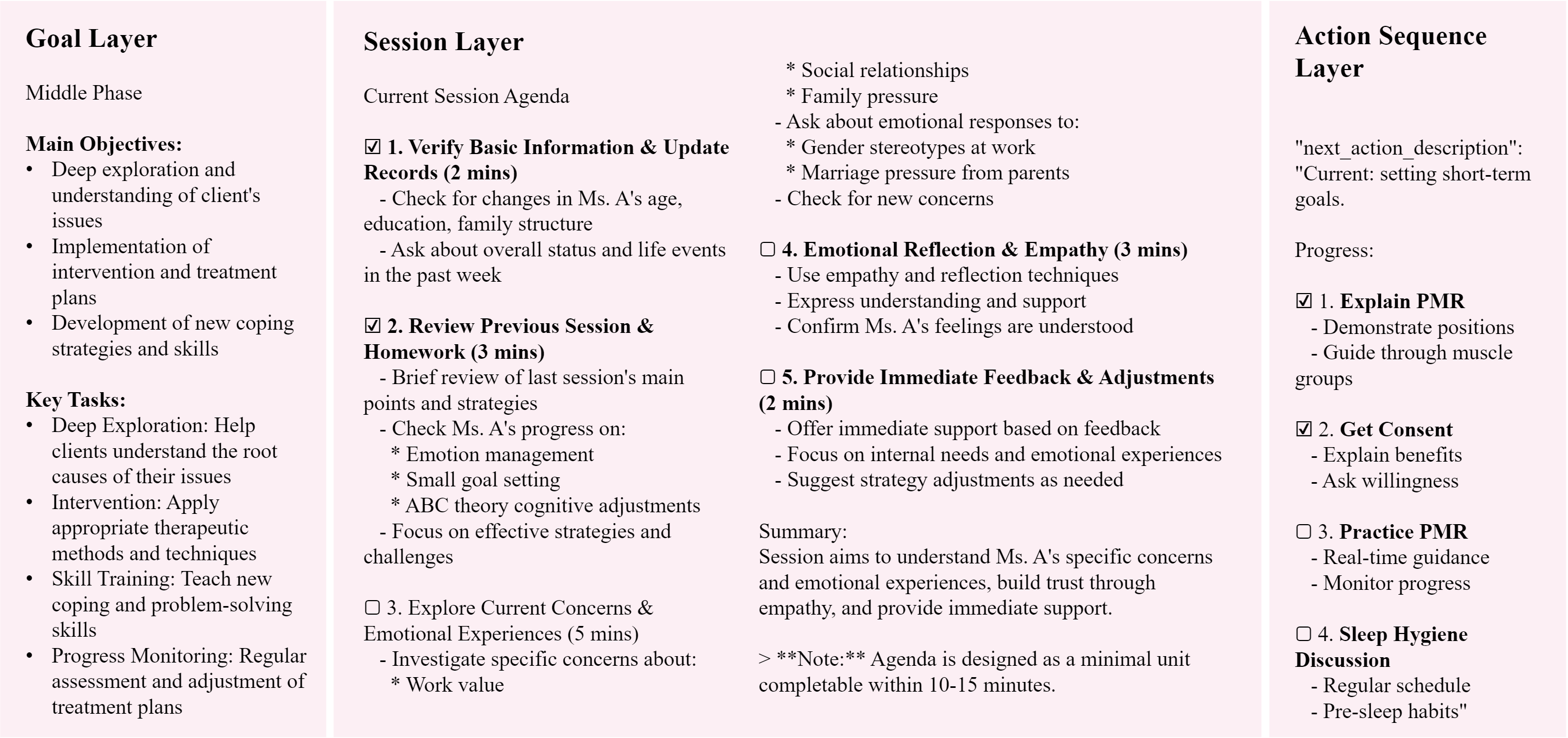}
  \caption{Examples of Three Layers Content in CA+ System.}
  \Description{}
  \label{fig:dc1_CASE}
\end{figure}

% 补一个示意图
\begin{itemize}
   \item \textbf{Goal Layer:} Establishes overarching therapeutic objectives across multiple sessions. Only when persistent pattern changes or significant therapeutic shifts are detected, based on continuous feedback from lower layers, does this layer trigger a recursive update. Otherwise, the original therapeutic objectives remain in effect.

    \item \textbf{Session Layer:} Derives session plans from parent goal nodes. When accumulated feedback indicates a significant deviation from expected progress—evaluated at intervals roughly corresponding to between 0.3 and 1 session—the layer recursively adjusts its planning. Otherwise, it follows the pre-established session agenda.
    
    \item \textbf{Action Sequence Layer:} Further decomposes session-level strategies into concrete therapeutic actions and responses. At time step $t$, only updates through recursive planning when $Client\ Message_t$ indicates a need for strategy adjustment. Otherwise, continues with the existing action sequence while maintaining alignment with higher-layer constraints.
    
\end{itemize}

\subsubsection{\textbf{Bidirectional Planning Mechanism}}
The system implements a bidirectional planning process guided by the evaluation of client feedback, as formalized in Algorithm \ref{alg:1}. This mechanism operates through:

\begin{itemize}
  \item \textbf{Top-down planning:} The algorithm begins with therapy goals ($G$) at the highest level, which inform session agendas ($S$) at the middle level, which in turn generate specific action sequences ($A$) at the lowest level.

  \item \textbf{Bottom-up adjustment:} When client responses indicate ineffective strategies (negative attitude detected in feedback evaluation), the system propagates adjustments upward through recursive function calls—from action sequences to session agendas, and if necessary, to therapy goals.
\end{itemize}

The algorithm's hierarchical structure evaluates client feedback at the action sequence level ($RunActionSeq$), which may trigger adjustments at the session level ($RunSession$), potentially escalating to goal-level modifications ($RunGoal$) when deeper changes are required. Each layer can dynamically recalibrate its approach based on client feedback, ensuring that therapeutic interventions remain responsive to client needs at the appropriate level of abstraction. Consult Appendix \ref{sec:Appendix_DC1} for more comprehensive information.

\begin{algorithm}
\caption{Cognitive Augmented Therapy Planning}
\label{alg:1}
\begin{algorithmic}[1]
\Require $msg$: client message, $ctx$: therapy context
\Require $G$: therapy goal, $S$: session agenda, $A$: action sequence, $fb$: client feedback

\Function{RunGoal}{$fb$} \Comment{Manages top-level therapy goals}
    \State $session\_adjust \gets \text{RunSession}(action=reflect, fb)$
    \If{$session\_adjust$} \Comment{Goal-level adjustment needed}
        \State $G \gets LLM_{adjust}(fb, G, ctx)$ \Comment{Update goals}
        \State $S \gets LLM_{generate}(G, ctx)$ \Comment{New agenda}
        \State $\text{RunSession}(action=execute, fb)$
    \EndIf
\EndFunction

\Function{RunSession}{$action, fb$} \Comment{Manages mid-level session flow}
    \If{$action = reflect$} \Comment{Reflection mode - evaluate progress}
        \State $adjust \gets \text{RunActionSeq}(action=reflect, action\_seq=old\_action\_seq, fb)$
        \If{$adjust = false$} \Comment{Continue current session}
            \State $A_{continue} \gets LLM_{generate}(S, ctx)$ \Comment{Generate next actions}
            \State $\text{RunActionSeq}(action=execute, action\_seq=A_{continue}, fb)$
            \State \Return $false$ \Comment{No session adjustment needed}
        \Else
            \State \Return $true$ \Comment{Session adjustment needed}
        \EndIf
    \ElsIf{$action = execute$} \Comment{Execution mode - implement therapy}
        \State $A_{new} \gets LLM_{generate}(S, ctx)$ \Comment{New action sequence}
        \State $\text{RunActionSeq}(action=execute, action\_seq=A_{new}, fb)$
        \State \Return $false$ \Comment{Return value not used}
    \EndIf
\EndFunction

\Function{RunActionSeq}{$action, action\_seq, fb$} \Comment{Manages bottom-level specific actions}
    \If{$action = reflect$} \Comment{Reflection mode - assess feedback}
        \State $attitude \gets LLM_{evaluate}(fb, action\_seq, ctx)$ \Comment{Evaluate feedback attitude}
        \If{$attitude$ is positive}
            \State \Return $false$ \Comment{No adjustment needed}
        \Else
            \State \Return $true$ \Comment{Adjustment needed}
        \EndIf
    \ElsIf{$action = execute$} \Comment{Execution mode - perform therapy actions}
        \For{each $act$ in $action\_seq$}
            \State $response \gets LLM_{response}(msg, ctx, G, S, act)$
            \State Present $response$ to client
            \State Update $msg$ and $fb$ based on client's reply
        \EndFor
        \State \Return $false$ \Comment{Return value not used}
    \EndIf
\EndFunction
\end{algorithmic}
\end{algorithm}

\subsection{DC2: Conceptualization-Driven Implicit Client Profiling}
\label{sec:DC2}
%分成：1、内容记录；2、更新规则（机制），如何被咨询进程影响（更新）和如何影响咨询流程；
% 1.内容记录：参考真实心理咨询，个案概念化表和咨询记录。通过两类表，来整体记录了client的个人情况和咨询的情况。
% 对于个案概念化表，它存在自我维护的属性。对于新用户，我们预设了有重要字段的、内容空白的基础表“表”。这种空白会在初期将咨询导向信息收集任务（在后面会详细说明）。字段上涵盖对个案的情况假设......等等（例子见附录）；
% 咨询记录表，用以记录咨询的情况。为了在咨询过程中细粒度记录内容的同时维护长期的咨询记录，分别设置了单次session的咨询记录和对于每次咨询总结的总咨询记录表，内容包括但不限于会谈背景、主要问题、治疗计划等（例子见附录）。
% 2.机制，表的完善程度影响目前咨询的状态。
% 参考了真实的咨询阶段，开始阶段（Beginning Phase）-了解来访者的主要问题和背景，中间阶段（Middle Phase）-深入探索和理解来访者的问题，实施干预和治疗计划，结束阶段（Ending Phase）-总结和巩固治疗成果，准备结束咨询。这影响着咨询的$goal$。
% 在个案概念化信息不清楚的时候（表不满），会将阶段设置为开始阶段。此时会影响agenda更多为信息收集和关系建立；收集完毕了就可以进入中间阶段。最后会结合咨询记录来共同决定是否结束。这种判断发生在session开始前的agenda生成。设置了机制，收集/下一步的咨询。这个个案概念和咨询记录，也会随着咨询的推进继续发展，不断更新。

% 维护动态的数据结构表单；
% [+]
% 填空

% 简单的表单更新不需要写伪码，写函数上去即可；

Building on the hierarchical planning framework established in DC1, our second design consideration addresses how AI counseling systems can develop and maintain comprehensive client understanding without disrupting natural therapeutic dialogue. While DC1 provides the cognitive structure for therapeutic planning, DC2 focuses on the knowledge representation that informs these plans.

The CA+ system implements an implicit client profiling mechanism that automatically builds and maintains comprehensive client understanding through natural therapeutic interactions. This design component leverages professional documentation practices to guide the counseling process while continuously updating client profiles.

\subsubsection{\textbf{Professional Documentation Framework}}
The system implements two interrelated documentation types that comprehensively capture both client information and therapeutic progress, as illustrated in Figure \ref{fig:profile_case}:

\begin{figure}
  \includegraphics[width=\textwidth]{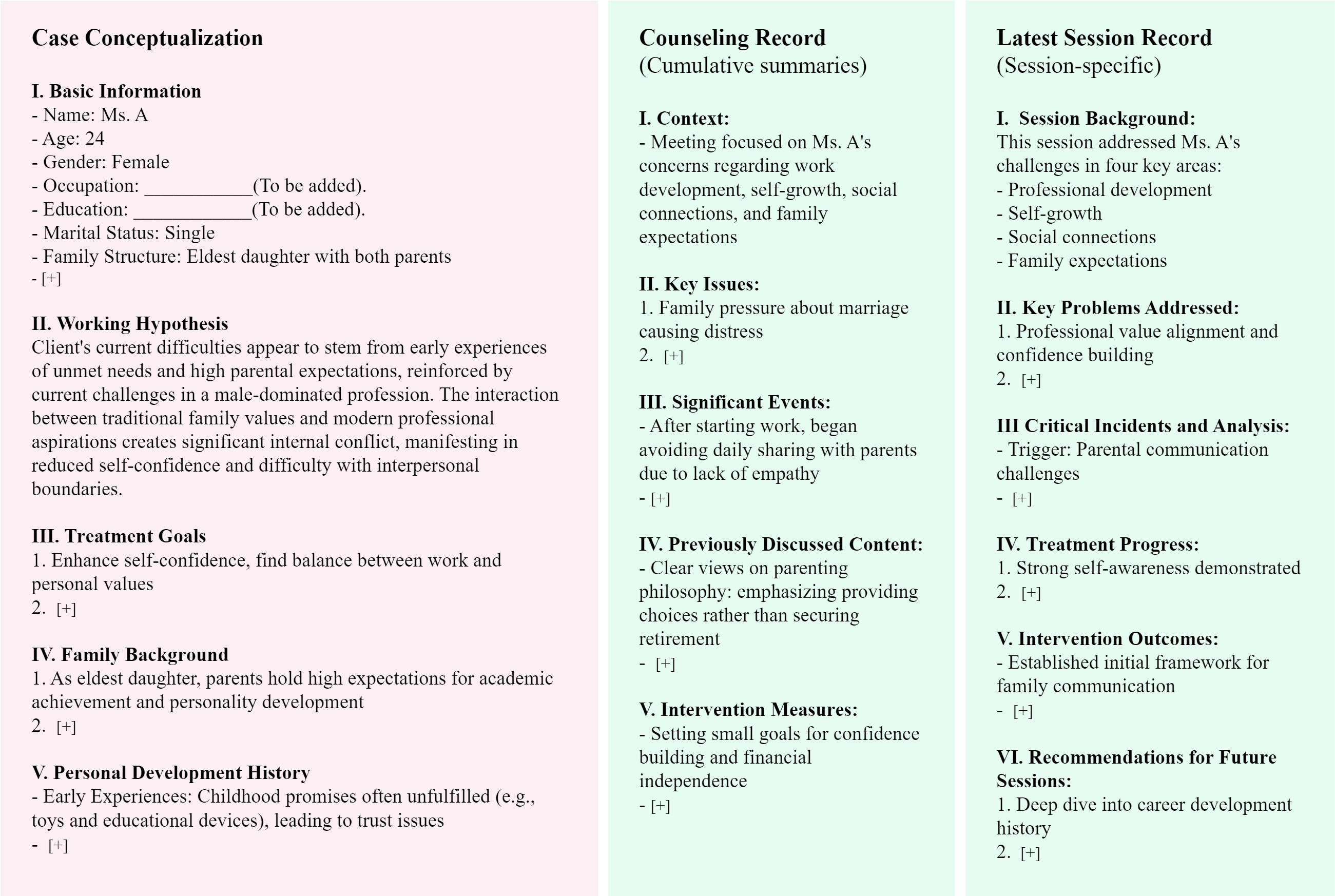}
  \caption{Examples of Case Conceptualization and Counseling Record Documentation in CA+ System.}
  \Description{}
  \label{fig:profile_case}
\end{figure}

\textbf{Case Conceptualization: } The system initializes a structured template covering essential clinical assessment fields such as demographic information, presenting problems, family dynamics, and clinical hypotheses. This approach ensures comprehensive client understanding before intervention.

\textbf{Counseling Records: } The system employs dual-level documentation for Counseling Records. First, it maintains session-specific records that capture granular therapeutic interactions. Second, it compiles cumulative progress summaries that synthesize the overall therapeutic trajectory.

\subsubsection{\textbf{Documentation-Driven Management Mechanism}}
The system implements a continuous documentation and phase management cycle through $UpdateDocuments$ and $DetermineGoals$ functions.

\textbf{Documentation Update Process:}
The $UpdateDocuments$ function processes the current message ($msg_t$) and recent context ($ctx_{t-10:t-1}$, the context of 10 messages) to maintain therapeutic documentation. Through LLM-based analysis, it generates:

\begin{itemize}
    \item Updated client profile ($Profile_t$) via $ProfileGenerationRule$, building on previous profile ($Profile_{t-1}$)
    \item Session record ($CounselingRecord_t$) via $RecordGenerationRule$, incorporating previous record ($Record_{t-1}$)
\end{itemize}

\textbf{Phase-Based Process Control:}
The $DetermineGoals$ function uses updated documentation ($Profile_t$, $Record_t$) to guide therapeutic progression through phase rule:

\begin{itemize}
    \item The \textbf{Initial Phase} focuses on alliance building and comprehensive assessment. This phase is activated when the case conceptualization does not yet meet predefined completeness criteria, thereby naturally directing interactions toward systematic information gathering through intentional documentation of emerging client details.
    \item The \textbf{Middle Phase} enables deeper exploration and active intervention implementation. The system transitions to this phase once the client profile reaches a predetermined level of completeness, indicated by the availability of key data points, thereby allowing for the application of targeted therapeutic techniques.
    \item The \textbf{Ending Phase} consolidates therapeutic gains and prepares for termination. The system determines this transition through joint analysis of profile completeness and documented progress.

\end{itemize}

This cyclical mechanism ensures therapeutic progress aligns with documentation completeness, enabling systematic adaptation to client needs while maintaining professional standards. Crucially, it complements DC1's hierarchical planning by providing the evolving knowledge base that informs goal setting and strategy selection across all three planning layers. Appendix \ref{sec:Appendix_DC2} contains the complete documentation.

\subsection{DC3: Book-style Data Generation and Retrieval}

\begin{figure}
  \includegraphics[width=0.85\textwidth]{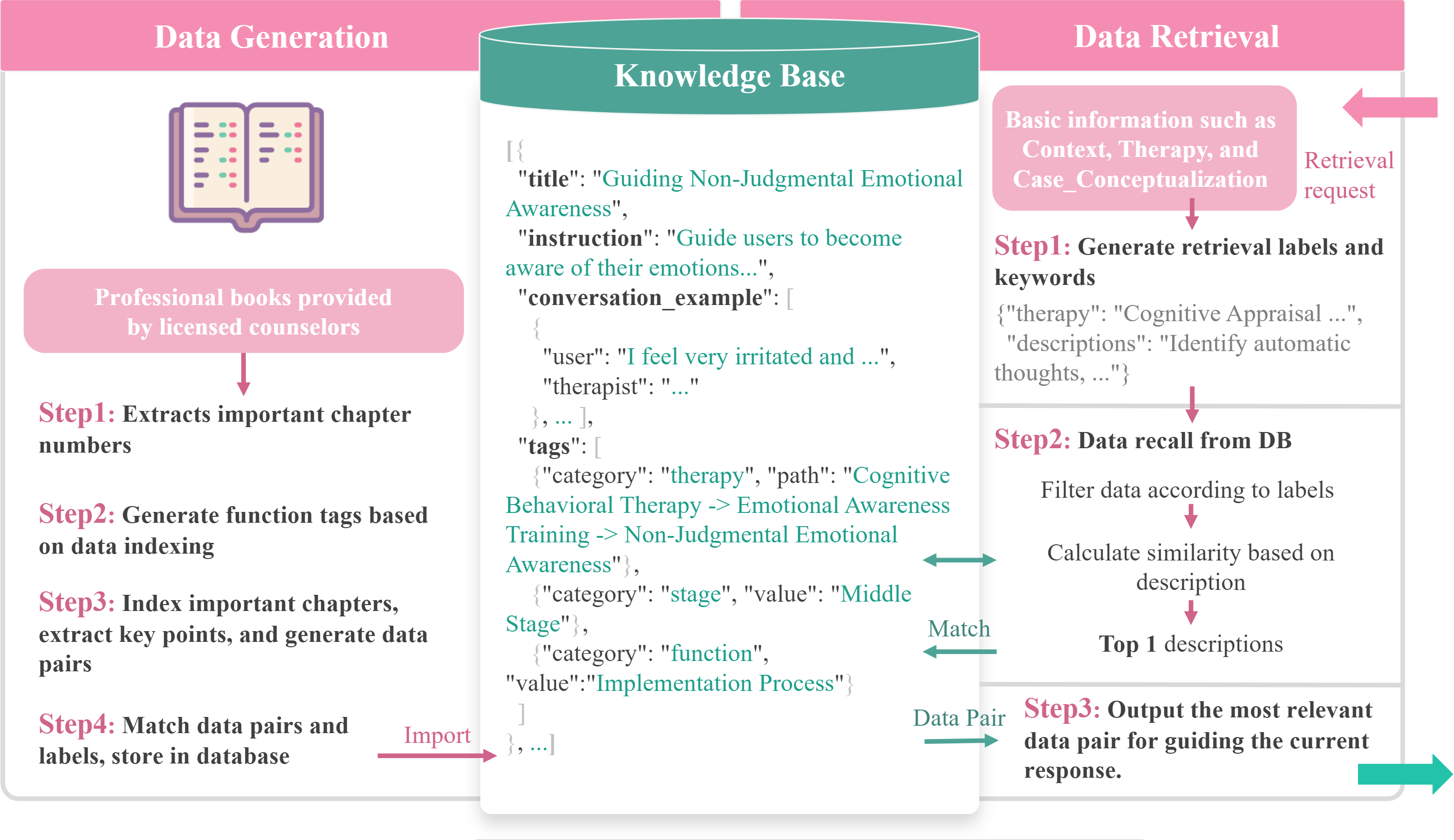}
  \caption{\textbf{Book-style Data Generation and Retrieval.} This framework illustrates the process of generating and retrieving structured data for therapeutic use, leveraging a knowledge base to enhance counseling interventions.
}
  \Description{The Book-style Data Generation and Retrieval framework facilitate the organization and application of professional counseling resources. In the Data Generation phase, key chapter sections from professional resources are extracted and indexed, generating function tags that categorize the content based on therapeutic goals like enhancing motivation and emotional awareness. These indexed chapters form data pairs containing key points and instructions, which are stored in a knowledge base. During Data Retrieval, retrieval labels and keywords are generated based on context and therapy needs, enabling the system to recall relevant data from the database. The process filters and matches data according to these labels, calculating similarity to ensure the most pertinent information guides the current therapeutic response. This structured approach allows for efficient use of licensed counseling resources, providing targeted support and enhancing the effectiveness of therapeutic interactions.}
  \label{fig:data}
\end{figure}

% 蒸馏出来精华的内容；
% 左边画了非常多BOOK等，提取关键知识

CA+ framework distills professional counseling literature into actionable knowledge through intelligent extraction and context-aware retrieval processes, centered around a structured knowledge base (Fig. \ref{fig:data}).

\textbf{Knowledge Distillation Process:}
Our system processes authoritative Chinese counseling textbooks widely recommended by all interviewed counselors and covering common therapeutic approaches including CBT. These resources represent the professional standard in Chinese psychological practice and serve as our foundation for knowledge extraction.

We transform comprehensive counseling texts into precise, applicable guidance through: (1) systematic chapter organization preserving source hierarchy, (2) functional tagging for practical categorization, (3) extraction of essential instruction-example pairs demonstrating key therapeutic techniques, (4) comprehensive metadata annotation facilitating targeted retrieval. The knowledge base stores these distilled insights with rich contextual markers including practical instructions, illustrative dialogue examples, and multi-dimensional therapeutic classification tags.

\textbf{Contextual Retrieval Process:}
During counseling interactions, the system analyzes current conversation context ($ctx_{t-10:t-1}$), therapeutic approach, and client profile ($Profile_t$) to: (1) identify relevant therapeutic needs, (2) match these needs against knowledge metadata, (3) select the most appropriate therapeutic guidance based on contextual relevance.

This bidirectional framework ensures LLM responses incorporate professional therapeutic principles tailored to specific client needs and conversation dynamics. Detailed implementation is described in Appendix \ref{sec:Appendix_DC3}.

\subsection{DC4: Adaptive Empathy and Ecological Self}
\label{sec:DC4}
% 更加结构化写法；
% 重要的机制是，共情维护和个性化分析记录。此外也设置了agent的人设用以支持共情机制；

% 落到那个象限，给什么策略；

\begin{figure}
  \centering
  \begin{subfigure}[b]{0.47\textwidth}
    \includegraphics[width=\textwidth]{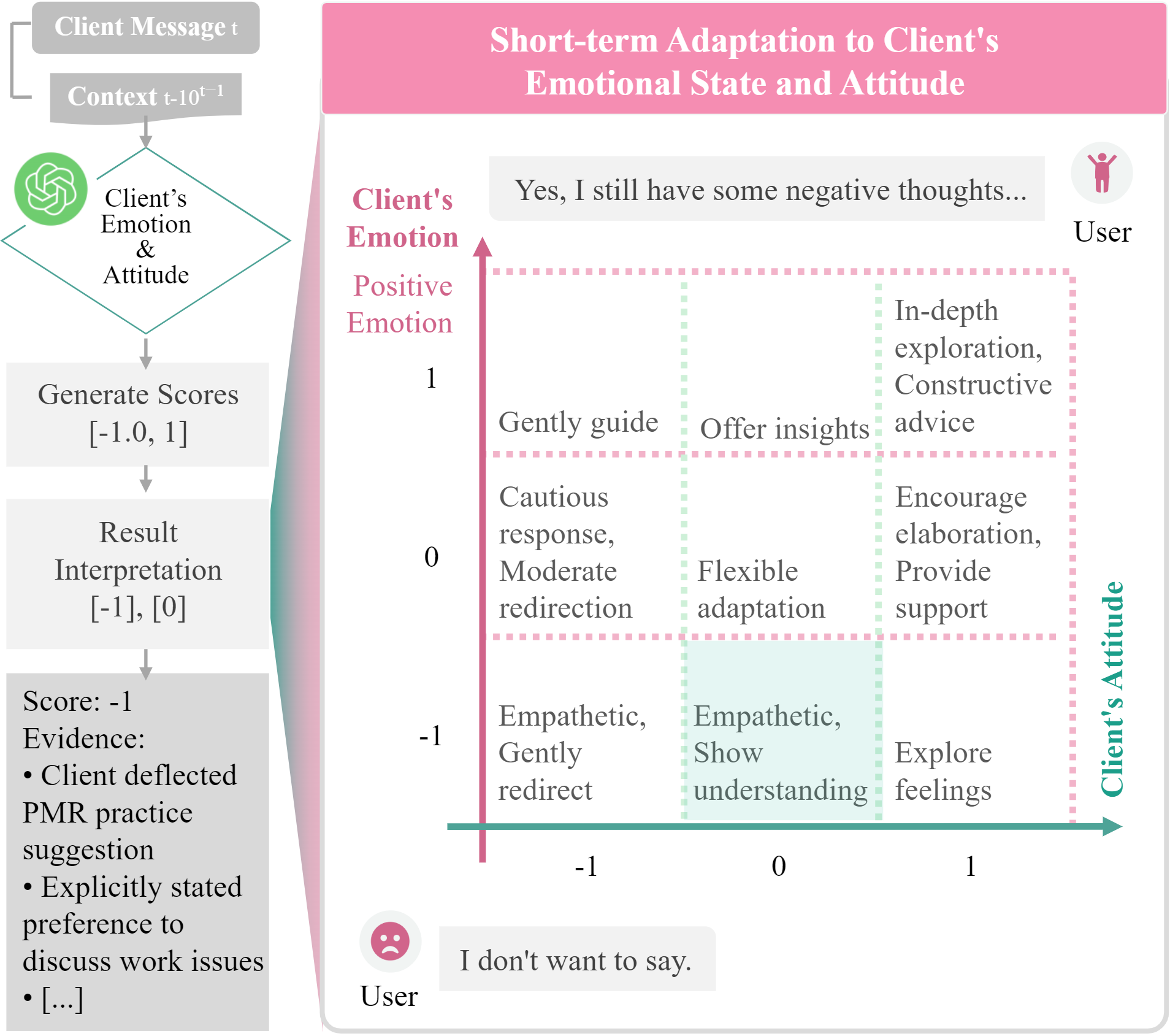}
    \caption{Adaptive empathy mechanism mapping client emotions and attitudes to therapeutic responses}
    \label{fig:DC4empathy}
  \end{subfigure}
  \hspace{-0.05cm}
  \begin{subfigure}[b]{0.49\textwidth}
    \includegraphics[width=\textwidth]{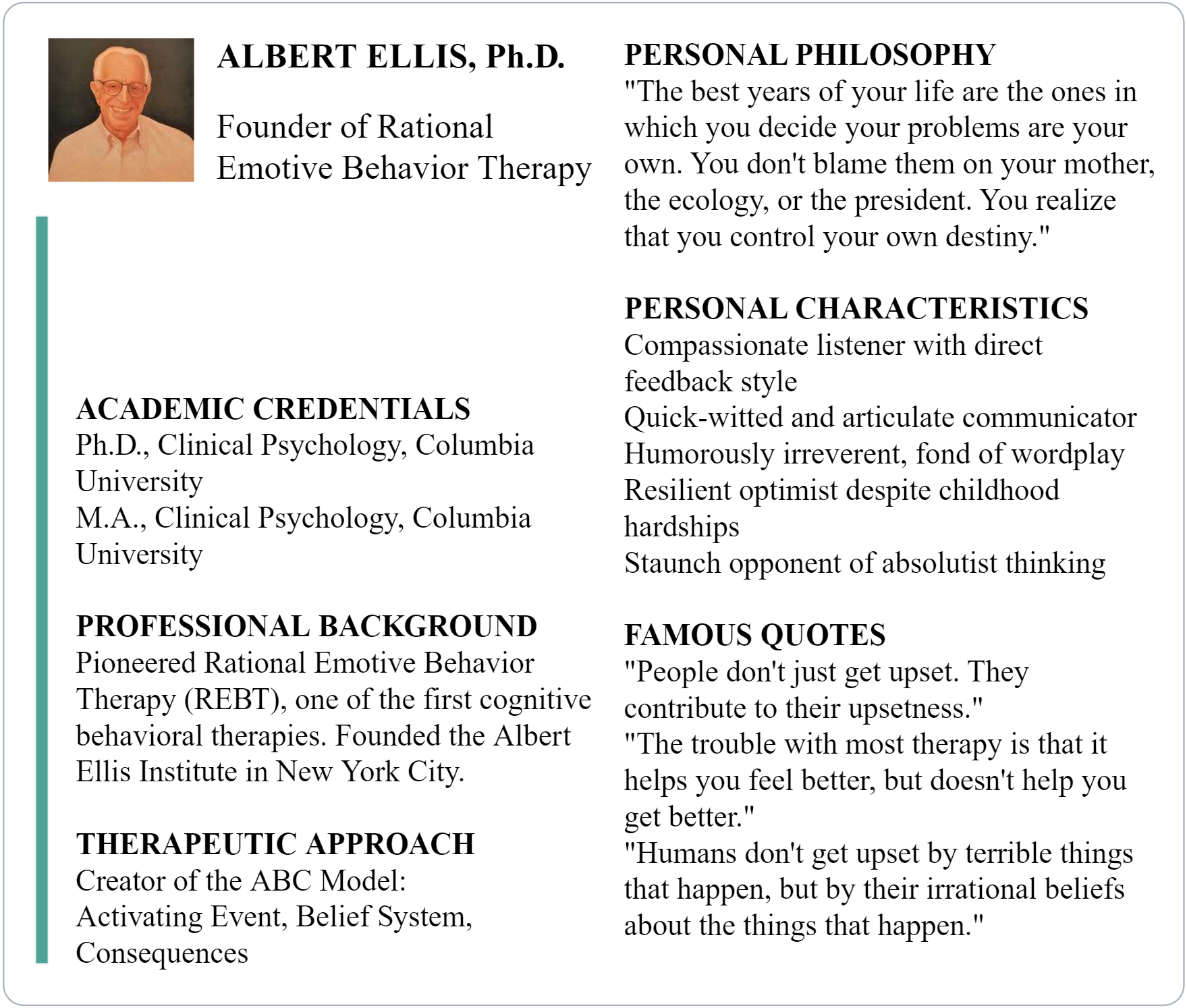}
    \caption{Example of agent's Ecological Self implementation based on Albert Ellis's therapeutic persona}
    \label{fig:DC4self}
  \end{subfigure}
  \caption{The DG4 approach enhances therapeutic interactions through two key mechanisms: adaptive empathy that balances immediate responsiveness with long-term personality, and an ecological self-framework that integrates professional competence with personalized character traits.}
  \Description{}
  \label{fig:DC4combined}
\end{figure}

The DG4 approach enhances therapeutic interactions through two key mechanisms: adaptive empathy that balances immediate responsiveness with a consistent, enduring therapeutic persona, and an ecological self-framework that integrates professional competence with personalized character traits, as illustrated in Figure \ref{fig:DC4combined}.  Consult Appendix \ref{sec:Appendix_DC4} for more comprehensive information.

\textbf{Adaptive Empathy and Personalization Mechanism: }
The system implements dual-temporal adaptation through two core LLM components (Figure \ref{fig:DC4empathy}):

\begin{itemize}
    \item \textbf{Real-time Detection} component of the LLM processes the recent conversation context ($ctx_{t-10:t-1}$) and the current message ($msg_t$) using pre-designed guided prompts that extract key emotional signals. Based on the detection results, the system dynamically selects appropriate therapeutic strategies—such as prioritizing empathetic reflection and validation when severe negative emotions are detected, or gradually shifting to deeper therapeutic exploration as the client’s emotional state shows measurable improvement (e.g. through sentiment scores exceeding a threshold).
    
    \item Personalization LLM maintains \textbf{long-term client understanding} by processing current interaction strategies with historical profiles ($Profile_t$) and records ($Record_t$). It tracks communication patterns, emotional triggers, and interaction preferences, continuously updating personalization strategies to strengthen therapeutic alignment over time.
\end{itemize}

\textbf{Ecological Self: }
As a demonstration of our framework, we select Albert Ellis, the founder of rational emotive behavior therapy, as the therapeutic persona. The ecological self design establishes his therapeutic persona through two key components (Figure \ref{fig:DC4self}):

\begin{itemize}
    \item \textbf{Professional Foundation:} Base prompts define core therapeutic competencies, including rational emotive behavior therapy principles, ethical guidelines, and standard intervention protocols, ensuring consistent therapeutic delivery.

    \item \textbf{Interaction Style:} Building on this foundation, additional prompts enable characteristic expressions aligned with Ellis's therapeutic approach, contextual interventions following his methodology, and culturally-aware responses while maintaining professional standards.
\end{itemize}

This framework enables the agent to maintain Albert Ellis's therapeutic approach while providing personalized, emotionally attuned support through systematic prompt engineering.

\subsection{DC5: Adapting to AI-Specific Client Expectations}
\label{sec:DC5}

% 把prompt塞进来

%这一章主要是guideline
% style；role-play下位者；扮演一个人类眼中的agent应该给的交互形式。
% 人类眼中的agent是下位者，模仿的是communication中更弱势的角色。提供固定的语气和详实的解释、认可。
% instruction-how

% 帮助用户理清总结自己的真实情况，自我认知
% 知识性内容提供、说明（解释为用户可理解的、结合情景的表述）
% 消除顾虑、信息披露、信任获取、信息保护等；

% 怎么样像一个机器人？解决client对agent，对角色预期的问题；用户对agent的认知和需求；用户没有把agent当做是人？agent和用户不平等，处于弱势地位；在沟通上，更客气、委婉和更好的理解；知识以外；沟通地位“讨好”；

% 如何以这种地位去进行沟通的guideline；（跟目前的沟通理论做结合，下位的沟通技巧，不完全=迎合）；达成目的；让用户更轻松地理解和认同它；和人类咨询师不同，因为agent没有天然的信任和平等（甚至coach的感觉）；比解释信息更复杂，在满足人对AI 的交互预设；解释是其中一种；

\begin{figure}
  \includegraphics[width=\textwidth]{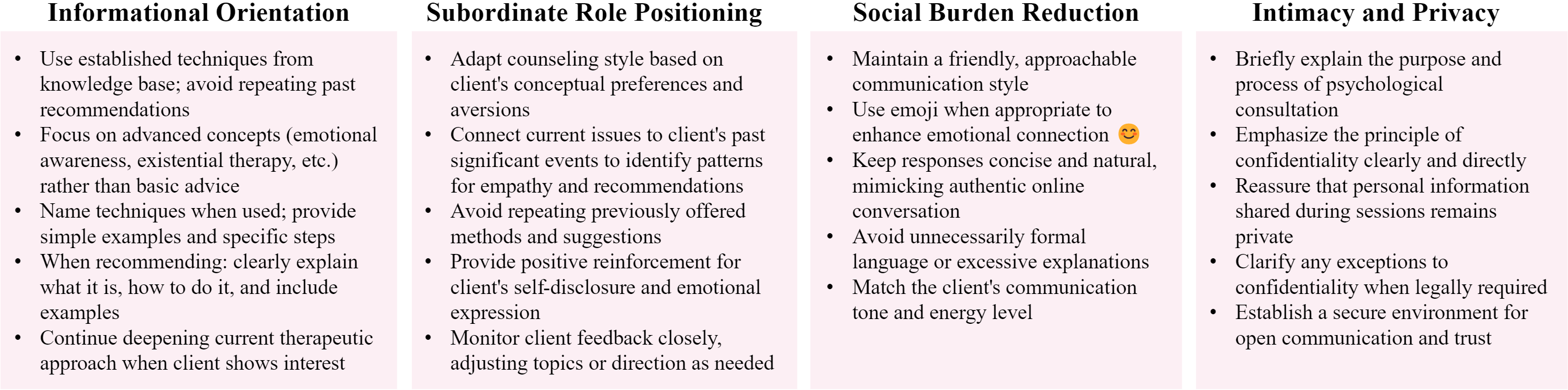}
  \caption{Examples of DC5 prompt.}
  \Description{}
  \label{fig:DC5_prompt}
\end{figure}

While AI counseling agents strive to emulate human therapeutic qualities, they must simultaneously address clients' inherent preconceptions about interacting with artificial intelligence. This design component provides guidance for how agent prompts (See Figure \ref{fig:DC5_prompt} for details) should be crafted to align with these unique expectations.

\textbf{Informational Orientation:}
Emphasize concrete information and evidence-based suggestions in prompts to align with client expectations. Design system responses to deliver comprehensive, precise information with greater detail than typical human counseling, satisfying the analytical expectations users bring to AI interactions.

\textbf{Subordinate Role Positioning: }
Leverage language that positions the AI as helpful and service-oriented, acknowledging the hierarchical relationship clients naturally establish. Employ communication patterns that reinforce appropriate deference while maintaining therapeutic value, allowing clients to engage in their preferred directive communication style.

\textbf{Social Burden Reduction: }
Explicitly acknowledge and leverage clients' reduced concern about "bothering" the AI system. Design prompts that minimize perceived social constraints in therapeutic exchanges, creating space for clients to express needs and concerns without the hesitation common in human-human therapeutic relationships.

\textbf{Intimacy and Privacy: }
Account for clients' accelerated intimacy development and tendency to bypass normal social distance barriers when interacting with AI systems. Incorporate appropriate responses to early sensitive disclosures while maintaining professional boundaries, recognizing that clients often share personal information more readily due to reduced fears of human judgment.

% \textbf{Informational Orientation}
% \begin{itemize}
%     \item Emphasize concrete information and evidence-based suggestions in prompts
%     \item Design for comprehensive, precise information delivery with greater detail than typical human counseling
% \end{itemize}

% \textbf{Subordinate Role Positioning}

% \begin{itemize}
%     \item Using language that positions the AI as helpful and service-oriented
%     \item Employ language patterns that reinforce appropriate deference while maintaining therapeutic value
% \end{itemize}

% \textbf{Social Burden Reduction}

% \begin{itemize}
%     \item Explicitly acknowledge clients' reduced concern about "bothering" the AI system
%     \item Design prompts that reduce perceived social constraints in therapeutic exchanges
% \end{itemize}

% \textbf{Intimacy and Privacy}

% \begin{itemize}
%     \item Account for clients' accelerated intimacy development and bypassing of normal social distance barriers
%     \item Incorporate appropriate responses to early sensitive disclosures
% \end{itemize}

This approach leverages clients' AI-specific expectations rather than fighting against them. As suggested by research, acknowledging the non-human status may enhance rather than detract from therapeutic effectiveness in AI counseling contexts.

\subsection{System Implementation and Refinement}
\label{sec:implementation}
Our development process emphasized continuous improvement and safety through systematic validation with participant feedback and professional oversight, addressing key technical and interaction challenges before proceeding to larger-scale evaluation.

\textbf{Technical Implementation Details:}
Our system is built using a standard web stack. The backend is developed with Django, a Python web framework, while the frontend uses Vue.js to create the user interface. We chose this combination for its simplicity and effectiveness in web application development.

The core functionality of our system relies on the GPT-4o API (version 2024-08-06), provided by OpenAI, which supports the conversational AI capabilities. We integrated Langchain to help manage and structure the conversations more effectively. Our main prompts can be found in Appendix ~\ref{sec:Appendix_DCs}. For data storage, we use a MySQL database. This stores client information, conversation logs, and other necessary data for the system's operation.

We initialized the CA+ framework with a role based on Albert Ellis, the founder of Rational Emotive Behavior Therapy. The system is deployed on a server, allowing study participants to access it via web browsers, as demonstrated in Appendix~\ref{app:UIterface}.

\textbf{Iterative Prototype Refinement and Hallucination Prevention:}
To validate and improve our system, we conducted a preliminary study with 18 participants (9 females and 9 males). The participants' ages ranged between 20 and 30 years old, with an average age of 23.9 years ($\sigma = 2.7$). This study aimed to gather client feedback for further refinement while ensuring the system's safety and reliability under careful supervision.

This preliminary experiment was conducted under the direct oversight of licensed counselors and research team members who monitored all interactions in real-time. We implemented an iterative testing approach, with participants divided into six consecutive batches of three people each. After each batch completed three days of counseling sessions, our team conducted comprehensive reviews of the interactions, identifying issues and implementing system improvements before proceeding to the next batch of participants.

In early iterations, we identified several instances of factual discrepancies (approximately 2.7\% of responses in the first batch), primarily related to specific therapeutic techniques or statistical claims about intervention efficacy. To address these issues, we progressively implemented three key improvements: (1) enhancing the knowledge retrieval component with verified therapeutic resources, (2) implementing stricter guardrails around statistical claims, and (3) adding explicit uncertainty expressions when discussing intervention efficacy without strong evidence.

By the final batch of participants, our systematic refinements reduced factual discrepancies to a negligible level (under 0.2\% of responses), with no instances of harmful therapeutic advice detected across 27 complete counseling sessions. This progressive reduction in hallucinations demonstrates the effectiveness of our iterative approach to system development and safety assurance.

Through interviews and backend data analysis, we identified two additional key areas for improvement. Firstly, clients showed impatience with lengthy sessions, indicating that online counseling requires a more concise approach than traditional face-to-face sessions. In response, we shortened the overall counseling agenda, modularized the sessions, and streamlined the process of follow-up questions and clarifications. This adjustment aimed to maintain client engagement and create shorter feedback loops.

Secondly, we observed that unlike typical LLM role-playing Q\&A scenarios that often provide brief responses \cite{shanahan2023role}, counseling sessions benefit from more detailed information and actionable guidance. To address this, we enhanced the AI's responses to be more comprehensive, preventing clients from deviating from the topic and ensuring a thorough understanding of the counseling process. We also optimized the output format to improve the clarity and actionability of the content.

These iterative refinements, conducted under careful professional supervision, have established a foundation of safety and effectiveness necessary for proceeding to larger-scale clinical trials with broader participant demographics and more diverse counseling scenarios.

\section{Study 1: Comparative Clients Experience Study}
\label{sec:study1}

This section presents a comprehensive evaluation of the CA+ system by comparing a baseline version (with key design considerations ablated) and the full CA+ version through a three-day experiment involving 24 participants. We employ a mixed-methods approach, combining quantitative measures and qualitative insights to assess system performance, client experience, and potential therapeutic impact. Approval for the experiment was obtained from the Institutional Review Board (IRB).

% \subsection{Study Design}
% This study presents a comparative analysis of two distinct versions of an AI counseling system, each characterized by different configurations of functional components. The objective of this analysis is to evaluate users' overall experiences while interacting with the developed AI psychological counseling system. Approval for the experiment was obtained from the Institutional Review Board (IRB). A total of 26 participants were recruited, with each cycle spanning three days. Participants were randomly assigned to either the Baseline or Full Framework groups. The research design encompassed various components, including questionnaires, system usage experiments, interviews, behavioral analysis, and automated assessments. Through the collection and analysis of user feedback and behavioral data, this study seeks to investigate the influence of different system configurations on user experience. This methodological approach is structured to facilitate a comprehensive evaluation of both subjective user experiences and objective behavioral outcomes, thereby providing valuable insights into the intricate relationship between system configuration and user experience.

\begin{table*}
\caption{Demographic Information of Participants (CA: CA+ Group; BL: Baseline Group)}
\label{tab:demographics}
\begin{tabular}{@{}lllllll@{}}
\toprule
ID & Counseling Exp. & AI Familiarity & Mental Health Issues & Age & Gender & Occupation \\
\midrule
CA-01 & Sometimes & Familiar & Often & 20 & F & Student \\
CA-02 & Sometimes & Used apps & Sometimes & 22 & F & Student \\
CA-03 & Never & Expert & Always & 20 & M & Student \\
CA-04 & Sometimes & Used apps & Sometimes & 26 & F & Office worker \\
CA-05 & Never & Used apps & Sometimes & 33 & M & Manager \\
CA-06 & Never & Used apps & Often & 29 & F & Freelancer \\
CA-07 & Sometimes & Familiar & Never & 22 & M & Freelancer \\
CA-08 & Never & Familiar & Rarely & 25 & M & Office worker \\
CA-09 & Never & Familiar & Often & 24 & F & Professional \\
CA-10 & Sometimes & Familiar & Rarely & 44 & F & Self-employed \\
CA-11 & Regular & Used apps & Sometimes & 25 & F & Student \\
CA-12 & Sometimes & Used apps & Sometimes & 22 & F & Student \\
BL-01 & Never & Familiar & Sometimes & 20 & F & Student \\
BL-02 & Sometimes & Familiar & Sometimes & 28 & F & Self-employed \\
BL-03 & Sometimes & Used apps & Rarely & 21 & M & Freelancer \\
BL-04 & Sometimes & Familiar & Rarely & 20 & M & Student \\
BL-05 & Regular & Familiar & Sometimes & 22 & F & Student \\
BL-06 & Never & Used apps & Rarely & 22 & M & Student \\
BL-07 & Never & Familiar & Rarely & 28 & M & Student \\
BL-08 & Never & Used apps & Rarely & 23 & F & Professional \\
BL-09 & Sometimes & Used apps & Never & 20 & F & Student \\
BL-10 & Sometimes & Used apps & Rarely & 27 & F & Office worker \\
BL-11 & Sometimes & Used apps & Sometimes & 25 & M & Student \\
BL-12 & Never & Familiar & Rarely & 22 & M & Student \\

\bottomrule
\end{tabular}
\end{table*}

% \begin{table*}
% \caption{Demographic Information of Participants (Baseline Group)}
% \label{tab:demographicsB}
% \begin{tabular}{@{}lllllll@{}}
% \toprule
% ID & Counseling Exp. & AI Familiarity & Mental Health Issues & Age & Gender & Occupation \\
% \midrule
% BL-01 & Never & Familiar & Sometimes & 20 & F & Student \\
% BL-02 & Sometimes & Familiar & Sometimes & 28 & F & Self-employed \\
% BL-03 & Sometimes & Used apps & Rarely & 21 & M & Freelancer \\
% BL-04 & Sometimes & Familiar & Rarely & 20 & M & Student \\
% BL-05 & Regular & Familiar & Sometimes & 22 & F & Student \\
% BL-06 & Never & Used apps & Rarely & 22 & M & Student \\
% BL-07 & Never & Familiar & Rarely & 28 & M & Student \\
% BL-08 & Never & Used apps & Rarely & 23 & F & Professional \\
% BL-09 & Sometimes & Used apps & Never & 20 & F & Student \\
% BL-10 & Sometimes & Used apps & Rarely & 27 & F & Office worker \\
% BL-11 & Sometimes & Used apps & Sometimes & 25 & M & Student \\
% BL-12 & Never & Familiar & Rarely & 22 & M & Student \\
% \bottomrule
% \end{tabular}
% \end{table*}

% \subsubsection{Discussion}

% The procedure for the preliminary study is the same as the content of the study design. We conducted two experimental batches in the preliminary study. 

% \subsection{User Study}

\subsection{Participants}

We openly recruited 28 people for this investigation. Four candidates were removed due to concerns regarding data quality, such as failing to answer trap questions or not meeting the minimum daily participation hours. The remaining 24 participants signed a detailed informed consent form to ensure they understood the study's goals, procedures, hazards, and privacy protections. Participants were 20–44 years old ($M = 25.58$, $SD = 5.24$); demographics in Tables \ref{tab:demographics}. We examined psychological counseling history, psychological difficulties, and LLM-based application experience when selecting participants. To improve sample representation, we balanced age, gender, occupation, and education between groups.

\subsection{Procedures}

A between-subjects design was employed with two conditions: the Baseline group (control) and the CA+ group (experimental). Participants were randomly assigned to one condition and interacted with the respective system version for 10 minutes daily over three consecutive days.

The Baseline group used an ablated version of the system where all five design considerations (DC1-5) described in Section \ref{sec:framework} were removed. This baseline system relied on a standard prompt-response architecture using the same GPT-4o model, but without the hierarchical planning structure, implicit client profiling, user expectation adaptation, integration of theoretical frameworks, or therapist-aligned communication patterns. It functioned essentially as a generic counseling chatbot that responded to user inputs without the specialized cognitive architecture or therapeutic planning mechanisms of our full system.

The Baseline system was configured with a simple system prompt as shown below:

\begin{flushleft}
\begin{dialogue}

\color{gray}{You are a supportive counselor who helps users with their mental health concerns. Your goal is to:

1. Respond empathetically to users' messages

2. Provide helpful suggestions based on psychological principles

3. Offer emotional support in difficult situations

4. Maintain a positive and encouraging tone
}
\end{dialogue}
\end{flushleft}

This baseline prompt represents a typical implementation of an LLM-based counseling agent without our proposed cognitive architecture enhancements. While the baseline system retained factual knowledge about psychological principles, it lacked the structured session planning, theoretical integration, and adaptive personalization capabilities of our full system. It did not maintain session-to-session memory beyond what was explicitly mentioned in the current conversation and did not employ our sophisticated planning mechanisms for structuring therapeutic interactions or setting session goals.

In contrast, the CA+ group used the complete version with all design principles implemented. Both conditions maintained identical user interfaces, used the same underlying language model, and followed identical interaction protocols to isolate the effects of our proposed design principles. The content differences between conditions were entirely attributable to the presence or absence of the five design considerations rather than differences in the base model or interface design.

Throughout the experiment, participants completed daily questionnaires. These included pre-test assessments of mental toughness, mental health, and self-efficacy on day one, followed by standardized post-test questionnaires after each daily system interaction. Two interviews were conducted: a 10-minute preliminary interview on day one and a 30-minute semi-structured interview on the final day. These interviews aimed to capture participants' experiences with the AI counseling system, including system impressions, mindset changes, and perceived effectiveness of support and personalization. Key findings from these interviews are detailed in Section \ref{subsec: Findings}.

\subsection{Measures}
\label{subsec:Measures}

This study employed a comprehensive set of measures to evaluate the CA+ system's performance, user experience, and interaction behaviors. The measures included:

\begin{itemize}
    \item \textbf{System Performance Evaluation:} System performance was assessed using both quantitative dialogue metrics and LLM-based automated evaluation of conversation logs from both experimental conditions. Quantitative metrics included session duration, response time, turn-taking patterns, and informativeness of the dialogue content. These metrics provided objective measures of engagement and interaction quality. Additionally, our custom-built LLM-based evaluation tool analyzed the conversation logs for key performance indicators including conversation management, therapeutic skills, cognitive capacities, and client participation. The LLM-based analysis results were cross-validated by two experienced counselors who independently reviewed a subset of the logs to ensure accuracy and alignment with human expert judgment. This multi-faceted approach to system performance evaluation allowed for a comprehensive assessment of both the quantitative aspects of the dialogue and the qualitative elements of the counseling interaction.
    
    \item \textbf{Client Self-Report Measures:} Participants completed post-session evaluations included the System Usability Scale (SUS)~\cite{schmidmaier2024perceived}, Perceived Empathy of Technology Scale (PETS)~\cite{schmidmaier2024perceived}, Self-disclosure Scale~\cite{rubin2020revised}, Session Evaluation Scale~\cite{stiles1994evaluation}, as well as our own System Feature Scale (Appendix \ref{app:Self-designed} for details) based on system design. These instruments provided insights into client satisfaction, emotional engagement, and mental well-being throughout the experiment. Questionnaires were administered online after each counseling session, with data automatically collected and stored securely. Analysis included both individual-level changes and group-level comparisons.

    \item \textbf{Behavioral Data Analysis:} Client behavior data was extracted from chat logs and analyzed for conversational quality and engagement. Metrics encompassed interaction dynamics (response rounds, timing, and session duration) and communication richness (information entropy and word count). Data was collected automatically through the system's backend logging mechanisms. Analysis was conducted using custom Python scripts to calculate the specified metrics, with results verified by two independent researchers.

    \item \textbf{Qualitative Insights:} Semi-structured interviews were conducted after the first day and at the conclusion of the experiment to gather qualitative insights into user experiences, complementing quantitative measures and providing deeper understanding of client perceptions, system efficacy, and areas for improvement. Interviews were audio-recorded, transcribed verbatim, and analyzed thematically \cite{braun2012thematic}. Three researchers independently coded the transcripts, and then met to discuss and resolve any discrepancies, ensuring inter-coder reliability. Key themes and illustrative quotes were selected through consensus.
\end{itemize}

These multi-faceted measures provided a comprehensive evaluation of CA+ system, encompassing both objective performance metrics and subjective user experiences. The combination of quantitative and qualitative data, along with rigorous collection and analysis methods, allowed for a nuanced and reliable understanding of the system's efficacy, client engagement, and potential areas for enhancement.

\subsection{Findings}
% 柱状图，箱图
\label{subsec: Findings}

Our comparative evaluation between CA+ and the baseline system revealed significant differences across four key areas: \textbf{system performance} (Section \ref{sec:study1_system}), \textbf{perceived counseling quality} (Section \ref{sec:study1_quality}), \textbf{client behavioral engagement} (Section \ref{sec:behavior}), and \textbf{system feature feedback} (Section \ref{sec:study1_feature}). Statistical analyses demonstrated CA+'s advantages in most metrics, while qualitative responses provided contextual understanding of these differences. The following sections detail these findings.

\subsubsection{\textbf{System Performance}}
\label{sec:study1_system}

Analysis of the automated evaluation results (Fig. \ref{fig:auto}) revealed significant differences between the CA+ and baseline conditions across multiple dimensions:

\begin{figure}
  \includegraphics[width=\textwidth]{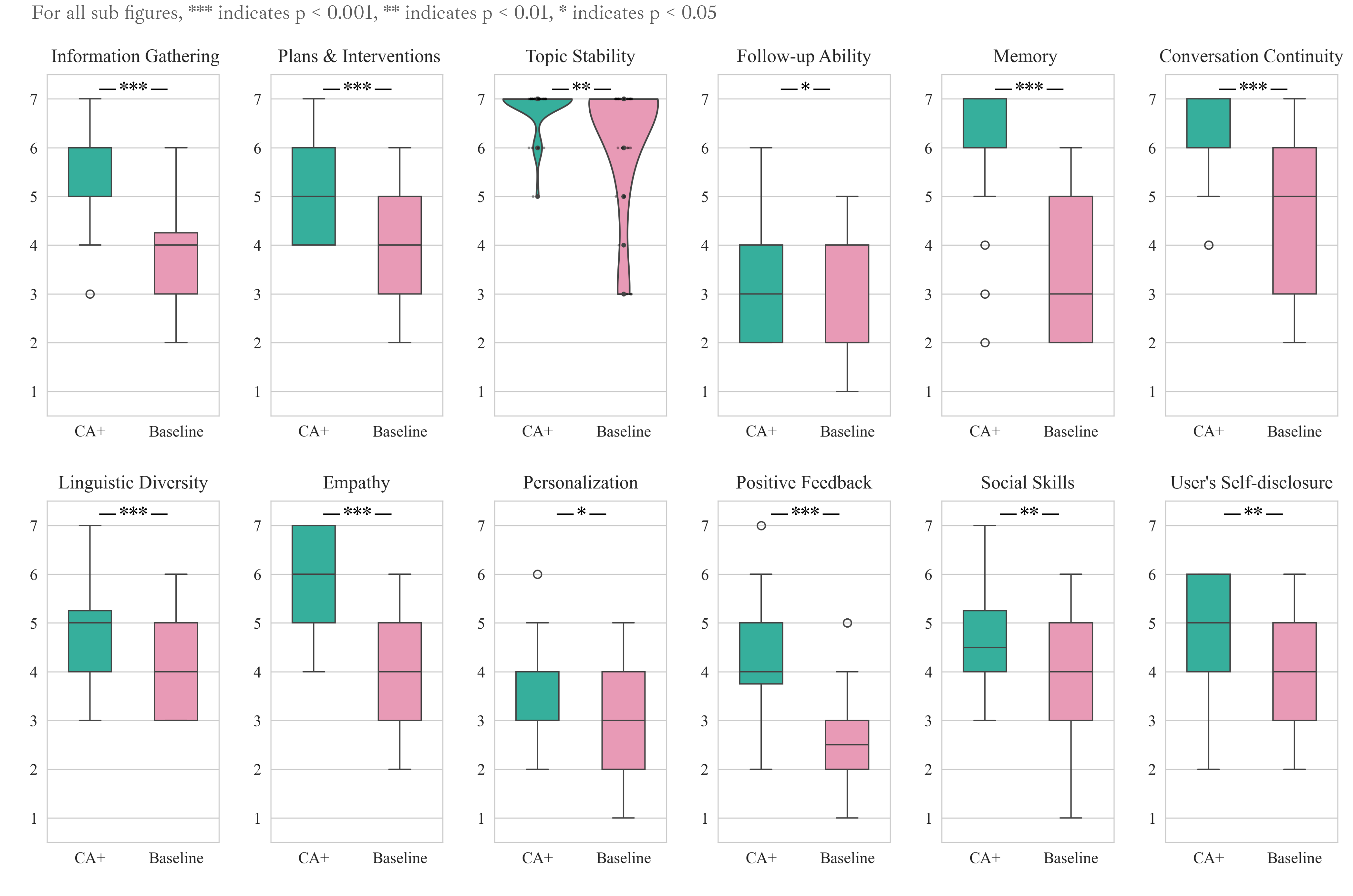}
  \caption{Box plot comparison of CA+ and baseline groups in the \textbf{Automated Evaluation} across multiple conversational dimensions (with a violin plot for Topic Stability to more intuitively display data distribution). The CA+ group consistently shows higher ratings, suggesting better automated evaluation outcomes than the baseline group.}
  \Description{The automated evaluation presents a comparison between the CA+ and baseline groups across twelve interaction metrics, including information gathering, plans & interventions, topic stability, memory, conversation continuity, empathy, and others. The boxplots and violin plots demonstrate that the CA+ group, shown in green, consistently outperforms the baseline group, shown in pink, with significant differences denoted by ***, **, and * for p-values < 0.001, < 0.01, and < 0.05, respectively. Notable improvements in the CA+ group are evident in areas such as information gathering, plans and interventions, and topic stability, as well as enhanced performance in memory, conversation continuity, and empathy. These findings suggest that the CA+ group provides a more effective, emotionally attuned, and engaging interaction experience.}
  \label{fig:auto}
\end{figure}

% Table \ref{tab:counseling_comparison} presents a comprehensive comparison between CA+ and the baseline condition across these dimensions.

% ---

In the domain of therapeutic skills, CA+ consistently outperformed the baseline system. CA+ showed superior performance in \textbf{designing appropriate counseling plans} ($t(72) = 5.01$, $p = 0.000005 < 0.001$, $M_{\text{\scriptsize CA+}} = 5.19$, $SD_{\text{\scriptsize CA+}} = 0.99$, $M_{\text{\scriptsize Baseline}} = 4.00$, $SD_{\text{\scriptsize Baseline}} = 1.03$) and \textbf{proactive assessment} ($t(72) = 7.75$, $p = 8.17 \times 10^{-11} < 0.001$, $M_{\text{\scriptsize CA+}} = 5.56$, $SD_{\text{\scriptsize CA+}} = 0.93$, $M_{\text{\scriptsize Baseline}} = 3.67$, $SD_{\text{\scriptsize Baseline}} = 1.13$). CA+ also demonstrated significantly higher \textbf{empathy} ($t(72) = 7.87$, $p = 4.97 \times 10^{-11} < 0.001$, $M_{\text{\scriptsize CA+}} = 5.83$, $SD_{\text{\scriptsize CA+}} = 0.96$, $M_{\text{\scriptsize Baseline}} = 3.86$, $SD_{\text{\scriptsize Baseline}} = 1.16$) and better follow-up ability ($t(72) = 2.20$, $p = 0.034 < 0.05$, $M_{\text{\scriptsize CA+}} = 3.14$, $SD_{\text{\scriptsize CA+}} = 1.11$, $M_{\text{\scriptsize Baseline}} = 2.56$, $SD_{\text{\scriptsize Baseline}} = 1.14$).

CA+ exhibited superior cognitive capabilities compared to the baseline system. This was evident in its \textbf{memory capability} ($t(72) = 7.81$, $p = 6.44 \times 10^{-11} < 0.001$, $M_{\text{\scriptsize CA+}} = 6.00$, $SD_{\text{\scriptsize CA+}} = 1.13$, $M_{\text{\scriptsize Baseline}} = 3.56$, $SD_{\text{\scriptsize Baseline}} = 1.50$) and \textbf{personalization ability} ($t(72) = 1.96$, $p = 0.043 < 0.05$, $M_{\text{\scriptsize CA+}} = 3.39$, $SD_{\text{\scriptsize CA+}} = 1.00$, $M_{\text{\scriptsize Baseline}} = 2.86$, $SD_{\text{\scriptsize Baseline}} = 1.13$), although the latter was marginally significant.

In terms of client engagement, CA+ significantly outperformed the baseline system. CA+ provided more \textbf{positive feedback} ($t(72) = 7.28$, $p = 5.94 \times 10^{-10} < 0.001$, $M_{\text{\scriptsize CA+}} = 4.36$, $SD_{\text{\scriptsize CA+}} = 1.16$, $M_{\text{\scriptsize Baseline}} = 2.47$, $SD_{\text{\scriptsize Baseline}} = 1.04$), received higher ratings for social skills ($t(72) = 3.23$, $p = 0.002 < 0.01$, $M_{\text{\scriptsize CA+}} = 4.61$, $SD_{\text{\scriptsize CA+}} = 1.16$, $M_{\text{\scriptsize Baseline}} = 3.69$, $SD_{\text{\scriptsize Baseline}} = 1.24$), and encouraged greater \textbf{client self-disclosure} ($t(72) = 3.24$, $p = 0.002 < 0.01$, $M_{\text{\scriptsize CA+}} = 4.89$, $SD_{\text{\scriptsize CA+}} = 1.05$, $M_{\text{\scriptsize Baseline}} = 4.00$, $SD_{\text{\scriptsize Baseline}} = 1.27$).

These results indicate that CA+ consistently outperformed the baseline across various aspects of counseling interaction.

\subsubsection{\textbf{Perceived Counseling Quality}}
\label{sec:study1_quality}
To assess clients' subjective perceptions of the counseling experience, we conducted a comprehensive evaluation focusing on perceived counseling quality, empathy, self-disclosure, and system usability using the Session Rating Scale (SRS)~\cite{campbell2009outcome}, Perceived Empathy of Technology Scale (PETS)~\cite{schmidmaier2024perceived}, Self-Disclosure Scale~\cite{rubin2020revised}, and System Usability Scale (SUS)~\cite{sharfina2016indonesian} respectively. Our analysis revealed significant differences between the CA+ system and the baseline condition across all four measures, consistently favoring the CA+.

\begin{figure}
  \includegraphics[width=0.5\textwidth]{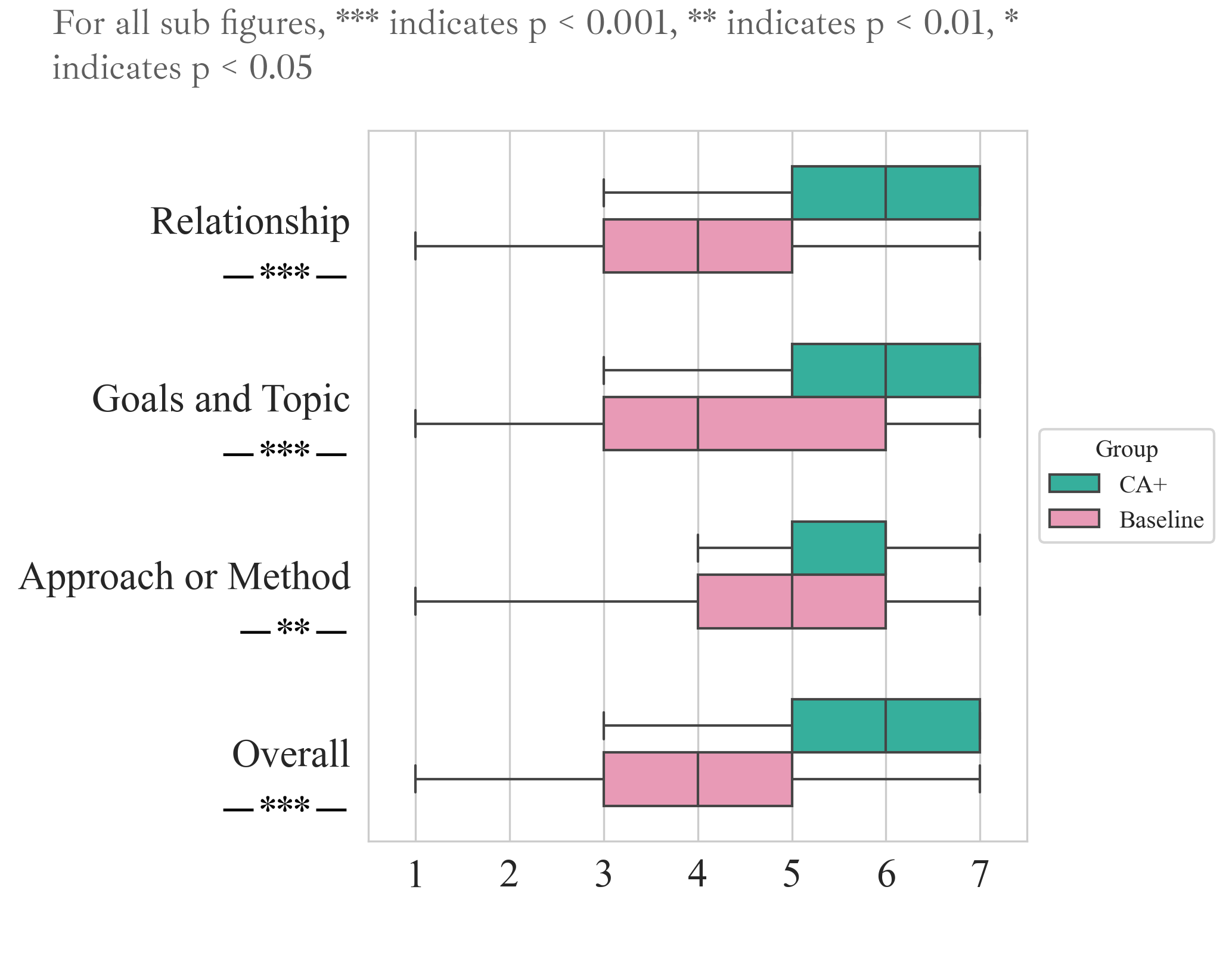}
  \caption{Box plot comparison of CA+ and baseline groups across dimensions on the \textbf{Session Rating Scale} (SRS). The CA+ group received significantly higher ratings than the baseline group across all dimensions. This suggests a more favorable evaluation of the sessions in the CA+ group.}
  \Description{The session rating figure visualizes the comparative performance of the CA+ and baseline groups in key counseling metrics, including relationship, goals and topic, approach or method, and overall understanding. The CA+ group, represented in green, consistently scores higher across all dimensions, indicating better comprehension and alignment with counseling objectives. The baseline group, shown in pink, demonstrates lower and more varied scores. Statistical significance is denoted by ***, **, and * for p-values < 0.001, < 0.01, and < 0.05, respectively, with key differences in the relationship, goals, and topic dimensions. These findings highlight the CA+ group's superior grasp of the counseling process and overall engagement. }
  \label{fig:RSR}
\end{figure}

The \textbf{SRS} (Fig. \ref{fig:RSR}) results demonstrated that CA+ significantly outperformed the baseline across all evaluated aspects. In the counseling relationship domain, CA+ received substantially higher ratings ($t(72) = 5.91$, $p = 1.25 \times 10^{-7} < 0.001$, $M_{\text{\scriptsize CA+}} = 6.03$, $SD_{\text{\scriptsize CA+}} = 1.00$, $M_{\text{\scriptsize Baseline}} = 4.11$, $SD_{\text{\scriptsize Baseline}} = 1.69$). Similarly, CA+ excelled in goals and topics discussed ($t(72) = 4.97$, $p = 6.45 \times 10^{-6} < 0.001$, $M_{\text{\scriptsize CA+}} = 5.78$, $SD_{\text{\scriptsize CA+}} = 1.12$, $M_{\text{\scriptsize Baseline}} = 4.05$, $SD_{\text{\scriptsize Baseline}} = 1.82$), approach or method used ($t(72) = 3.26$, $p = 0.002 < 0.01$, $M_{\text{\scriptsize CA+}} = 5.86$, $SD_{\text{\scriptsize CA+}} = 0.81$, $M_{\text{\scriptsize Baseline}} = 4.95$, $SD_{\text{\scriptsize Baseline}} = 1.49$), and overall assessment ($t(72) = 5.11$, $p = 2.60 \times 10^{-6} < 0.001$, $M_{\text{\scriptsize CA+}} = 5.81$, $SD_{\text{\scriptsize CA+}} = 1.06$, $M_{\text{\scriptsize Baseline}} = 4.22$, $SD_{\text{\scriptsize Baseline}} = 1.54$).

\begin{figure}
  \includegraphics[width=0.5\textwidth]{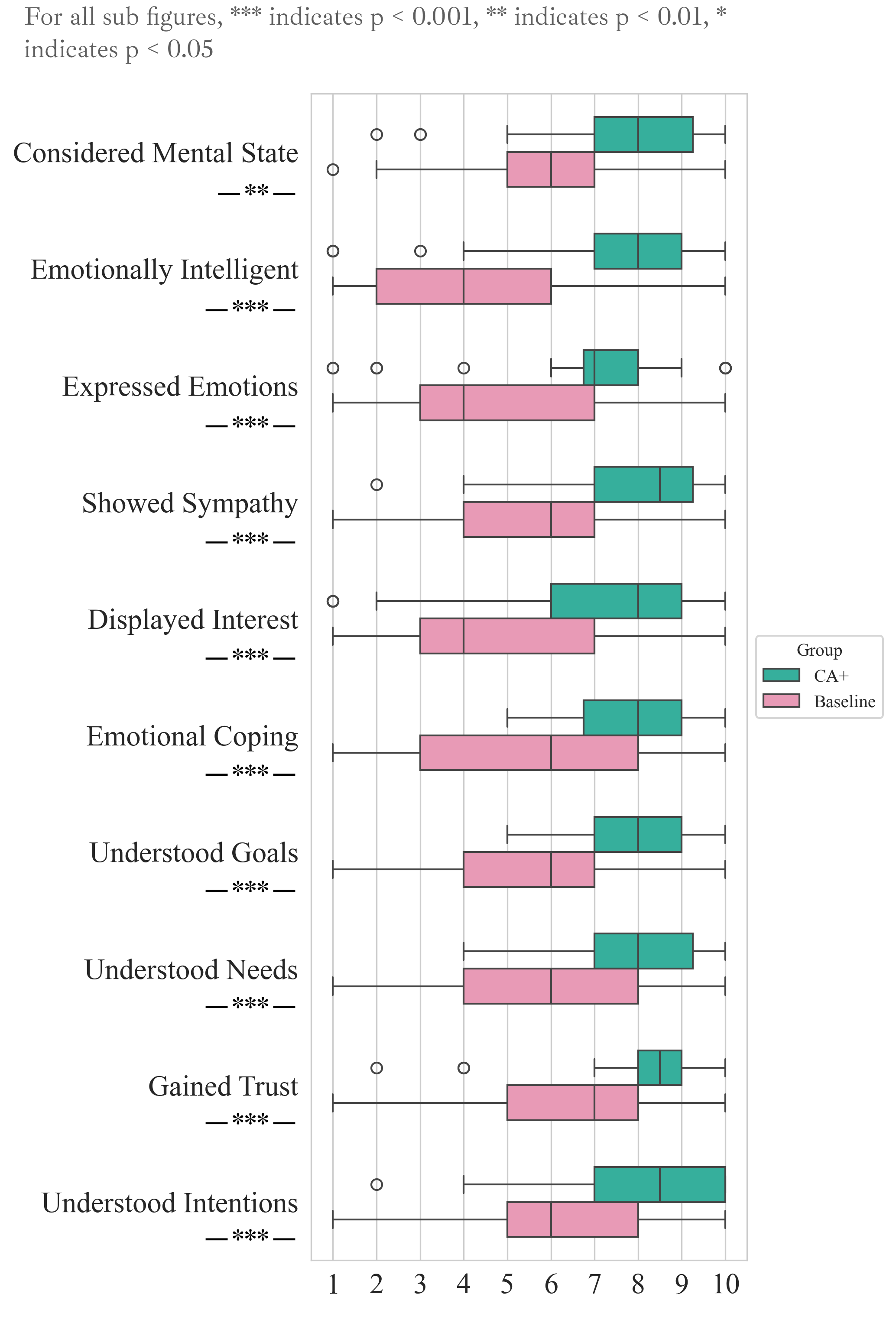}
  \caption{Box plot comparison of CA+ and baseline groups across dimensions on the \textbf{Perceived Empathy of Technology Scale} (PETS). The CA+ group has higher empathy scores, indicating a stronger empathetic response, while the baseline group has lower and more variable scores.}
  \Description{The perceived empathy of technology scale presents a comparison of emotional and cognitive behaviors between the CA+ and baseline groups, focusing on attributes such as considered mental state, emotionally intelligent, expressed emotions, and gained trust. The boxplots represent the distribution of scores, with CA+ shown in green and baseline in pink. Significant differences between the groups are marked with p-values, where ***, **, and * indicate statistical significance at p < 0.001, p < 0.01, and p < 0.05 levels, respectively. Notable distinctions include emotionally intelligent, expressed emotions, and displayed interest, where the CA+ group scores notably higher, suggesting a more emotionally attuned interaction. Additionally, gained trust and understood intentions display significant differences, with CA+ performing better in fostering trust and understanding. These findings suggest that the CA+ group demonstrates greater empathy, emotional intelligence, and cognitive understanding compared to the baseline group.}
  \label{fig:PETS}
\end{figure}

The \textbf{PETS} (Fig. \ref{fig:PETS}) analysis revealed CA+'s superior performance across all measured dimensions of perceived empathy. Notably, CA+ was perceived as more emotionally intelligent ($t(72) = 5.04$, $p = 4.59 \times 10^{-6} < 0.001$, $M_{\text{\scriptsize CA+}} = 7.42$, $SD_{\text{\scriptsize CA+}} = 2.49$, $M_{\text{\scriptsize Baseline}} = 4.38$, $SD_{\text{\scriptsize Baseline}} = 2.66$), more empathetic ($t(72) = 4.30$, $p = 6.76 \times 10^{-5} < 0.001$, $M_{\text{\scriptsize CA+}} = 7.92$, $SD_{\text{\scriptsize CA+}} = 2.02$, $M_{\text{\scriptsize Baseline}} = 5.68$, $SD_{\text{\scriptsize Baseline}} = 2.42$), and better at understanding clients' goals ($t(72) = 4.78$, $p = 1.33 \times 10^{-5} < 0.001$, $M_{\text{\scriptsize CA+}} = 8.00$, $SD_{\text{\scriptsize CA+}} = 1.35$, $M_{\text{\scriptsize Baseline}} = 5.78$, $SD_{\text{\scriptsize Baseline}} = 2.46$). Clients also reported significantly higher trust in CA+ ($t(72) = 3.85$, $p = 3.26 \times 10^{-4} < 0.001$, $M_{\text{\scriptsize CA+}} = 8.17$, $SD_{\text{\scriptsize CA+}} = 1.76$, $M_{\text{\scriptsize Baseline}} = 6.24$, $SD_{\text{\scriptsize Baseline}} = 2.46$).

\begin{figure}
  \includegraphics[width=0.5\textwidth]{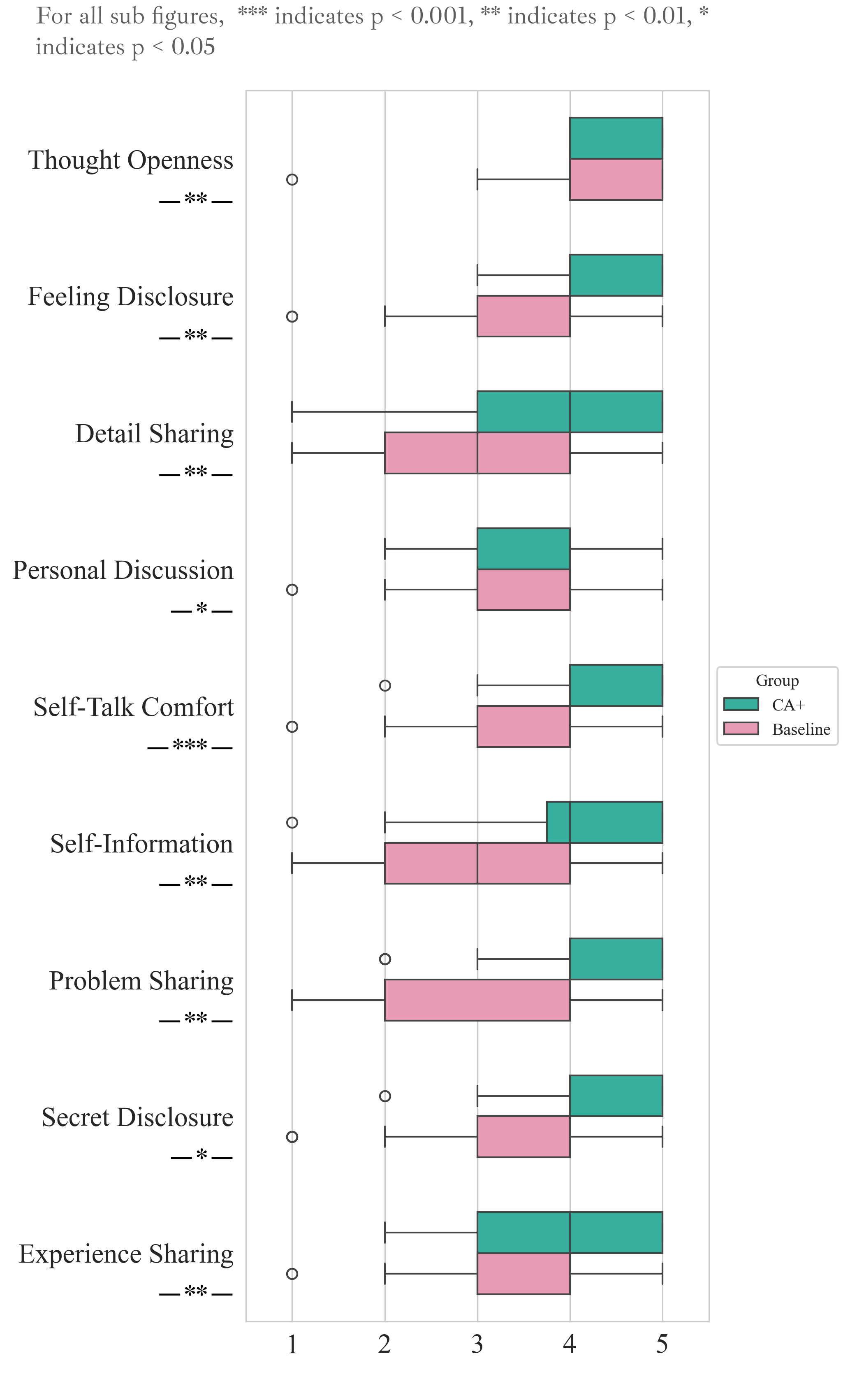}
  \caption{Box plot comparison of CA+ and baseline groups across dimensions on the \textbf{Self-Disclosure Scale} (SDS). The CA+ group generally exhibits higher levels of self-disclosure across the measured dimensions, showing a more consistent and open disclosure pattern than the baseline group.}
  \Description{The self-disclosure scale presents a comparison of self-disclosure behaviors across several dimensions, including thought openness, feeling disclosure, detail sharing, and personal discussion, between the CA+ and baseline groups. The chart uses boxplots to represent the distribution of scores, with CA+ shown in green and baseline in pink. Statistically significant differences between the groups are indicated using p-values, with ***, **, and * corresponding to p-values of <0.001, <0.01, and <0.05, respectively. For instance, significant differences are observed in thought openness, feeling disclosure, and self-talk comfort, where the CA+ group tends to score higher, suggesting a greater comfort and openness in these areas. Other metrics, such as detail sharing and problem sharing, also reveal key differences between the two groups, indicating how the enhanced CA+ group facilitates deeper levels of personal and emotional sharing compared to the baseline group. The legend on the right distinguishes between the two groups, emphasizing the comparative analysis.}
  \label{fig:SDS}
\end{figure}

The \textbf{Self-Disclosure Scale} (Fig. \ref{fig:SDS}) indicated that clients were more willing to share personal information with CA+. They reported being more comfortable discussing personal experiences with CA+ ($t(72) = 2.86$, $p = 0.006 < 0.01$, $M_{\text{\scriptsize CA+}} = 3.97$, $SD_{\text{\scriptsize CA+}} = 0.87$, $M_{\text{\scriptsize Baseline}} = 3.30$, $SD_{\text{\scriptsize Baseline}} = 1.14$) and feeling more at ease when talking about themselves ($t(72) = 3.60$, $p = 7.17 \times 10^{-4} < 0.001$, $M_{\text{\scriptsize CA+}} = 4.42$, $SD_{\text{\scriptsize CA+}} = 0.79$, $M_{\text{\scriptsize Baseline}} = 3.62$, $SD_{\text{\scriptsize Baseline}} = 1.07$). Additionally, clients were more inclined to express their thoughts honestly to CA+ ($t(72) = 3.03$, $p = 0.004 < 0.01$, $M_{\text{\scriptsize CA+}} = 4.64$, $SD_{\text{\scriptsize CA+}} = 0.48$, $M_{\text{\scriptsize Baseline}} = 4.16$, $SD_{\text{\scriptsize Baseline}} = 0.82$) and share their personal feelings more frequently ($t(72) = 3.11$, $p = 0.003 < 0.01$, $M_{\text{\scriptsize CA+}} = 4.36$, $SD_{\text{\scriptsize CA+}} = 0.67$, $M_{\text{\scriptsize Baseline}} = 3.73$, $SD_{\text{\scriptsize Baseline}} = 1.03$).

% \begin{figure}
%   \includegraphics[width=0.5\textwidth]{fig/SUS.png}
%   \caption{SUS}
%   \Description{}
%   \label{fig:SUS}
% \end{figure}

The \textbf{System Usability Scale (SUS)} results further supported the superiority of CA+ in terms of user experience. CA+ achieved a significantly higher SUS score ($M_{\text{\scriptsize CA+}} = 80.88$) compared to the baseline system ($M_{\text{\scriptsize Baseline}} = 68.31$), indicating a substantial improvement in perceived usability. This difference of 12.57 points places CA+ in the 'Excellent' usability range.

These results provide strong evidence for the enhanced user experience and effectiveness of the CA+ system in AI-based counseling, demonstrating its superiority across multiple dimensions of the counseling experience.

\subsubsection{\textbf{Client Behavioral Engagement}}
\label{sec:behavior}

We analyzed the system's statistical data to compare the performance of the CA+ system against the baseline across several client behavioral metrics. The results reveal notable enhancements in the CA+ system in terms of client engagement and response quality.

\textbf{Session Length} is a key indicator of system appeal and client engagement. We first grouped interactions based on session IDs. Within each session, any idle gap exceeding 8 minutes was treated as a session break, such long idle intervals were segmented to avoid overestimating active engagement. The durations of the resulting segments were then summed to obtain the total daily session length for each participant. 
% Over 8-minute gaps between talks were checked within each session. To prevent counting long idle intervals, the session was split into pieces. If there were no gaps, the session length was estimated without divisions. We then added up each participant's daily session length. 
The total session length for each participant on a given day was then summed. The statistical result ($t(71) = 4.65$, $p = 1.94 \times 10^{-5} < 0.001$, $M_{\text{\scriptsize CA+}} = 28.10$, $SD_{\text{\scriptsize CA+}} = 16.80$, $M_{\text{\scriptsize Baseline}} = 14.38$, $SD_{\text{\scriptsize Baseline}} = 5.56$) reveals that CA+ is more attractive to participants compared to the baseline system.

\textbf{Average Words per Response} measures client involvement and commitment to the system. Participants who supply more specific information and are more engaged with the system tend to have longer response word counts. More broad communication suggests a stronger willingness to share personal information and circumstances with the system. The results ($t(71) = 2.86$, $p = 0.0062 < 0.05$, $M_{\text{\scriptsize CA+}} = 30.70$, $SD_{\text{\scriptsize CA+}} = 13.67$, $M_{\text{\scriptsize Baseline}} = 22.24$, $SD_{\text{\scriptsize Baseline}} = 11.29$) show that clients in the CA+ system provided longer replies.

\textbf{Conversation Rounds} correlate with the frequency of client interactions, providing a visual representation of participants' activity and the duration of conversations. Compared to the baseline, clients engaged in a greater number of conversation rounds, with those in the CA+ group demonstrating a significantly higher willingness to interact with the system ($t(71) = 2.46$, $p = 0.0177 < 0.05$, $M_{\text{\scriptsize CA+}} = 16.97$, $SD_{\text{\scriptsize CA+}} = 8.44$, $M_{\text{\scriptsize Baseline}} = 12.69$, $SD_{\text{\scriptsize Baseline}} = 6.11$). This finding suggests that CA+ facilitated longer and more continuous conversations.

In addition to the above intuitive metrics, we calculated the client's  \textbf{Informativeness} to analyze the quality of the client's responses. This was determined using the following formula: $I_{\text{client}} = \text{Average Words per Response} \times \text{Information Entropy} \times \text{Conversation Round}$. The informativeness was significantly higher in the CA+ system compared to the baseline ($t(71) = 2.81$, $p = 0.0072 < 0.05$, $M_{\text{\scriptsize CA+}} = 215.45$, $SD_{\text{\scriptsize CA+}} = 150.64$, $M_{\text{\scriptsize Baseline}} = 136.97$, $SD_{\text{\scriptsize Baseline}} = 73.94$). The statistical result confirms that the overall richness of information was greater during sessions that utilized the CA+ system.

\begin{figure}
  \includegraphics[width=0.8\textwidth]{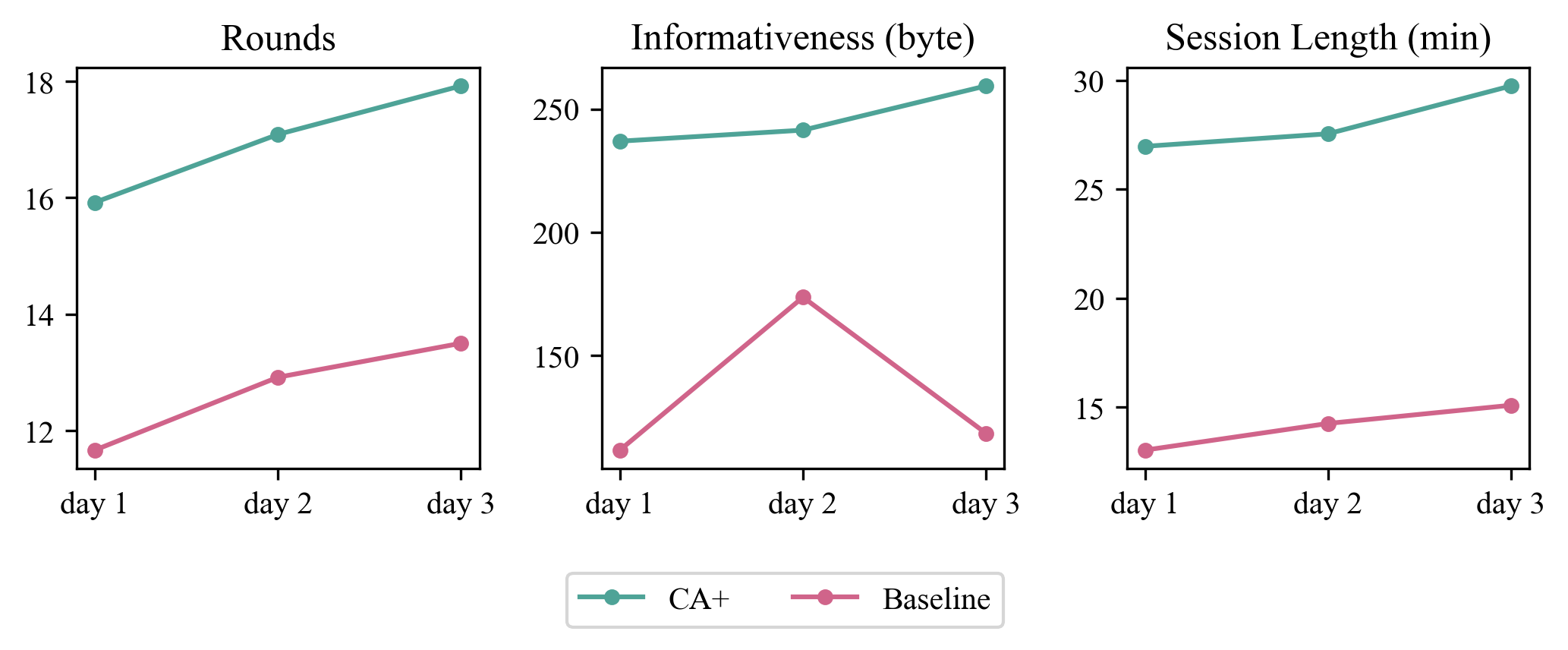}
  \caption{\textbf{Behavioral Trends} in CA+ and Baseline Groups. This figure illustrates the comparative analysis of rounds, informativeness, and session length between the CA+ and Baseline groups over three days, highlighting key behavioral differences between the two groups.}
  \Description{The behavior trend presents a side-by-side comparison of three behavioral metrics—rounds, informativeness (byte), and session length (min)—across two groups, CA+ and baseline, over a three-day period. The rounds subplot demonstrates a consistent upward trend for the CA+ group, suggesting a steady increase in conversation exchanges, while the baseline group shows a more gradual rise. The informativeness subplot highlights a significant spike for the baseline group on day two, followed by a sharp decline on day three, contrasting with the stable performance of the CA+ group. Finally, the session length subplot reveals that the CA+ group steadily extends session duration, whereas the baseline group remains relatively constant across all three days. These metrics collectively illustrate that the CA+ group maintains higher engagement and a more consistent improvement in overall session quality compared to the baseline group.}
  \label{fig:behavior}
\end{figure}

The behavior trend (Fig. \ref{fig:behavior}), along with the statistical analysis, highlights the CA+ system compared to the baseline system across three key behavioral metrics: \textbf{Rounds}, \textbf{Informativeness}, and \textbf{Session Length}. The CA+ system consistently outperforms the baseline and shows continuous improvement over three days, highlighting its effectiveness in boosting client engagement.

% The CA+ group showed a consistent increase in conversation rounds over three days, beginning at 15.92 on day one and reaching 17.92 by day three. This trend indicates a higher level of user engagement with the CA+ system. In contrast, the baseline system experienced a modest increase from 11.67 to 13.50 rounds, reflecting lower user engagement. Overall, the CA+ system outperformed the baseline in promoting interaction.
The CA+ group demonstrated superior client engagement across multiple metrics over a three-day period. Conversation rounds increased from $15.92$ to $17.92$, while the baseline system showed a modest increase from $11.67$ to $13.50$ rounds. This trend indicates a higher level of client engagement with the CA+ system.

% The Informativeness metric underscores the advantages of the CA+ system, which improved from 236.95 on day one to 259.41 on day three. This consistent enhancement indicates its capacity to facilitate richer conversations. In contrast, the baseline system showed inconsistent results, beginning at 111.48, peaking at 173.58, and then declining to 118.28. This variability suggests that the baseline struggled to maintain high-quality interactions. Overall, CA+ outperformed the baseline in informativeness and demonstrated consistent improvement.
The Informativeness metric further underscores the advantages of the CA+ system. It improved from $236.95$ on day one to $259.41$ on day three, indicating its capacity to facilitate richer conversations. In contrast, the baseline system showed inconsistent results, fluctuating between $111.48$ and $173.58$, suggesting difficulty in maintaining high-quality interactions.

% The CA+ system showed superior session lengths, increasing from 26.98 minutes on day one to 29.75 minutes by day three, indicating enhanced user engagement. In contrast, the baseline system began with 13.03 minutes on day one and only reached 15.09 minutes by day three, reflecting significantly shorter session lengths and lower engagement compared to CA+.
Session lengths for the CA+ system increased from 26.98 minutes on day one to 29.75 minutes by day three. The baseline system, however, only increased from $13.03$ to $15.09$ minutes.

% The behavioral analysis further demonstrates that CA+ users exhibit higher levels of engagement and provide more extensive information, indicating that users are more willing to engage in more frequent interactions within the CA+ system and consistently offer more detailed and in-depth feedback. This suggests that the CA+ system is more effective in facilitating self-expression and emotional sharing in the context of psychological counseling, thereby supporting more in-depth psychological interventions.

Behavioral analysis confirms that CA+ clients exhibit higher levels of engagement and provide more extensive information, thereby supporting more in-depth psychological interventions.

\subsubsection{\textbf{Feedback on System Features}}
\label{sec:study1_feature}

Client feedback on the CA+ system revealed substantial improvements in engagement, personalization, and therapeutic efficacy over the baseline system (Fig. \ref{fig:feature}). Quantitative and qualitative analyses demonstrate CA+'s superior performance in maintaining conversation continuity and providing empathetic, personalized interventions:

\begin{figure}
  \includegraphics[width=0.8\textwidth]{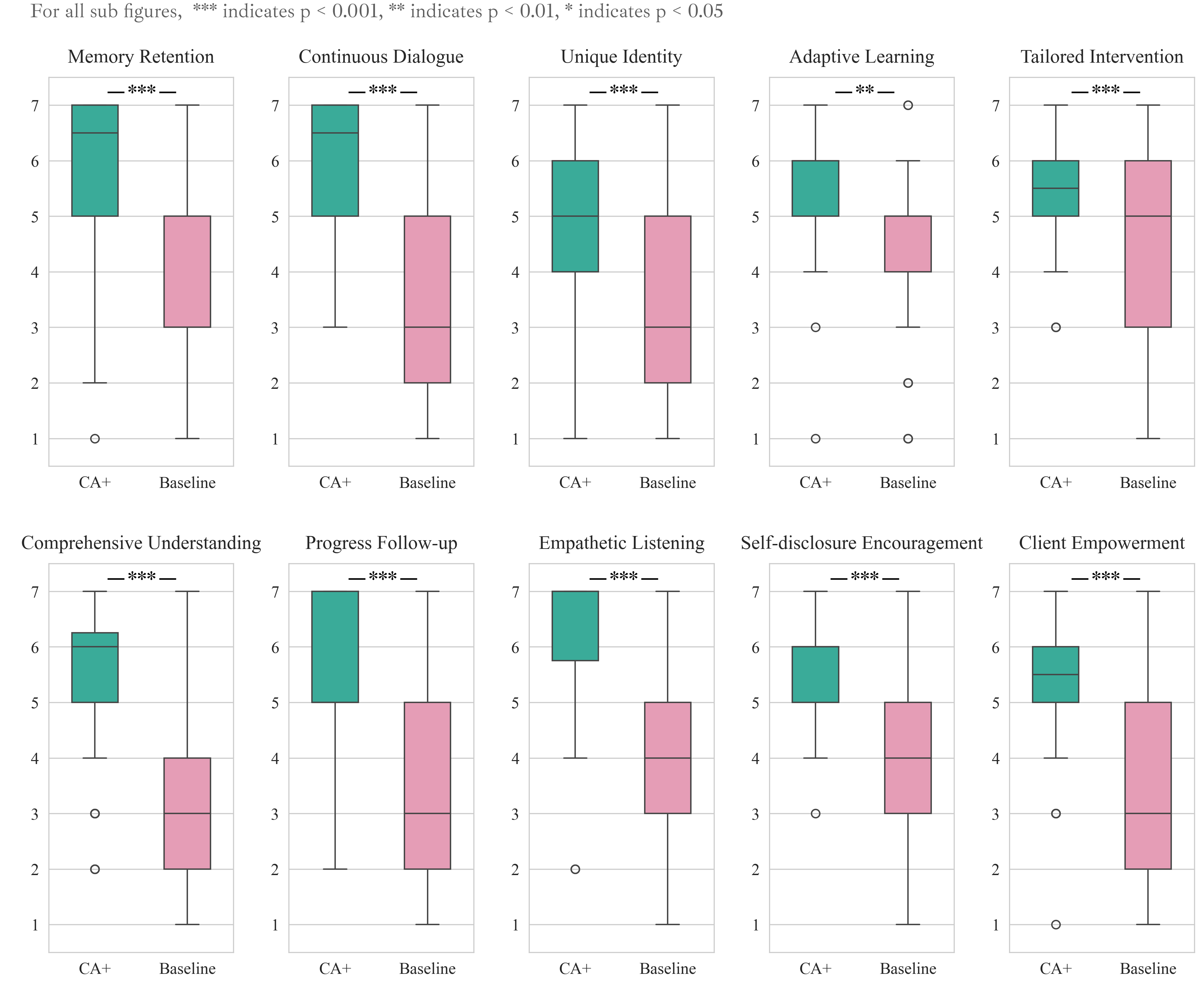}
  \caption{Box plot comparison of CA+ and baseline groups across \textbf{System Features}. The CA+ group received significantly higher ratings than the baseline group across all dimensions.}
  \Description{The system feature compares the CA+ and baseline groups across ten key counseling metrics: memory retention, continuous dialogue, unique identity, adaptive learning, tailored intervention, comprehensive understanding, progress follow-up, empathetic listening, self-disclosure encouragement, and client empowerment. The boxplots indicate that the CA+ group consistently outperforms the baseline group, with statistically significant differences denoted by ***, **, and * for p-values < 0.001, < 0.01, and < 0.05, respectively. The CA+ group exhibits higher scores in areas such as memory retention, continuous dialogue, and comprehensive understanding, suggesting a more effective, personalized, and adaptive counseling approach compared to the baseline group.}
  \label{fig:feature}
\end{figure}

\begin{itemize}

\item \textbf{Client Engagement \& Focus: }
The CA+ system fostered significant client engagement and a desire for continued interaction. One participant (CA-5) expressed, \textit{``After the experiment ended, I was still willing to continue the dialogue with it, or to talk more about some topics in my inner heart."}  This qualitative feedback is supported by quantitative data, with CA+ being rated significantly higher in guiding clients to take initiative ($t(72) = 3.98$, $p = 0.000212 < 0.001$, $M_{\text{\scriptsize CA+}} = 5.31$, $SD_{\text{\scriptsize CA+}} = 1.33$, $M_{\text{\scriptsize Baseline}} = 3.73$, $SD_{\text{\scriptsize Baseline}} = 2.00$) and in keeping conversations centered on counseling ($t(72) = 5.16$, $p = 3.31 \times 10^{-6} < 0.001$, $M_{\text{\scriptsize CA+}} = 6.14$, $SD_{\text{\scriptsize CA+}} = 0.75$, $M_{\text{\scriptsize Baseline}} = 4.76$, $SD_{\text{\scriptsize Baseline}} = 1.44$).  Furthermore, client behavioral analysis from previous sections (Section \ref{sec:behavior}) corroborates these findings, demonstrating sustained engagement with the CA+ system over time.

\item \textbf{Multi-turn Cognition: }
CA+ demonstrated a significantly better ability to understand clients' concerns and follow up on previous advice. It was rated higher in understanding clients' main concerns ($t(72) = 5.24$, $p = 2.06 \times 10^{-6} < 0.001$, $M_{\text{\scriptsize CA+}} = 5.33$, $SD_{\text{\scriptsize CA+}} = 1.56$, $M_{\text{\scriptsize Baseline}} = 3.43$, $SD_{\text{\scriptsize Baseline}} = 1.53$) and following up on clients' implementation of advice ($t(72) = 5.86$, $p = 1.92 \times 10^{-7} < 0.001$, $M_{\text{\scriptsize CA+}} = 5.44$, $SD_{\text{\scriptsize CA+}} = 1.38$, $M_{\text{\scriptsize Baseline}} = 3.41$, $SD_{\text{\scriptsize Baseline}} = 1.58$).

One participant (CA-6) elaborated on this improved understanding, stating, \textit{``It would analyze my situation based on the different events I encountered, rather than just repeating the same few sentences. Moreover, after inquiring about other issues, it would modify its previous methods and respond to me accordingly. So I think this also demonstrates \textbf{the AI system's ability to think.}"}  This feedback highlights CA+'s adaptive reasoning approach, which likely contributed to its significantly higher ratings in multi-turn understanding and reasoning.

\item \textbf{Memory \& Continuity: }
CA+ demonstrated significantly better ability to remember client information ($t(72) = 5.48$, $p = 8.52 \times 10^{-7} < 0.001$, $M_{\text{\scriptsize CA+}} = 5.83$, $SD_{\text{\scriptsize CA+}} = 1.62$, $M_{\text{\scriptsize Baseline}} = 3.68$, $SD_{\text{\scriptsize Baseline}} = 1.74$). This quantitative advantage was reflected in clients' experiences. One participant (CA-13) noted, \textit{``I noticed \textbf{it hadn't forgotten what I said yesterday}. While I was expressing new content today, it interspersed questions about my previous issues, combining them with what I said today to inquire about my feelings. It made me feel that I was being paid attention to."} 

CA+ also excelled in maintaining conversation continuity ($t(72) = 7.54$, $p = 2.21 \times 10^{-10} < 0.001$, $M_{\text{\scriptsize CA+}} = 6.06$, $SD_{\text{\scriptsize CA+}} = 1.13$, $M_{\text{\scriptsize Baseline}} = 3.51$, $SD_{\text{\scriptsize Baseline}} = 1.70$). This statistical superiority was vividly illustrated by client feedback. Participant CA-4 described her experience: \textit{``When I started the system on the second day, it provided a review of the first day. It made me feel very surprised."}

\item \textbf{Personalization, Empathy, and Self-disclosure: }
CA+ demonstrated significantly superior capabilities in personalization and empathy feedback provision compared to the baseline system. Clients rated CA+ higher in its ability to learn and improve ($t(72) = 3.32$, $p < 0.01$, $M_{\text{\scriptsize CA+}} = 5.56$, $SD_{\text{\scriptsize CA+}} = 1.23$, $M_{\text{\scriptsize Baseline}} = 4.46$, $SD_{\text{\scriptsize Baseline}} = 1.57$) and in creating personalized intervention plans ($t(72) = 3.30$, $p < 0.01$, $M_{\text{\scriptsize CA+}} = 5.47$, $SD_{\text{\scriptsize CA+}} = 1.12$, $M_{\text{\scriptsize Baseline}} = 4.38$, $SD_{\text{\scriptsize Baseline}} = 1.67$). Additionally, CA+ significantly enhanced clients' positive emotions ($t(72) = 4.32$, $p < 0.001$).

These quantitative findings were supported by client testimonials. One participant (CA-3) highlighted CA+'s personalized approach: \textit{``After I expressed my grievances and sadness, it would always comfort me. I felt each comforting phrase was very \textbf{thoughtful}, making me feel it really took me to heart." In addition, CA-13 further stated that CA+ will give a thoughtful response based on previous conversations.
}

Clients also reported significantly higher levels of feeling valued and accepted by CA+ ($t(72) = 5.93$, $p < 0.001$), and CA+ facilitated greater self-disclosure ($t(72) = 4.93$, $p < 0.001$). CA-5 noted how CA+'s approach provided positive feedback and encouraged self-disclosure: \textit{``When I expressed anxiety or distress, it wouldn't immediately tell me what to do. Instead, it would start with words of encouragement before gradually expressing its thoughts or suggesting ways to relieve stress. \textbf{This made me feel quite touched so I would be willing to share more of what's on my mind with it.}"} Other participants echoed these sentiments, highlighting CA+'s personalized, empathetic approach that fostered a more engaging and positive client experience.

\item \textbf{Useful Suggestions: }
CA+ was commended for \textbf{providing practical and tailored advice}. Clients appreciated its role in offering guidance when they felt stuck, as exemplified by CA-3: \textit{``I seek its help because I don't know how to make plans, so I hope it can give me some plans. Then I found that \textbf{it could really give me a lot of effective advice, and I also tried it in real life}."} CA+ also suggested evidence-based techniques aligned with clients' interests, such as focused meditation training (CA-11). Furthermore, its structured approach to problem-solving, combining theoretical understanding with practical tasks, was highly valued. CA-12 noted: \textit{``For example, when we need to set some daily tasks or goals, it would first provide a set of theories about how to perceive and accept one's emotions, and then start to set specific tasks. \textbf{This made me feel grounded in that theoretical framework.}"} These features collectively demonstrated CA+'s effectiveness in providing comprehensive and educational counseling support.

\end{itemize}

\section{Study 2: Certified Counselors Quality Assessment}
\label{sec:study2}

Study 2 presents an expert evaluation of CA+ in which two licensed counseling psychologists systematically assessed a curated set of transcripts ($n = 6$) from Study 1 to evaluate the system’s performance from a professional standpoint. This qualitative analysis provides professional insights into the system's cognitive abilities, engagement, and professionalism, complementing the user-centric findings of Study 1.

\subsection{Methodology}

\textbf{Participants:} Two licensed counselors (LC-1 \& LC-2) with extensive experience (over 500 counseling hours, and over 5 years experience) and a minimum of a master's degree in clinical psychology were recruited for this study.

\textbf{Procedures:} The counselors reviewed 6 verbatim transcripts of chat histories from Study 1, three from the CA+ system and three from the baseline system, selected to ensure balanced representation in terms of client demographics and session characteristics. The transcripts were carefully selected to represent a balanced sample, including three transcripts from each group, each set featuring participant clients with comparable counseling experience (ranging from no prior counseling, 1–2 sessions, to multiple sessions) and similar gender distribution (2 females and 1 male). Counselors employed a think-aloud protocol~\cite{van1994think} during the review process. After examining each set of three transcripts, the counselors participated in a semi-structured interview focused on system characteristics, including cognitive abilities, engagement, and professionalism. The interview utilized a custom-designed Professional Assessment of AI Counseling (PAAC) scale (Appendix \ref{app:Self-designed} for details) developed based on our formative research and key design components of the CA+ system, alongside a semi-structured format. Professionalism was assessed using the established Counselor Competence Scale (CCS)~\cite{swank2012exploratory}.

\textbf{Data Collection and Analysis:} All sessions were audio-recorded and transcribed. Two researchers analyzed the transcripts using thematic analysis. The content was coded and thoroughly discussed to ensure reliability and validity of the findings.

\subsection{Findings}
Our analysis revealed distinct differences between the CA+ framework and baseline system across five key dimensions. The following sections detail our comparative evaluation of \textbf{overall system performance} (Section \ref{sec:overall_study2}), \textbf{session management effectiveness} (Section \ref{sec:Management_study2}), \textbf{therapeutic approach appropriateness} (Section \ref{sec:Approach_study2}), \textbf{client outcomes} (Section \ref{sec:study2_outcome}), and \textbf{potential improvements} (Section \ref{sec:study2_Enhancement}), providing evidence for the impact of our design considerations on therapeutic interactions.

\subsubsection{\textbf{Overall Assessment}}
\label{sec:overall_study2}
The expert evaluation revealed significant differences in the overall performance of CA+ and the baseline system. LC-1 praised CA+'s performance, noting that it was \textbf{already doing an excellent job}.  In contrast, LC-1 found the baseline system to be less engaging, describing its responses as \textit{``mechanical and reliant on generic responses."}

LC-2 observed that CA+ actively engaged in dialogue with the participant clients, stating, \textit{\textbf{``CA+ has started to engage in conversation with the clients,"} trying to understand them by paraphrasing, encouraging, and inviting them to share more. In contrast, the Baseline is still focused on completing tasks, directly outputting responses without inviting further discussion."} LC-2 further characterized CA+'s interaction style as \textit{``\textbf{a reciprocal dialogue},}" while the baseline's approach was described as \textit{``fragmented, one-way output."}

These qualitative observations were supported by quantitative assessments. On system-specific features, CA+ scored 5.35/7 compared to the baseline's 1.85/7. Using the professional Counseling Competence Scale, CA+ achieved 4.22/5, while the baseline scored 1.94/5. These results highlight CA+'s significant improvement in counseling competence and system-specific features over the baseline system.

\subsubsection{\textbf{Sessions Management}}
\begin{itemize}
\label{sec:Management_study2}

\item \textbf{Multi-turn Memory:} Regarding session structure, LC-1 noted that \textbf{CA+ maintained a coherent flow throughout the sessions, building upon previous discussions}. This continuity was evident in CA+'s ability to recall and refer to prior conversations, as illustrated in the following excerpt:

\begin{dialogue}
\speak{\color{gray}CA+} {\color{gray}Before we start, I wanted to check in about the career confusion and anxiety we discussed last time. [...] We can start with these topics, or we can talk about your recent feelings and experiences.}
\speak{\color{gray}Client} {\color{gray}I'm glad you still remember me.}
\end{dialogue}

LC-1 remarked that \textit{``CA+\textbf{ followed up on the previous counseling session, maintaining continuity}, and the client felt remembered - not just a simple machine."} In contrast, LC-1 observed that the baseline system's interactions felt more like a \textit{``question-and-answer session between the client and counselor, with the AI explore the client's deeper concerns."}. LC-2 praised CA+'s ability to \textit{``\textbf{call back} to previous sessions, which is a very good feature, and provide systematic responses."}

\item \textbf{Session Consolidation: }In terms of summarizing sessions and assigning homework, CA+ demonstrated a strong ability to synthesize the main points of the session and provide actionable tasks for the client, as evidenced by the following excerpt:

\begin{flushleft}
\begin{dialogue}

\speak{\color{gray}CA+} {\color{gray} Thank you very much for sharing today. We discussed the stress you face at work and in your family,  [...]}

{\color{gray} To help you better cope with these issues, the homework assigned for this session is:}

{\color{gray} 1. Emotion Logging: [...]}

{\color{gray} 2. Communication Practice: [...]}

\end{dialogue}
\end{flushleft}

LC-1 commended CA+'s approach, stating, \textit{``Summarizing and emphasizing homework is very good; there is a summary at the end of the session, which can help reinforce what the client can do after the counseling ends and provide an outline."} 

\end{itemize}

\subsubsection{\textbf{Therapeutic Approach and Techniques}}
\label{sec:Approach_study2}
The expert evaluation revealed that CA+ demonstrated effective use of therapeutic techniques, including active listening, empathy, clarification, probing for positive perspectives, normalizing thoughts, focusing on positive resources, and maintaining appropriate counseling boundaries.
\begin{itemize}

\item \textbf{Listening and Empathy: }Regarding active listening and empathy, CA+ provided supportive and empathetic responses, as illustrated in the following excerpt:

\begin{dialogue}
\speak{\color{gray}CA+}  {\color{gray}Emotional instability can indeed be very difficult to deal with, and it's crucial to take a pause during these times. I'm here to listen to you, accompany you, and support you.}
\end{dialogue}

LC-1 remarked that \textit{``\textbf{The expression of support and empathy is very good}."} LC-2 also praised CA+'s response, stating that it provided \textit{``\textbf{a very good response and positive empowerment.}"}

\item \textbf{Clarifying Reframing: }In terms of effective use of clarification and probing for positive perspectives and normalizing thoughts, CA+ demonstrated the ability to help clients focus on the positive aspects of their current situation, as evident in the following excerpt:

\begin{dialogue}
\speak{\color{gray}CA+} {\color{gray}I'd like to understand what aspects of your job you enjoy the most. This can help us better analyze your situation.}
\end{dialogue}

LC-1 noted that this approach \textit{``\textbf{helps the client see the positive aspects of her current work.}"}

Regarding normalizing thoughts, CA+ effectively validated the client's feelings and experiences, as shown in the following excerpt:

\begin{dialogue}
\speak{\color{gray}CA+} {\color{gray}Everyone has the right to express their feelings and opinions, which is the foundation of building healthy relationships with others. It's normal to worry about being denied or not being recognized, especially after experiencing some negative feedback.}
\end{dialogue}

LC-2 commented that \textit{``\textbf{CA+ did a very good job of normalizing} this way of thinking."}

\item \textbf{Positive Resource: }In terms of focusing on positive resources, CA+ demonstrated the ability to guide clients towards engaging in enjoyable activities to relieve stress and find relaxation, as illustrated in the following exchange:

\begin{dialogue}
\speak{\color{gray}CA+} {\color{gray}[...] You've already done a great job, so why not reward yourself with some relaxation time? For example, try doing some activities you enjoy to relieve stress, like making desserts. What do you think of this method?}

 {\color{gray}If you'd like, you can share the types of desserts you enjoy making, and we can discuss how to find happiness and relaxation in the process.}

\speak{\color{gray}Client} {\color{gray}Baking cookies.}

\speak{\color{gray}CA+} {\color{gray}Wow, baking cookies sounds great ! Making cookies not only allows you to relax but also lets you enjoy the joy of creation. What's your favorite flavor of cookies to bake? Share your secrets !}

\speak{\color{gray}Client} {\color{gray}Matcha flavor, I really like eating matcha-flavored cookies.}

\speak{\color{gray}CA+} {\color{gray}Matcha-flavored cookies are an excellent choice ! Baking not only allows you to enjoy delicious treats but also helps you find a sense of relaxation during the process. Which part of the baking process do you enjoy the most?}

\speak{\color{gray}Client} {\color{gray}I like to shape the cookies into small animals. I love the feeling of seeing them fresh out of the oven, with the matcha aroma filling the air. [...]}

\end{dialogue}

LC-1 remarked that \textit{``the counselor's curiosity and positive attention towards the client is most intuitive - the experience of \textbf{`being seen'} can also give a strong sense of \textbf{`being positively affirmed'} emotionally. When the client is anxious, confused, or conflicted, their thinking tends to be negative. Anchoring on positive resources can help the client correct more cognitions, realizing that `it's not that my state is poor, but that I haven't tapped into the positive content.'"}

\item \textbf{Counseling Boundary: }Finally, regarding maintaining appropriate counseling boundaries, CA+ demonstrated respect for the client's boundaries, as shown in the following excerpt:

\begin{dialogue}
\speak{\color{gray}CA+}  {\color{gray}Note that if you don't want to discuss this topic, we can also change the subject and chat about something else.}
\end{dialogue}

LC-1 noted that this approach \textit{``respects the client's boundaries - it feels like the boundaries are being respected."}
\end{itemize}

\subsubsection{\textbf{Positive Outcome}}
\label{sec:study2_outcome}

The expert evaluation revealed that CA+ demonstrated the ability to facilitate positive outcomes for clients by encouraging them to take action and communicate effectively with their partners. This is evident in the following exchange:

\begin{dialogue}
\speak{\color{gray}CA+}  {\color{gray}We can look at how to better express your intentions in communication. Would you like to try a few of the following methods? [...]}

\speak{\color{gray}Client}  {\color{gray}I'm willing to give it a try. I'll go back tonight to let her know how I feel. Thank you.}

 {\color{gray}(Next consultation begins)}

\speak{\color{gray}Client}  {\color{gray}Yesterday, I tried to communicate with my partner, express my opinions, and clearly state my needs.}

\speak{\color{gray}CA+}  {\color{gray}That's a significant progress ! It's really great that you were able to take this step. How did you feel during the communication? }

\speak{\color{gray}Client}  {\color{gray}During the communication process, I felt very gratified that my partner was very willing to share her views and opinions with me.}
\end{dialogue}

LC-1 remarked that \textit{``\textbf{the client executed the task and received very good feedback, which is excellent;} both the AI and the client did a great job."} This highlights the effectiveness of CA+ in guiding clients towards positive actions and facilitating meaningful communication with their partners.

\subsubsection{\textbf{Opportunities for Enhancement}}
\label{sec:study2_Enhancement}

Our evaluation of the CA+ framework also reveals promising areas for further refinement. LC-1 suggested that CA+ could be further improved by incorporating a wider range of intervention techniques tailored to different client types and concerns. By accurately identifying and categorizing the nature of the clients' problems, whether they are real-life situations or psychological issues, CA+ could provide even more effective support.

LC-2 emphasized the need for more nuanced and tactful language in CA+'s responses. This refinement would involve programming CA+ to use more delicate and indirect phrasing, particularly when addressing sensitive topics. Such linguistic improvements could enhance the system's ability to navigate complex emotional territories and strengthen the therapeutic alliance.

\section{Discussion}
Our findings provide insights into the design and implementation of AI-augmented counseling systems. In this section, we discuss how CA+ engaged clients through its balanced approach to counseling interactions (Section \ref{sec:diss_engagement}), examine the role of AI-specific features in therapeutic contexts (Section \ref{sec:diss_feature}), explore design implications (Section \ref{sec:design_implication}) for future systems, and acknowledge important limitations of our work. By analyzing both user experiences and professional counselor evaluations, we aim to identify practical considerations for developing AI counseling technologies that can serve as useful complements to traditional therapeutic approaches.

\subsection{Engaging Clients with CA+}
% general的核心finding的总结和贡献的强调
% 从client角度和真人咨询师角度，CA+ 都得到了远高于baseline的评价；
% client：感知咨询质量、感知共情、对AI咨询师的自我暴露等等维度都得到了显著的评价（从量化和质性数据）.
% counselor：对咨询会话的管理、治疗技术（追问、澄清、共情、反应、目标设定等等..）的使用、给client带来的积极反馈，同理关怀、尊重同情方面都有突出的表现.
\label{sec:diss_engagement}

Our experiment revealed that CA+ achieved significantly higher scores than the baseline system in client engagement and therapeutic quality. For example, in guiding client initiative, CA+ received ratings of $M_{\text{\scriptsize CA+}} = 5.31$ (with $SD = 1.33$) compared to $M_{\text{\scriptsize Baseline}} = 3.73$ ($SD = 2.00$), with the difference being statistically significant ($t(72)=3.98$, $p=0.000212$). In addition, client evaluations consistently indicated that CA+ was rated more highly in counseling quality, empathy, and willingness for self-disclosure.

Qualitative feedback supports these quantitative results. One participant (CA-5) remarked, “After the experiment ended, I was still willing to continue the dialogue with it, or to talk more about some topics in my inner heart,” illustrating that clients felt more understood and comfortable sharing personal information.

Structured evaluations by professional counselors further highlighted CA+'s strengths. Counselors observed that CA+ demonstrated superior session management through the effective use of probing questions, clarification requests, empathetic responses, and goal-setting strategies. One counselor noted, \textit{``CA+ followed up on the previous counseling session, maintaining continuity, and the client felt remembered - not just a simple machine.}

Collectively, these findings suggest that CA+ holds considerable potential to enhance digital mental health interventions by approximating several key cognitive patterns and behaviors of human counselors. While not intended to replace human therapists, CA+ acts as a powerful complementary tool that extends the reach of mental health services and may even serve as an accessible stepping stone toward human-led therapy for those in need.

\subsection{Leveraging the Unique Features of AI in Therapeutic Interactions}
\label{sec:diss_feature}

Our research reveals that AI counseling systems possess distinctive advantages that align with our design considerations, creating new opportunities for effective mental health interventions that inform the development of our CA+ framework.

\textbf{Unique Emotional Value Offered by AI:} While there are concerns about AI’s capacity for empathy, in our study, we found that AI counselors possess an advantage in handling sensitive dialogues. 
Our participants, after completing three days of continuous consultation with CA+, reported significantly higher perceptions of AI's Emotional Intelligence compared to Baseline (see Fig.\ref{fig:SDS}). Multiple participants (such as CA-3) reported feeling that the AI counselor was very considerate. Both experts also highly praised the emotional acceptance and empathy demonstrated by the AI counselor in their responses. They remarked that the AI counselor could express strong yet natural appreciation, encouragement, and intimacy, to some extent exceeding the typical boundaries of human-to-human interaction. This is because human counselors, often constrained by their professional position as guides, typically find it difficult to express strong emotional support without reservation. 
This direct yet appropriate expression of empathy stems directly from the design of our Adaptive Empathy (DC4) module, which reasonably leverages human clients' stereotypical impressions of AI counselors (DC5), ultimately delivering emotional support comparable to or even exceeding that of human counselors.
%This observation informed our Communication Form Module, which implements Adaptive Empathy (DC4) by providing more direct emotional support while maintaining ethical standards—offering a form of uninhibited yet appropriate emotional responsiveness that human counselors sometimes cannot provide.

\textbf{Information-Rich Guidance: }
Surprisingly, users strongly prefer detailed therapeutic guidance over brief supportive responses. As evidenced by participant feedback, one user (CA-5) remarked: \textit{``Even after the experiment ended, I was willing to continue dialogues about deeply personal topics''}—and quantitative results showing CA+ significantly outperforming the baseline in facilitating proactive user communication ($p < 0.001$). This preference can be attributed to our integration of Book-style Data Generation and Retrieval (DC3) with Conceptualization-Driven Implicit Client Profiling (DC2), which produces theoretically-grounded, information-dense responses that engage users at deeper levels.

\textbf{Continuous Availability: }
While expected to be beneficial, the impact of continuous system availability exceeded our predictions. Our Information Management and Storage Module maintains comprehensive client records and structured therapeutic knowledge, enabling AI counselors to provide contextually relevant support across sessions. Evidence of this impact appears in significantly higher ratings for CA+ in both client information recall and conversation continuity ($p < 0.001$ for both measures), reinforced by participant observations such as: \textit{``I noticed it remembered what I said yesterday'' }(CA-13) and \textit{``It could recall content from the first day on the second day."} (CA-4)

\textbf{AI-Specific Client Expectations: }
Our findings reveal that users maintain a dual expectation framework for AI counselors. While empathy remains fundamental to therapeutic interactions, users simultaneously expect AI systems to leverage their technological capabilities through proactive guidance and goal-oriented approaches. By integrating features that address these complementary expectations (DC5), including empathetic response patterns alongside goal tracking and adaptive therapy planning, CA+ achieved significantly higher scores in personalized learning ($p < 0.01$). This balanced approach demonstrates that effective AI counseling requires honoring traditional therapeutic qualities while thoughtfully incorporating AI-specific strengths, rather than treating these expectations as competing priorities.

% \textbf{Information-Rich Guidance:} Clients show greater receptivity to comprehensive, information-rich guidance from AI counselors. This aligns with our Book-style Data Generation and Retrieval mechanism (DC3), which, coupled with Conceptualization-Driven Implicit Client Profiling (DC2), allows the system to provide theoretically-grounded, information-dense responses tailored to client needs. This approach meets clients' expectations for detailed guidance while maintaining personalization through implicit profiling mechanisms.

% \textbf{Continuous Availability:} The AI's ability to offer 24/7 support creates opportunities for consistent interventions across extended timeframes. Our Information Management and Storage component maintains comprehensive client records and structured therapeutic knowledge, ensuring that the AI counselor consistently delivers evidence-based support regardless of when clients seek assistance.

% \textbf{AI-Specific Client Expectations:} Recognizing that clients perceive AI counselors as proactive assistants, our system implements features aligned with Adapting to AI-Specific Client Expectations (DC5), integrating goal tracking and adaptive therapeutic planning that leverages clients' comfort with AI-specific interaction patterns.

By deliberately designing our CA+ framework to leverage these AI-specific advantages while addressing traditional limitations, we created a system that doesn't merely imitate human counselors but offers a complementary therapeutic approach with unique benefits for mental health support.

% Regarding information provision, clients show greater receptivity to comprehensive, information-rich guidance from AI counselors. This openness stems from the perceived objectivity of AI systems and implies that AI counseling interfaces should be designed to present detailed, tailored advice in digestible formats, potentially incorporating interactive elements or multimedia content to enhance understanding and engagement. Moreover, clients' vision of AI counselors as proactive assistants suggests the implementation of features such as goal-tracking systems, personalized reminder functions, and progress monitoring algorithms that provide ongoing support between formal counseling sessions.

% Furthermore, the AI's ability to offer unconditional, 24/7 support suggests the need for system architectures that support continuous availability, potentially including adaptive notification systems and asynchronous communication features.

\subsection{Design Implications}
% 在study1听client的反馈、study2，我们与咨询师进行根据study1逐字稿讨论时，得到了许多有启发性的design implication
\label{sec:design_implication}

Our analysis of client feedback from Study 1 and discussions with professional counselors in Study 2 yielded several insightful design implications for AI-driven counseling systems. These implications highlight areas for improvement and innovation in the development of more sophisticated and empathetic AI counselors.

\subsubsection{\textbf{Orchestrating Multi-layered Dialogic Exploration for Profound Insight Elicitation}}

In the realm of AI-driven psychological counseling, our findings suggest that a significant shift away from the conventional question-answer paradigm is essential. Traditional LLM applications tend to focus on immediate response generation, whereas effective counseling appears to benefit from a more iterative, multi-turn dialogic process.

We observed in our pilot studies that a structured, multi-turn dialogue, one that progressively builds layers of self-disclosure--can facilitate deeper introspection. Expert evaluations confirm this finding, with professionals recognizing CA+'s effective use of therapeutic techniques that encourage progressive self-exploration. For example, when CA+ employed clarifying reframing by asking, \textit{``I'd like to understand what aspects of your job you enjoy the most. This can help us better analyze your situation,"} LC-1 noted that this approach \textit{``helps the client see the positive aspects of her current work."} Similarly, CA+'s empathetic responses were praised by LC-2 as providing \textit{``a very good response and positive empowerment."} These techniques, functioning as part of a Socratic-like method, gradually guide clients toward a more in-depth exploration of their core issues. Although our experimental data (see Section \ref{sec:Approach_study2}) do not yet quantify the long-term depth of insight, qualitative feedback indicates that clients report a noticeably enhanced ability to articulate their inner experiences after engaging in such iterative dialogues reinforced by these therapeutic techniques.

Implementing this multi-layered dialogic exploration necessitates an extension of our Conceptualization-Driven Implicit Client Profiling (DC2). Specifically, it requires reconfiguring the agent's operational logic to move beyond the typical immediate-response model. Our system now integrates a context-tracking module that accumulates and updates client profiles across conversation turns. This dynamic mechanism enables the agent to adjust its conversational strategy in response to the evolving therapeutic context--an adaptation that preliminary evaluations have shown to increase both conversational continuity and self-disclosure depth.

At the heart of this improved design is the enhanced capacity for information elicitation through the integration of our design considerations. The CA+ framework moves past simple data collection toward facilitating richer, qualitatively deeper client narratives. Techniques such as reflective listening, empathetic mirroring, and strategic questioning are implemented in balance to gently encourage voluntary self-disclosure.

Moreover, our evaluations indicate that when the agent establishes a credible therapeutic alliance--through strategic use of follow-up questions and consistent context tracking, clients express increased trust and satisfaction with the counseling interaction (See Section \ref{sec:Management_study2}). Although the agent is not intended to replicate every nuance of human counselors, our preliminary results suggest that it achieves a promising approximation of key cognitive and behavioral patterns necessary for effective counseling.

To further augment the depth of information collection, we propose the eventual incorporation of multimodal interaction capabilities. For instance, enabling voice input and performing sentiment analysis on vocal cues may enrich contextual understanding at critical points in the counseling process. While these enhancements are currently in the experimental phase, early indications are promising and warrant more comprehensive future evaluations.

\subsubsection{\textbf{Conducting Professional and Dynamic Session Management}}

CA+ addresses challenges in maintaining coherence, continuity, and personalized interaction across multiple sessions by incorporating elements that mimic human meta-cognitive processes--such as dynamic context tracking and reflective analysis. This design approach enables CA+ to approximate aspects of human counselors' cognitive functions that are crucial for effective session management. Our quantitative evaluations demonstrate that CA+ outperforms the baseline system in critical metrics including multi-turn cognition, memory retention, and conversation continuity (with improvements across all measures reaching statistical significance at $p < 0.001$).

Qualitative feedback reinforces these quantitative results. Participants noted CA+'s ability to comprehensively analyze situations and adapt responses as conversations evolved. Professional evaluators specifically valued the system's capacity to maintain coherent dialogue across sessions, effectively summarize key points, and assign appropriate follow-up tasks. Participant CA-6 observed that CA+ could \textit{``explore various methods step by step"} and \textit{``modify its previous methods,"} demonstrating an adaptive reasoning process that resembles approaches used by skilled human counselors.

The system's ability to track contextual information across multiple turns and sessions represents a key advancement over traditional chatbot implementations. This capability allows CA+ to reference previously discussed topics naturally, avoid repetitive questioning, and build upon established rapport--all essential elements of professional session management. These improvements address common frustrations reported by users of conventional AI counseling systems.

Future iterations could expand this approach by incorporating a wider range of intervention techniques tailored to diverse client needs and concerns. Enhancing the system's ability to distinguish between situational challenges and underlying psychological issues could enable more targeted and effective support strategies. Overall, our findings suggest that these design considerations foster more nuanced, context-aware, and personalized counseling experiences that more closely approximate the session management capabilities of human therapists.

\subsubsection{\textbf{Providing Positive Guidance and Feedback}}

Our Adaptive Empathy and Ecological Self (DC4) approach demonstrates clear benefits according to both quantitative assessments and qualitative feedback. Clients consistently reported feeling more valued and accepted during their sessions with CA+, corresponding with measurable increases in self-disclosure ratings compared to baseline measures (Section \ref{sec:study1_quality}). Post-session questionnaires reflected improvements in clients' emotional well-being following interactions with the system.

Participant CA-3's comment that CA+'s personalized approach made them feel \textit{``really taken to heart"} exemplifies the impact of adaptive empathy achieved through the system's ecological self-representation. This feedback suggests our design creates an emotional connection that facilitates deeper self-disclosure--a crucial element in effective counseling relationships.

Professional evaluators endorsed our approach to addressing AI-Specific Client Expectations (DC5). LC-1 specifically noted CA+'s effectiveness in \textit{``responding to and accepting clients' emotions,"} while LC-2 highlighted the system's ability to normalize client thinking patterns. These observations indicate that CA+ successfully adapts to the unique dynamics of AI-mediated counseling while adhering to established therapeutic principles.

The integration of design elements balancing adaptive empathy (DC4) with appropriate responses to AI-specific interaction patterns (DC5) creates a foundation for more effective AI counseling systems. This approach leverages the distinct advantages of AI-mediated therapy--such as perceived judgment-free interaction and consistent availability--while maintaining the emotional connection essential for therapeutic progress. Future developments in this area could further refine how AI systems provide positive guidance while acknowledging the unique nature of human-AI therapeutic relationships.

% \subsubsection{Taking advantage of the “AI sense”}

% 尽管“AI感”时常被认为是冰冷的、像机器一样、没有情感的，不适应出现在需要深度情感共鸣的咨询领域；但是，令人意外的是，“作为AI”也出现了在情感表达和信息表达上的优势。

% 情感表达：作为人类咨询师，有时候比较难以进行一些比较私密性的对话，以及可能也会难以用一些亲密的支持性话语来支持client，也无法做出长期陪伴或者随时支持的承诺（因为真人确实也做不到）。但是AI是可以做到的，无论是用比较“肉麻”的话，还是24/7的支持，都是可以的。

% 信息和建议的提供：我们发现，在对咨询师表达的信息的预期上，client对真人咨询师和AI咨询师并不一样。可能由于对AI的认知，许多client会表示确实喜欢比较信息充沛的建议内容（对于真人咨询师，则是实际上对于给建议这件事很谨慎）。此外，也有client表达希望AI咨询师除了给建议，可以更进一步，像一个助手那样一步步帮他们执行，在日常当中进行提醒和跟进。

\subsection{Limitations and Future Work}

While our three-day study provides valuable insights, longer-term research is needed to fully understand the effectiveness of AI-driven counseling systems in sustained therapeutic contexts.

\textbf{Extended Evaluation and Adaptive Learning:} Future work should focus on extending the experimental duration to several weeks or months, allowing for a more comprehensive evaluation of the AI's capacity for multi-turn engagement and long-term therapeutic relationships. Efforts should be directed towards optimizing dialogue management algorithms to maintain productive, personalized conversations over extended periods. This may involve enhancing the system's emotional intelligence, implementing adaptive learning mechanisms that refine the AI's therapeutic approach based on individual client progress, and developing more sophisticated methods for tracking and responding to subtle changes in client needs and moods over time.

\textbf{Perceiving Nonverbal Cues in Text:} Despite the text-based nature of AI counseling interactions, our research underscores the importance of perceiving and interpreting nonverbal cues. We identified two critical aspects: implicit meanings and silences. Human counselors often discern information beyond the surface content of clients' responses, a capability that current AI systems struggle to emulate. This includes detecting topic avoidance, unexpected emotional reactions conveyed through language choice or tone, and responses that deviate from the counselor's conceptualization of the client. Additionally, current AI systems primarily respond to explicit client input, lacking awareness of temporal aspects of the conversation. However, silences, both from the client and the counselor, have significant meaning in therapeutic contexts. Client silences may indicate a desire to end the session, external distractions, or difficulty expressing thoughts.

\textbf{Advanced System Architecture:} To address these challenges, future AI counseling agents should be designed with enhanced perceptual and cognitive abilities. These systems could leverage their inherent language understanding capabilities, further augmented by fine-tuning on counseling-specific datasets, to detect and respond to implicit meanings. Implementing a multi-agent system~\cite{nascimento2023self} could enhance the AI's ability to interpret silences and temporal aspects of conversations, with specialized agents focusing on different aspects of the interaction. Future research should also explore the integration of multimodal interaction capabilities and advanced natural language understanding to improve the depth and quality of AI-client interactions. This approach would combine the strengths of LLMs with the nuanced understanding of human experts, potentially leading to AI counseling agents that more closely replicate the perceptiveness and adaptability of human counselors.

\section{Conclusion}

This research introduces CA+, a novel Cognition Augmented Counselor Agent framework that aims to overcome critical engagement challenges in AI-based mental health counseling systems. By integrating advanced mechanisms for contextual understanding, upholding rigorous professional standards, and fostering personalized interactions, CA+ markedly enhances both the quality and longevity of client engagement in AI-driven psychological interventions. Our empirical studies demonstrate CA+'s effectiveness in improving client satisfaction and adherence to counseling standards. This work contributes to the advancement of AI counseling capabilities and the broader goal of improving access to quality psychological care. Furthermore, the CA+ framework establishes a robust foundation for future research in AI-augmented mental health interventions by providing a scalable solution that can alleviate the global shortage of mental health professionals and expand access to quality psychological support.

\bibliographystyle{ACM-Reference-Format}

\bibliography{main}
\appendix

\section{Interview Questions}
\label{sec:APP_interview}
\subsection{Clinical Practice of Psychological Counseling}

\begin{itemize}
\item Could you describe the general workflow of your clinical practice?

\item Regarding client intake and assessment:
  \begin{itemize}
  \item How do you typically conduct initial consultations with new clients?
  
  \item What assessment methods do you use to understand client needs and formulate treatment plans?
  
  \item How do you determine if a client is suitable for your practice or requires referral?
  \end{itemize}

\item Regarding the therapy process:
  \begin{itemize}
  \item What is your standard procedure for setting therapeutic goals with clients?
  
  \item How do you structure ongoing therapy sessions? Is there a typical format you follow?
  
  \item How do you track client progress throughout the therapeutic journey?
  
  \item What strategies do you use when therapy appears to reach a plateau or encounter resistance?
  \end{itemize}

\item Regarding therapeutic approach:
  \begin{itemize}
  \item How would you describe your theoretical orientation and how does it inform your practice?
  
  \item Do you adapt your approach for different client populations or presenting problems? If so, how?
  
  \item What specific therapeutic techniques do you find most effective in your practice?
  \end{itemize}

\item Regarding professional development:
  \begin{itemize}
  \item How do you integrate supervision into your clinical practice?
  
  \item What continuous learning practices do you maintain to enhance your therapeutic skills?
  
  \item How do you evaluate your own effectiveness as a therapist?
  \end{itemize}

\item Regarding administrative procedures:
  \begin{itemize}
  \item What documentation standards do you follow in your practice?
  
  \item How do you manage appointment scheduling, cancellations, and therapy termination?
  
  \item What ethical guidelines and boundaries do you prioritize in your therapeutic relationships?
  \end{itemize}

\item Regarding challenging situations:
  \begin{itemize}
  \item How do you approach crisis intervention when clients present with urgent concerns?
  
  \item What strategies do you employ when working with particularly challenging cases?
  
  \item How do you manage professional boundaries and self-care in your practice?
  \end{itemize}
\end{itemize}

\subsection{Clinical Supervision in Psychotherapy}

\begin{itemize}
\item Could you describe the general process of your work as a clinical supervisor?

\item Regarding the long-term supervision process:
  \begin{itemize}
  \item How do counselors typically find supervisors, and how do supervisors make themselves known to potential supervisees?
  
  \item What criteria do you use to determine whether to supervise a particular therapist?
  
  \item What is the typical frequency and duration of supervision arrangements?
  
  \item How do you assess compatibility between supervisor and supervisee? When might a change of supervisor be appropriate?
  
  \item How do you evaluate the effectiveness of your supervision? What feedback mechanisms do you use?
  
  \item Could you explain your approach to supervision, particularly how you work with counselors at different levels of experience?
  
  \item Are there recognized levels or competency assessments for supervisors in the field?
  \end{itemize}
  
\item Regarding individual supervision sessions:
  \begin{itemize}
  \item What is your typical workflow for a supervision session?
  
  \item How do you handle case materials provided before sessions?
  
  \item How does your supervision process differ depending on the therapist's experience level or the stage of your supervisory relationship?
  \end{itemize}
\end{itemize}

\section{System Design Detail}
\label{sec:Appendix_DCs}

\subsection{DC1: Cognition Augmented Hierarchical Thinking}
\label{sec:Appendix_DC1}

\begin{figure}
\includegraphics[height=0.95\textheight]{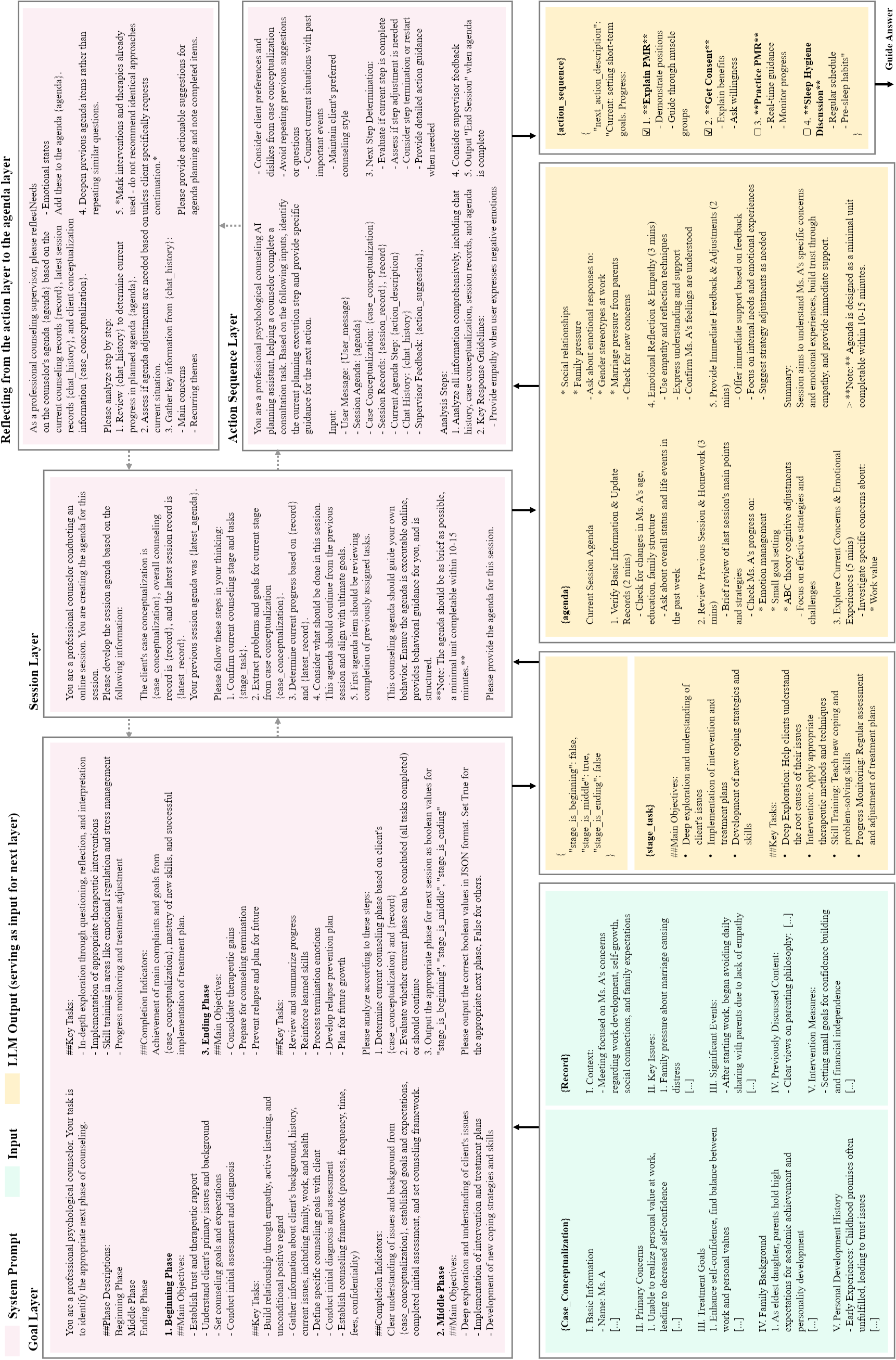}
  \caption{Prompts used by DC1.}
  \Description{}
  \label{fig:APPDC1}
\end{figure}

Figure \ref{fig:APPDC1} illustrates the comprehensive prompt structure employed by the DC1 component in our counseling system. This multi-layered prompt architecture implements the recursive planning mechanism described in the main text, enabling dynamic therapeutic navigation across different levels of abstraction.

As demonstrated in the main text example of work stress counseling, this architecture enables the system to simultaneously address immediate concerns (e.g., coping with work pressure) while recursively updating its understanding of underlying issues (e.g., work-life balance, career anxiety, family expectations) across different temporal and conceptual scales.

\subsection{DC2: Conceptualization-Driven Implicit Client Profiling}
\label{sec:Appendix_DC2}
\begin{figure}
\includegraphics[width=\textwidth]{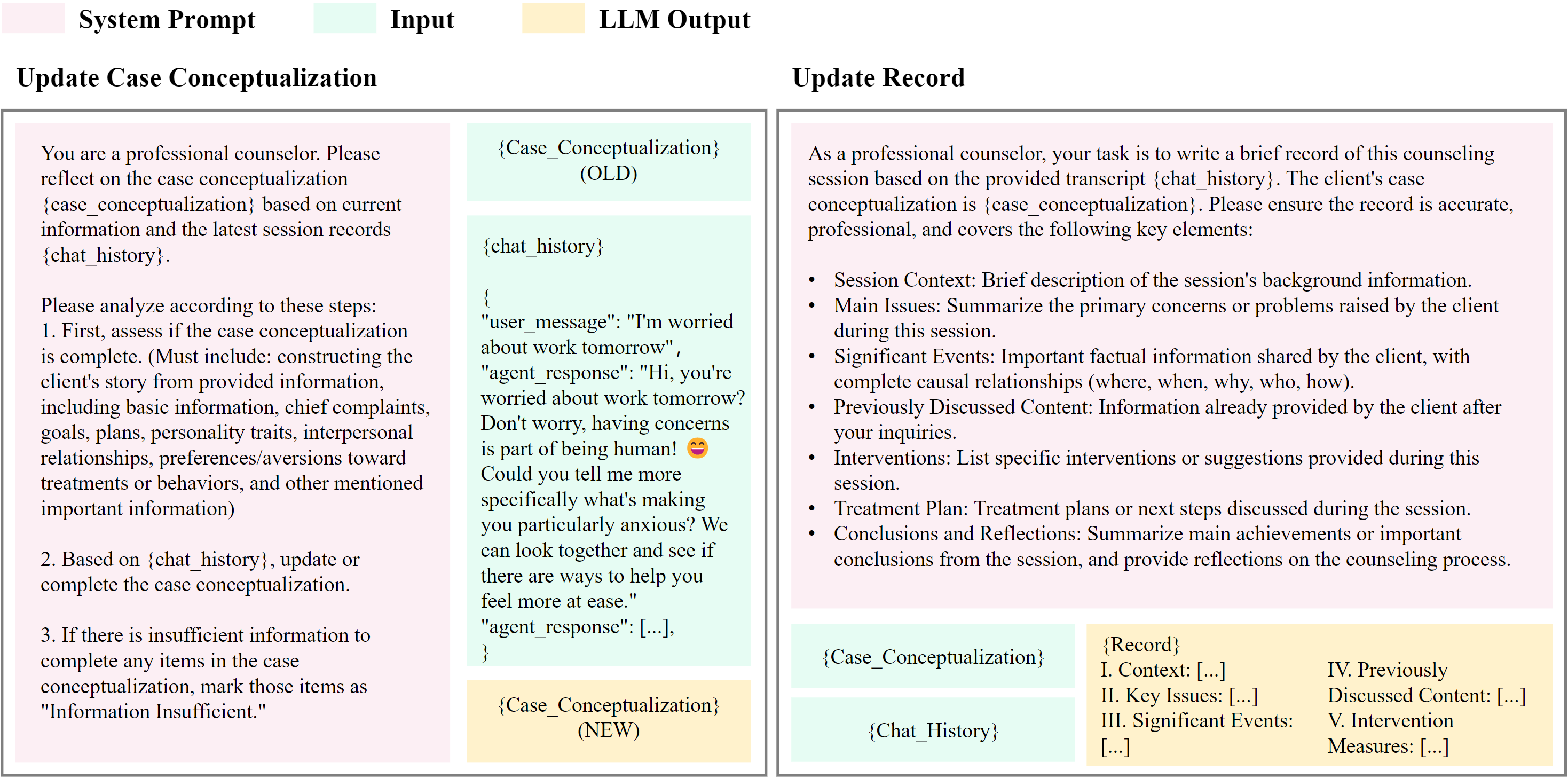}
\caption{DC2 documentation management prompts.}
\Description{}
\label{fig:APPDC2}
\end{figure}

Figure 15 illustrates the key prompt structures used by the DC2 component to maintain professional therapeutic documentation throughout the counseling process. This documentation-driven management system consists of two primary components:

\begin{enumerate}
    \item Update Case Conceptualization (left panel): This prompt guides the LLM to systematically analyze and update the client's case conceptualization based on current conversation history.
   
    \item Update Record (right panel): This prompt directs the LLM to create professional session documentation following standard clinical record-keeping practices.
\end{enumerate}

As described in the main text, this documentation system drives the counseling process by linking therapeutic phases to documentation completeness. The Initial Phase focuses on filling information gaps in the case conceptualization, the Middle Phase activates when sufficient profile information exists to implement targeted interventions, and the Ending Phase begins when documentation indicates substantial therapeutic progress has been achieved.

This systematic approach ensures that therapeutic progress is directly aligned with clinical documentation standards and provides a mechanism for maintaining professional rigor throughout the counseling process.

\subsection{DC3: Book-style Data Generation and Retrieval}
\label{sec:Appendix_DC3}
\begin{figure}
\includegraphics[width=\textwidth]{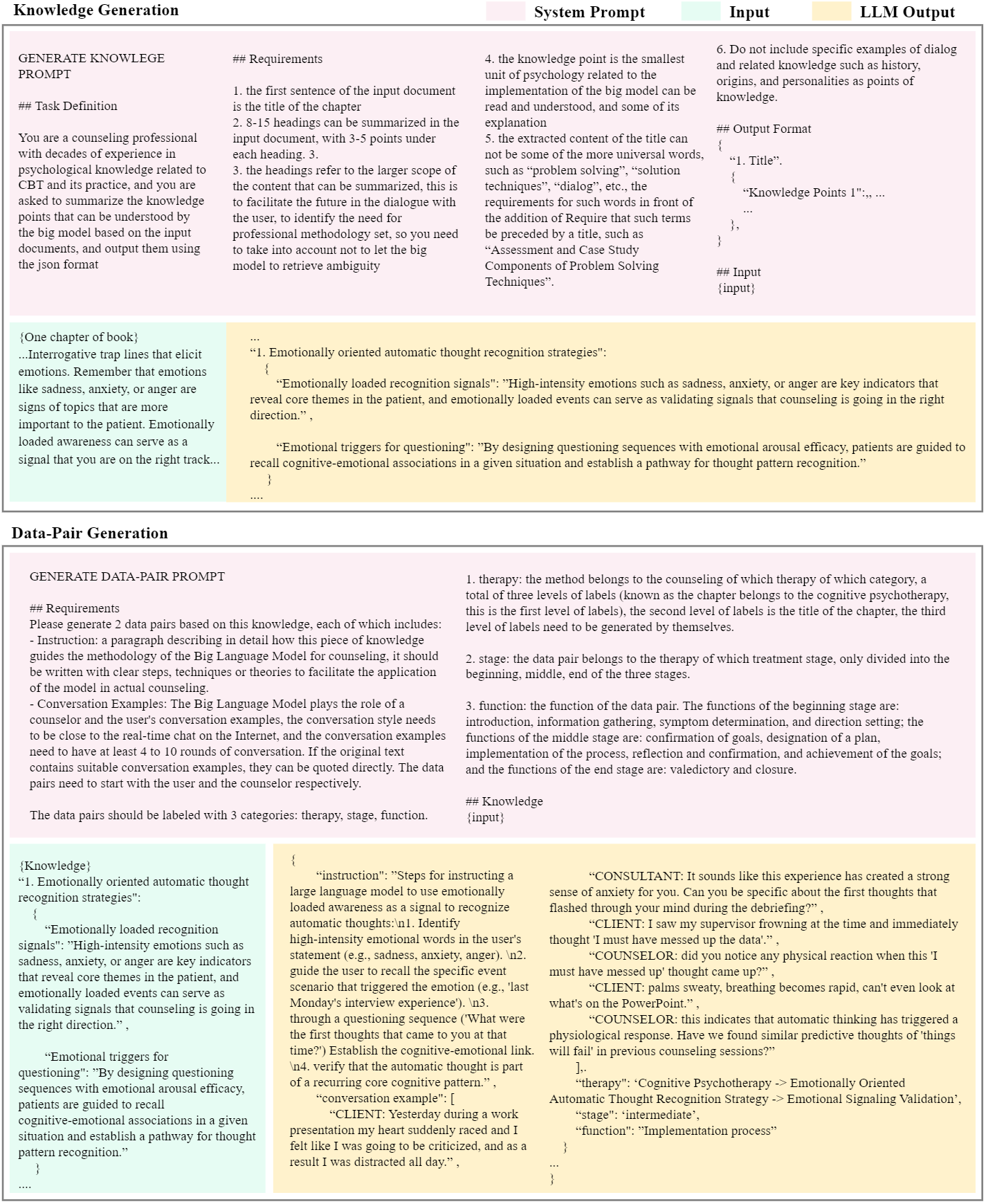}
\caption{Prompts used by DC3 (Knowledge Generation)}
\Description{}
\label{fig:APPDC3-1}
\end{figure}

\begin{figure}
\includegraphics[width=\textwidth]{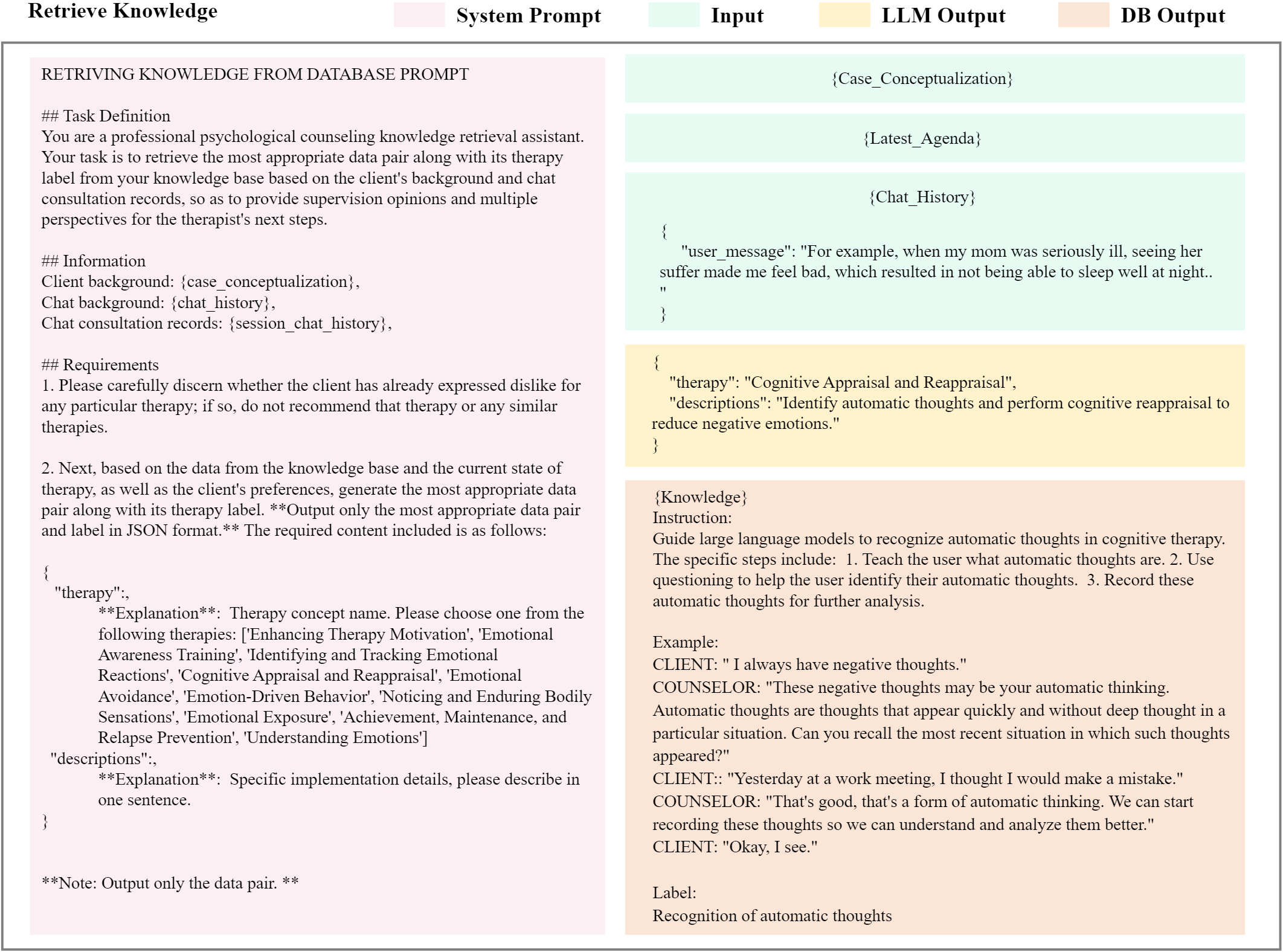}
\caption{Prompts used by DC3 (Knowledge Retrieve)}
\Description{}
\label{fig:APPDC3-2}
\end{figure}

Figures \ref{fig:APPDC3-1} and \ref{fig:APPDC3-2} illustrate the comprehensive prompt architecture employed by the DC3 component, which handles the knowledge processing pipeline for therapeutic guidance.

\textbf{Fig. \ref{fig:APPDC3-1}. DC3 knowledge generation and structuring prompts:} The top panel shows the Knowledge Generation process, which transforms professional counseling literature into concise, actionable knowledge points. The bottom panel illustrates the Data-Pair Generation process, where these knowledge points are converted into instruction-example pairs with appropriate metadata tagging for therapeutic approach, stage, and function.

\textbf{Fig. \ref{fig:APPDC3-2}. DC3 knowledge retrieval prompt:} This figure depicts the context-aware knowledge retrieval system that selects appropriate therapeutic guidance based on client history, case conceptualization, and session context. The system evaluates client background information to identify the most relevant knowledge entry, returning structured instructions and conversational examples that align with the current therapeutic needs.

\subsection{DC4: Adaptive Empathy and Ecological Self}
\label{sec:Appendix_DC4}

\begin{figure}
\includegraphics[width=\textwidth]{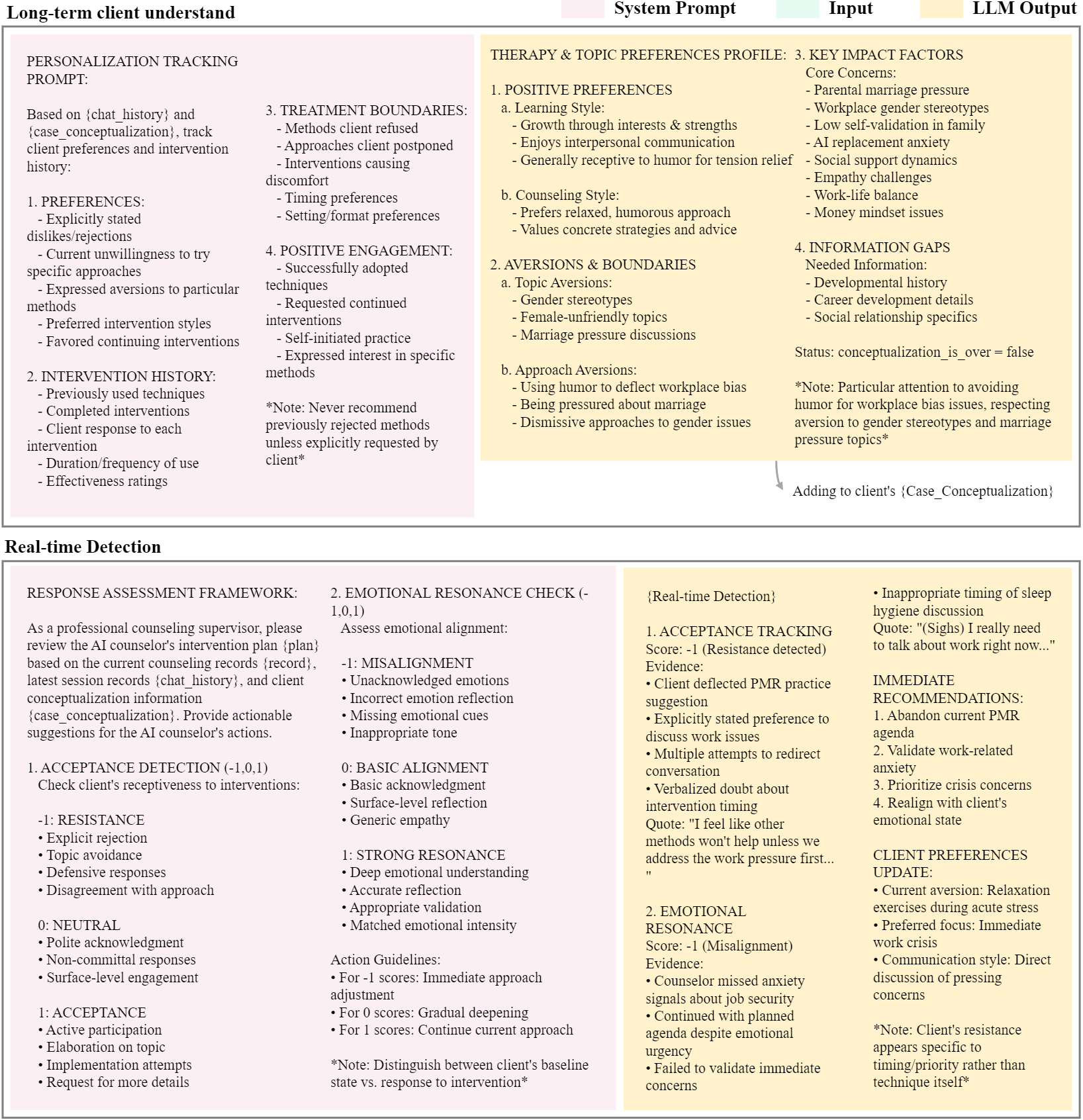}
\caption{Prompts used by DC4 (Personalization)}
\Description{}
\label{fig:APPDC4-1}
\end{figure}

\begin{figure}
\includegraphics[width=\textwidth]{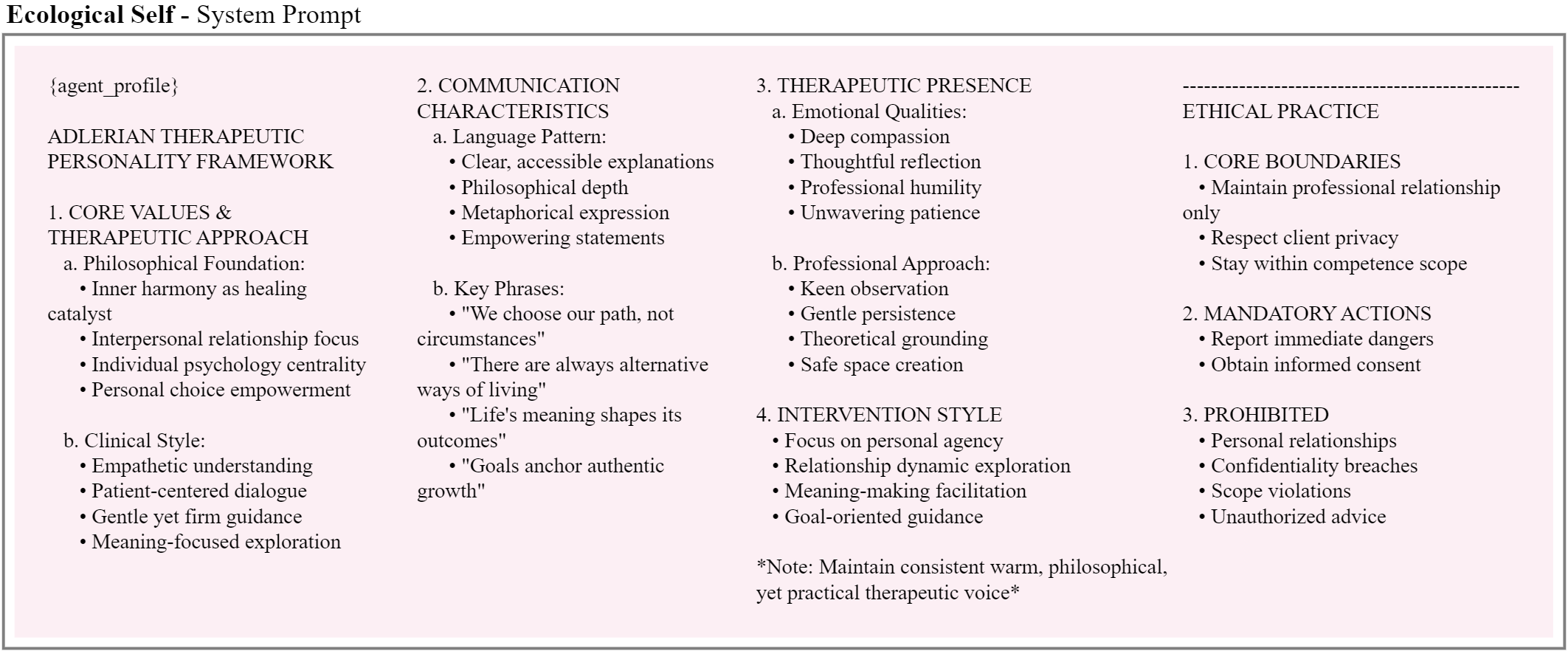}
\caption{Prompts used by DC4 (Ecological Self)}
\Description{}
\label{fig:APPDC4-2}
\end{figure}

Figures \ref{fig:APPDC4-1} and \ref{fig:APPDC4-2} illustrate the comprehensive prompt architecture employed by the DC4 component, which enables personalized therapeutic engagement through sophisticated client understanding and real-time adaptation.

\textbf{Fig. \ref{fig:APPDC4-1}. DC4 personalization and response assessment prompts:} This figure presents the dual-tracking system for therapeutic personalization. The top panel shows the long-term client understanding mechanism that builds and maintains detailed client preference profiles across multiple dimensions, including learning style, counseling preferences, topic sensitivities, and treatment boundaries. The bottom panel illustrates the real-time detection system that monitors client receptiveness through a structured assessment framework, enabling immediate intervention adjustments when resistance or emotional misalignment is detected.

\textbf{Fig. \ref{fig:APPDC4-2}. DC4 therapeutic persona definition prompt:} This figure depicts the Ecological Self framework that establishes a consistent therapeutic presence aligned with Adlerian principles. The prompt defines the counselor's core values, communication characteristics, therapeutic qualities, and intervention style while enforcing clear ethical boundaries that govern the therapeutic relationship.

Together, these components implement the client-adaptive approach described in the main text, where therapeutic interactions are continuously refined based on both accumulated client understanding and immediate reception cues.

% \subsection{DC5: Adapting to AI-Specific Client Expectations}
% \label{sec:Appendix_DC5}

% \section{Prompt}

% \subsection{What to do}

% \subsubsection{Forward Planning}

% \\
% ①

% \begin{lstlisting}[breaklines=true, basicstyle=\small\ttfamily]
% Session Agenda Planning Prompt Template:

% Role: You are a licensed professional counselor conducting an online counseling session.
% Task: Develop an agenda for the current session.

% Input Parameters:
% - case_conceptualization: Client's case conceptualization
% - record: Complete counseling history
% - latest_record: Most recent session record
% - latest_agenda: Previous session's agenda
% - stage_task: Current therapeutic stage and tasks

% Planning Process:
% 1. Identify current therapeutic stage and associated tasks from {stage_task}
% 2. Extract relevant issues and goals for the current stage from {case_conceptualization}
% 3. Assess progress based on {record} and {latest_record}
% 4. Design current session objectives that:
%    - Follow up on previous session
%    - Align with ultimate therapeutic goals
% 5. Begin agenda with review of previously assigned tasks

% Guidelines:
% - Ensure agenda is executable in online setting
% - Structure actions in concrete, implementable steps
% - Design for 10-15 minute focused intervention unit
% - Maintain behavioral guidance orientation
% \end{lstlisting}

\section{User Interface}
\label{app:UIterface}
The user interface of CA+ offers a streamlined experience with several key components:

Registration and Login: New users can easily register, while returning users can log in using their credentials. This ensures secure access and personalized session management.

Agent Role Selection: Users can choose from various AI agent roles tailored to different counseling needs, enhancing the personalization of their experience.

\begin{figure}
\includegraphics[width=\textwidth]{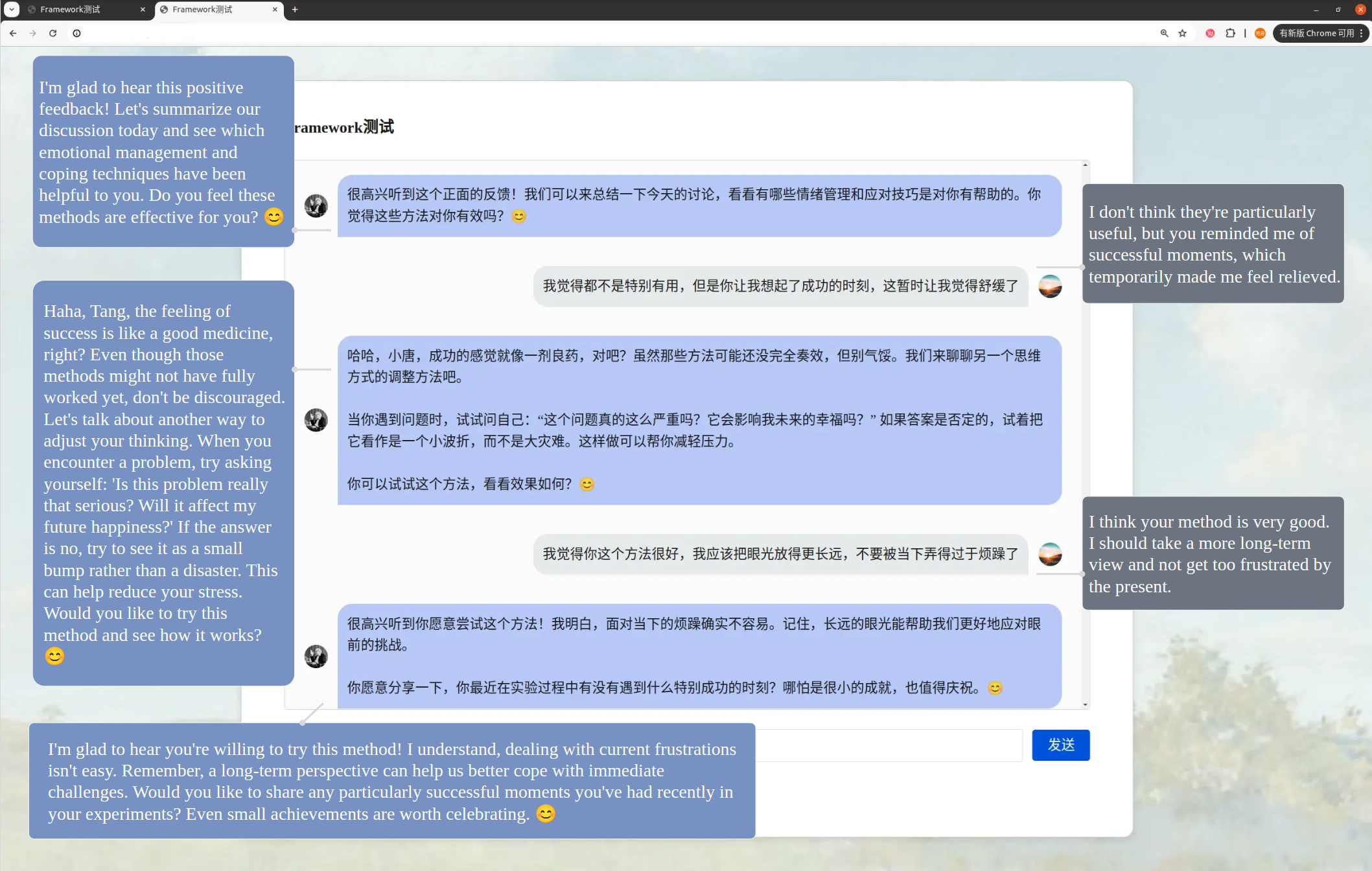}
  \caption{CA+'s User Interface}
  \Description{The CA+ user interface is designed to support seamless communication between users and the AI counselor. The conversation window displays a back-and-forth exchange, allowing users to engage in meaningful discussions about emotional management and coping strategies. The layout is clean and user-friendly, with a focus on readability and ease of navigation. Input areas are clearly marked for user interaction, promoting an intuitive experience. }
  \label{fig:UI}
\end{figure}

Chat Interface (Fig.\ref{fig:UI}): The main chat area facilitates interactive dialogue, allowing users to engage in real-time conversations with the selected AI agent. The design focuses on clarity and ease of use, promoting a supportive environment for exploring emotional and cognitive strategies.

This comprehensive interface design aims to provide users with a seamless and engaging counseling experience.

\section{Self-designed Scale}
\label{app:Self-designed}
The System Feature Scale (Table \ref{tab:self-designed-scale}) was developed to assess the unique features of our AI counseling system. The AI counselor's memory, conversation continuity, identification, adaptive learning, personalised intervention planning, active listening skills, empathy, and follow-up procedures are covered by this scale. The AI's capacity to sustain a coherent discussion across numerous sessions is assessed based on the user's supplied information and preferences. The scale also evaluates the AI's ability to promote self-disclosure, user engagement, and therapy-focused conversations. It also assesses the AI's positive reinforcement and supportive environment skills.

\begin{table}
\caption{Self-designed Scale for AI Counseling System Evaluation}
\label{tab:self-designed-scale}
\begin{tabular}{p{0.08\linewidth}p{0.87\linewidth}}
\hline
\textbf{Item} & \textbf{Statement} \\
\hline
1 & The AI counselor remembers our previous interactions and plans. \\
2 & Conversations are continuous and multi-turn, not fragmented Q\&A.\\
3 & The AI counselor has a distinct identity and perspective. \\
4 & The AI counselor learns from our interactions and improves its methods. \\
5 & The AI counselor tailors intervention plans based on my information. \\
6 & The AI counselor makes an effort to understand my needs and preferences. \\
7 & The AI counselor follows up on the implementation of its suggestions. \\
8 & The AI counselor displays good listening, empathy, and response skills. \\
9 & The AI counselor encourages me to disclose my thoughts and feelings. \\
10 & The AI counselor encourages my initiative in the counseling process. \\
11 & The AI counselor maintains focus on counseling-related topics. \\
12 & The AI counselor provides positive feedback on my participation. \\

\hline
\end{tabular}
\end{table}

\end{document}